\documentclass[11pt]{article}
\pdfoutput=1
\usepackage{jheppub}
\usepackage[usenames,dvipsnames,table]{xcolor}
\usepackage{graphicx,amsmath,amssymb,amsthm,multirow,array,bm,bbm,esint,physics}
\usepackage[mathscr]{eucal}
\usepackage{epsf,amsfonts}
\usepackage[normalem]{ulem}
\usepackage{slashed}
\usepackage[numbers,sort&compress]{natbib}
\usepackage{minibox}
\usepackage{booktabs}
\usepackage{pdflscape}
\usepackage{comment}
\usepackage{tikz,pgf}
\usepackage{enumerate}

\DeclareSymbolFontAlphabet{\mathbb}{AMSb}%

\usepackage[dvipsnames]{xcolor}

\usepackage{mathtools,upgreek}

\usepackage{slashed}

\def \sgn{\mbox{sgn\,}}

\def \Tr{\mbox{Tr\,}}

\def \tr{\mbox{tr\,}}

\def\+{{+\!\!\!+}}		

\newcommand{\cF}{\mathcal{F} }

\newcommand{\cH}{\mathcal{H} }

\DeclarePairedDelimiter{\absv}{\lvert}{\rvert}    
\DeclarePairedDelimiter{\sets}{\lbrace}{\rbrace}  
\DeclarePairedDelimiter{\pts}{(}{)}              
\DeclarePairedDelimiter{\ketv}{\lvert}{\rangle}   
\DeclarePairedDelimiter{\sq}{\lbrack}{\rbrack}   

\DeclarePairedDelimiterX\Sets[1]\{\}{
	
	#1
}
\DeclarePairedDelimiterX{\inner}[2]{\langle}{\rangle}{
	#1\,\delimsize\vert\,\mathopen{}#2
}
\DeclarePairedDelimiterX{\innerx}[2]{\langle}{\rangle}{
	#1\,,\,#2
}
\DeclarePairedDelimiterX{\braketv}[3]{\langle}{\rangle}{
	#1\delimsize\vert\mathopen{}#2\delimsize\vert\mathopen{}#3
}

\allowdisplaybreaks[4] 

\usepackage{cleveref}
\usepackage{tensor}
\makeatletter
\AddToHook{cmd/appendix/before}{\def\cref@section@alias{appendix}\def\cref@subsection@alias{appendix}}
\makeatother

\def\RD#1{{\color{blue}{\bf RdK: 1}}}

\title{Beyond the Fock Vacuum in Finite-$\bm N$ Bilocal Algebras of Vector Models}

\author[a,b]{Byoungjoon Ahn,}
\author[c,d]{Robert de Mello Koch,}
\author[a]{Vatsal Garg,}
\author[e,f]{Antal Jevicki,}
\author[a,b,g]{ and Junggi Yoon}
\affiliation[a]{Department of Physics, College of Science, Kyung Hee University, Seoul 02447, Republic of Korea}
\affiliation[b]{Research Institute for Basic Sciences, Kyung Hee University, Seoul 02447, Republic of Korea}
\affiliation[c]{School of Science, Huzhou University, Huzhou 313000, China}
\affiliation[d]{Mandelstam Institute for Theoretical Physics, School of Physics, University of the Witwatersrand, Private Bag 3, Wits 2050, South Africa}
\affiliation[e]{Department of Physics, Brown University,
	182 Hope Street, Providence, RI 02912, United States}
\affiliation[f]{Brown Center for Theoretical Physics and Innovation, Brown University,
	340 Brook Street, Providence, RI 02912, United States}
\affiliation[g]{International Research Institute for Quantum Materials, Kyung Hee University, Seoul 02447, Republic of Korea}
\date{September 2026}
\abstract{
	We discuss the finite-$N$ Hilbert space of bilocal operators that underlies bilocal holography of vector models, beyond the sector built on the Fock vacuum. If the microscopic oscillator Fock space carries a multiplicity-free dual action of a compact color group $G$ and a bilocal algebra, Howe duality implies that projection to the derived subgroup $[G,G]$ selects precisely the characters of the abelianization $G^{\rm ab}=G/[G,G]$ that occur in the Fock space. Thus $O(N)/SO(N)=\mathbb Z_2$ gives two sectors, $U(N)/SU(N)=U(1)$ gives integer determinant-charge sectors, while $Sp(N)$ is perfect and gives no additional sector; Bose/Fermi statistics and the number of modes determine which sectors are actually realized. We construct the epsilon and determinant backgrounds that generate these sectors, together with their finite-$N$ constraints and Casimirs, and find that the two bosonic $O(N)$ sectors are not distinguished by any Casimir. Using group-integral projectors, we derive explicit sector-resolved characters for the $O(N)$ and $U(N)$ theories. Around each new background we solve the constraints with auxiliary oscillators, which form a Fermi sea for bosonic matter. In the free vector model on $\mathbb R\times S^2$ the new sectors consist of baryonic primaries; we determine their energies, degeneracies and spin content exactly at finite $N$, together with their lowest energy at fixed angular momentum.
}

\emailAdd{bjahn123@khu.ac.kr}
\emailAdd{robert.demellokoch@gmail.com}
\emailAdd{vatsal\_garg@khu.ac.kr}
\emailAdd{Antal\_Jevicki@brown.edu}
\emailAdd{junggi.yoon@khu.ac.kr}

\begin{document}
	\maketitle
	
	\section{Introduction}
	\label{sec: introduction}
	
	The bilocal description of vector models~\cite{deMelloKoch:2010wdf,deMelloKoch:2014mos,Jevicki:2015sla,deMelloKoch:2018ivk} is naturally formulated in terms of operators which are invariant under a color group $G$.  At finite $N$ these bilocals are an overcomplete set of generators and obey non-trivial relations~\cite{deMelloKoch:2025ngs,deMelloKoch:2025cec,Jevicki:2012fh}.  Correspondingly, the physical Hilbert space is not an arbitrary Fock space of bilocals: it is a distinguished irreducible representation of the algebra generated by the bilocal operators.  The oscillator realization makes both statements concrete.
	
	Our starting point is a simple observation:  if one imposes $SO(N)$ rather than $O(N)$ invariance, Levi--Civita epsilon states appear as a new background in addition to the ordinary $O(N)$ singlet vacuum.  Likewise, imposing $SU(N)$ rather than $U(N)$ invariance produces towers of determinant backgrounds, while the analogous construction for $Sp(N)$ produces no genuinely new Fock-space sector. We include the $Sp(N)$ models to complete the list of classical color groups; the $Sp(N)$ model of anticommuting scalars has also been proposed as a dual of higher-spin gravity in de Sitter space~\cite{Anninos:2011ui}, although none of our results depends on this interpretation.
	
	In three-dimensional vector models these new states represent the baryon operators of the $SO(N)$ and $SU(N)$ theories. Their dimensions grow as $N^{3/2}$ for scalars and as $N$ for fermions~\cite{Shenker:2011zf,Aharony:2015mjs}, so they are invisible in the $1/N$ expansion around the Fock vacuum. For scalars, their spectrum on the sphere is that of a Fermi gas of $N$ particles, whose thermodynamics and possible bulk duals were studied recently in~\cite{Sundborg:2026cut}. For fermions, the lightest baryons form a single multiplet of maximal spin. A goal of this paper is to understand how such sectors arise within the bilocal description at finite $N$.
	
	We will argue that these facts are not separate accidents.  They are consequences of one general representation-theoretic statement.  Let $D=[G,G]$ be the derived subgroup of $G$ and $G^{\rm ab}=G/[G,G]$
	the abelianization of $G$. Howe duality~\cite{Joung:2014qya,Joung:2015jza,Basile:2020gqi,Howe:1989nad,864511ea-d90d-3d9a-8d58-650c63876c29,Kashiwara:1978kv} gives a multiplicity-free decomposition of the microscopic oscillator Fock space into representations of the color group and of the commuting bilocal algebra.  Projection onto $G$-singlets selects the bilocal representation paired with the trivial color representation. This is finite-$N$ Hilbert space constructed in~\cite{deMelloKoch:2026dfo,Ahn:2026ohj}. Projection only onto the singlets of $D=[G,G]$ instead selects the bilocal representations paired with all one-dimensional representations of $G$ which occur in the microscopic Fock space.  These one-dimensional representations are precisely the characters of $G^{\rm ab}$.
	
	Thus the appropriate notion of ``background'' is an \emph{abelianization sector}.  For the classical groups relevant here,
	\begin{equation}
		\frac{O(N)}{SO(N)}=\mathbb Z_2,\qquad
		\frac{U(N)}{SU(N)}=U(1),\qquad
		\frac{Sp(N)}{Sp(N)}=\{e\}\,.
	\end{equation}
	The first case gives an even and an odd determinant sector.  The second gives sectors labelled by an integer determinant charge $n$.  The third gives no additional sector. Physically, $G^{\rm ab}$ is a $\mathbb Z_2$ baryon parity for $SO(N)$ and the baryon number $U(1)$ for $SU(N)$, and the abelianization sectors are its charge sectors. 
	Bose/Fermi statistics do not change this classification, but they do determine which sectors actually occur for a finite number $K$ of one-particle modes.
	
	This new structure provides a natural organization for computations in the finite-$N$ Hilbert space. Once the Hilbert space decomposes into abelianization sectors,
	\begin{equation}
		\cF^{[G,G]}=\widehat{\bigoplus}_{\chi}\cH_\chi\,,
	\end{equation}
	where $\widehat{\bigoplus}$ denotes the completion of the algebraic direct sum,
	the natural quantity is not only the total singlet partition function, but also the \emph{sector-resolved character}, with fugacities $\mathbf z$ for the Cartan generators $\mathbf H$ of the bilocal algebra
	\begin{equation}
		Z_\chi=\Tr_{\cH_\chi}(\mathbf z^{\mathbf H})\,.
	\end{equation}
	Howe duality identifies this directly with the character of the bilocal representation paired with the color character $\chi$.  For $U(N)$ this gives $Z_n$; for $O(N)$ it gives $Z_+$ and $Z_-$; for $Sp(N)$ there is only the usual single character.  We construct the corresponding sector projectors explicitly in \cref{sec:sectorcharacters}.
	
	The organization of the paper is as follows. \Cref{sec:bilocalhowedualityfiniteN} serves as a review of the vector models we are working with, their bilocal algebras, the Howe decomposition, and fixes our conventions. In this section we also supply our conventions.  In \cref{sec:abelianizationtheorem} we derive the abelianization-sector decomposition and obtain the $O(N)$, $U(N)$ and $Sp(N)$ classifications, determine the $GL(K)$ content of each sector, and discuss their physical implications. In \cref{sec:extremalstates}, we  construct the extremal (or \emph{highest-weight}) states of every sector. In \cref{sec:constraintscentralchar}, we derive the constraints in every sector and compute all the Casimirs. We also explain what the center of the bilocal enveloping algebra can and cannot see, and we construct an exact observable that distinguishes the sectors beyond the center.  In \cref{sec:sectorcharacters} we compute the sector-resolved characters. We solve the constraints around the new backgrounds with auxiliary oscillators in \cref{sec:collectivenewbackground}, and identify the null operators whose quotient gives the physical modules. In \cref{sec: field theory} we apply these results to the free vector model on $\mathbb{R}\times S^2$, where the sectors become baryonic primaries, and determine their energies, degeneracies and dependence on angular momentum. We conclude our work in \cref{sec:discussion}, where we discuss our results and future directions. The appendices provide further conventions (\cref{appx:conventions}) and the details of the collective realization (\cref{appx:collective}).
	
	\section{Vector models, bilocal algebras and Howe duality at finite $N$}\label{sec:bilocalhowedualityfiniteN}
	In this section we list the oscillator models according to how Howe duality organizes their Fock spaces and fix a unified notation. Further conventions are collected in \cref{appx:conventions}.
	
	We consider $N$ colors and $K$ single-particle modes labelled by $p,q = 1,\ldots, K$. The color group $G$ acts on bosonic/fermionic oscillators in its defining representation. The $G$-invariant bilinears, bilocals, close into a Lie algebra $G^\ast$ referred to as the \emph{master algebra}, which together with $G$ forms a reductive dual pair in the sense of Howe~\cite{Howe:1989nad,864511ea-d90d-3d9a-8d58-650c63876c29,Basile:2023vyg}. The six cases we consider in this work are listed in \cref{tab:sixmodels}. Bosonic vector models are associated with the non-compact real form of $G^\ast$, and their bilocal Hilbert spaces are infinite-dimensional singlet modules. On the other hand, fermionic models are associated with compact real forms and their bilocal Hilbert spaces are finite-dimensional. The pairs ($O(N)$ bosons, $Sp(N)$ fermions) and ($O(N)$ fermions, $Sp(N)$ bosons) share the complexified algebras $\mathfrak{sp}\pts*{2K, \mathbb C}$ and $\mathfrak{so}\pts*{2K, \mathbb C}$ respectively. This is the origin of the $N\to -N$ correspondence~\cite{Das:2012dt} visible in the Casimirs of \cref{subsec:Casimirs}. 
	
	We assume that the microscopic oscillator Fock space $\cF$ carries commuting actions of $G$ and the master algebra $G^\ast$, and that the corresponding Howe decomposition is multiplicity free
	\begin{equation}
		\cF=\widehat{\bigoplus}_{\lambda\in\Lambda}V^G_\lambda\otimes\Theta(V^G_\lambda)\ .
		\label{eq:HoweDecomposition}
	\end{equation}
	Here $V^G_\lambda$ is an irreducible representation of the color group and $\Theta(V^G_\lambda)$ is its uniquely paired irreducible representation of the master algebra. For the dual pairs of \cref{tab:sixmodels}, see~\cite{Joung:2014qya,Joung:2015jza,Basile:2020gqi} for the oscillator realizations used in higher-spin theories. Projecting onto $G$-singlets selects the module paired with the trivial representation. This is the finite-$N$ Hilbert space of bilocal holography constructed in~\cite{deMelloKoch:2026dfo,Ahn:2026ohj}. Its finite-$N$ relations are quadratic constraints collected here and derived in every sector in \cref{sec:constraintscentralchar}.
	
	\begin{table}[t]
		\centering
		\renewcommand{\arraystretch}{1.25}
		\begin{tabular}{lllll}
			\toprule
			color & statistics & bilocal algebra & sectors & occurrence condition\\
			\midrule
			$O(N)$ & bosons & $\mathfrak{sp}(2K,\mathbb R)$, non-compact & $+,-$ & $-$ requires $K\ge N$\\
			$O(N)$ & fermions & $\mathfrak{so}(2K)$, compact & $+,-$ & both occur for $K\ge1$\\
			$U(N)$ & bosons & $\mathfrak{u}(K,K)$, non-compact & $n\in\mathbb Z$ & $n\neq0$ requires $K\ge N$\\
			$U(N)$ & fermions & $\mathfrak{u}(2K)$, compact & $-K\le n\le K$ & Pauli truncation\\
			$Sp(N)$ & bosons & $\mathfrak{so}^*(2K)$, non-compact & one sector & no abelianization sector\\
			$Sp(N)$ & fermions & $\mathfrak{sp}(2K)$, compact & one sector & no abelianization sector\\
			\bottomrule
		\end{tabular}
		\caption{The six oscillator models. The bilocal algebra is given with its real form. The last two columns list the abelianization sectors of \cref{sec:abelianizationtheorem} and the condition for each to occur with $K$ modes.}
		\label{tab:sixmodels}
	\end{table}

	\paragraph{Bilocal invariants.}
	The color oscillators satisfy $[a_i(p), a^\dagger_j(q)] = \delta_{ij}\delta_{pq}$ for bosons and $\sets{b_i(p),b^\dagger_j(q)} = \delta_{ij}\delta_{pq}$ for fermions, with all other (anti)commutators vanishing. 
	For $O(N)$ bosons, the bilocal invariants are 
	\begin{equation}
		\begin{gathered}
			A_{pq} = \sum_{i=1}^{N}a_i(p)a_i(q),\quad \bar A_{pq} = \sum_{i=1}^{N}a^\dagger_i(p)a^\dagger_i(q)\ ,\\
			B_{pq} = \sum_{i=1}^{N}a^\dagger_i(p)a_i(q)\,,
		\end{gathered}\label{eq:ONBilocals}
	\end{equation}
	and for $O(N)$ fermions
	\begin{equation}
		\begin{gathered}
			A_{pq} = \sum_{i=1}^{N}b_i(p)b_i(q),\quad \bar A_{pq} = \sum_{i=1}^{N}b^\dagger_i(p)b^\dagger_i(q)\ ,\\
			B_{pq} = \sum_{i=1}^{N}b^\dagger_i(p)b_i(q)\,.
		\end{gathered}\label{eq:ONFilocals}
	\end{equation}
	We also use the Weyl-ordered operators
	\begin{equation}
		B^{\rm W}_{pq} = \frac{1}{2}\sum_{i=1}^N a^\dagger_i(p)a_i(q) + a_i(q)a^\dagger_i(p) = B_{pq} + \frac N2\delta_{pq}\,,\label{eq:WeylOrderedBosons}
	\end{equation}
	\begin{equation}
		B^{\rm W}_{pq} = \frac{1}{2}\sum_{i=1}^N b^\dagger_i(p)b_i(q) - b_i(q)b^\dagger_i(p) = B_{pq} - \frac N2\delta_{pq}\,,\label{eq:WeylOrderedFermions}
	\end{equation}
	for bosons and fermions respectively. For $Sp(N)$, the bilocal invariants are defined in the same manner as in~\cite{Das:2012dt,Ahn:2026ohj}. 
	
	For $U(N)$ there is an additional species of antifundamental oscillators, $b^i(p)$ for bosons and $d^i(p)$ for fermions with $[b^i(p), b^\dagger_j(q)] = \delta^i_j\delta_{pq}$ and $\sets{d^i(p), d^\dagger_j(q)} = \delta^i_j\delta_{pq}$. The $U(N)$ bosonic bilocal invariants are
	\begin{equation}
		\begin{gathered}
			A_{pq}  = \sum_{i=1}^N b^i(p)a_i(q),\quad \bar A_{pq} = \sum_{i=1}^N a^{\dagger\,i}(p)b^\dagger_i(q)\ ,\\
			B_{pq}  = \sum_{i=1}^N a^{\dagger\,i}(p)a_i(q),\quad C_{pq} = \sum_{i=1}^N b^\dagger_i(q)b^i(p)\,.
		\end{gathered}\label{eq:UNBilocals1}
	\end{equation}
	The index order of $C_{pq}$ is transposed with respect to $B_{pq}$, so that
	\begin{equation}
		D_{pq} = \sum_{i=1}^N b^i(p)b^\dagger_i(q) = C_{pq} + N\delta_{pq}\,.\label{eq:UNBilocals2}
	\end{equation}
	The corresponding fermionic bilocals are
	\begin{equation}
		\begin{gathered}
			A_{pq}  = \sum_{i=1}^N d^i(p)b_i(q),\quad \bar A_{pq} = \sum_{i=1}^N b^{\dagger\,i}(p)d^\dagger_i(q)\ ,\\
			B_{pq}  = \sum_{i=1}^N b^{\dagger\,i}(p)b_i(q),\quad C_{pq} = \sum_{i=1}^N d^\dagger_i(q)d^i(p)\,
		\end{gathered}\label{eq:UNFilocals}
	\end{equation}
	\begin{equation}
		D_{pq} = \sum_{i=1}^N d^i(p)d^\dagger_i(q) = - C_{pq} + N\delta_{pq}\ .\label{eq:UNFilocals2}
	\end{equation}

	\paragraph{Finite-$\bm{N}$ constraints.}
	The four orthogonal and symplectic models can be treated simultaneously with the help of two parameters. In the vacuum Fock space, the finite-$N$ quadratic constraints take the form,
	\begin{equation}
		\bar A A - \pts*{\varepsilon\xi N - \varepsilon K - 1}B - \varepsilon B^2 = 0\,,
	\end{equation}
	where $\varepsilon$ is 1 for $\mathfrak{sp}(2K)$ and is $-1$ for $\mathfrak{so}(2K)$ algebras and $\xi$ is 1 for bosons and is $-1$ for fermions. For the unitary models, the bilocal algebra has the linear Casimir
	\begin{equation}
		Q = \frac{1}{N}\pts*{\Tr B - \Tr C}\,.\label{eq:UNBlinearCasimir}
	\end{equation} 
	On states with eigenvalue of $Q$ being $n$, the finite-$N$ constraints now take the following form,
	\begin{gather}
		\bar A A + (\xi K + n - N)B - \xi B^2 = 0\,,\\
		A\bar A - \xi(C + \xi N\mathbb{I})(C + (K + \xi n)\mathbb{I}) = 0\,.
	\end{gather}
	The Fock vacuum $\ketv 0$ and its bilocal excitations have $n=0$, sectors with $n\neq 0$ are constructed in \cref{sec:extremalstates}.

	\section{Abelianization sectors}\label{sec:abelianizationtheorem}
	Let $G$ be a compact Lie group, the color group, and let $D=[G,G]$ denote its commutator subgroup, also known as its \emph{derived subgroup}.  Define
	\begin{equation}
		G^{\rm ab}\equiv G/D.
	\end{equation}
	$G^{\rm ab}$ is an Abelian group, known as the \emph{abelianization} of $G$. We use the multiplicity-free Howe decomposition \cref{eq:HoweDecomposition}.
	
	Let $V$ be an irreducible complex representation of $G$.  Let $V^D$ denote the subspace of $V$ that is invariant under the action of $D$. Then $V^{D}\neq0$ if and only if $V$ is one-dimensional. Indeed, suppose that $0\neq v\in V^D$.  Since $D$ is normal, for any $g\in G$ and $d\in D$ we have
	\begin{equation}
		d(gv)=g(g^{-1}dg)v=gv.
	\end{equation}
	Thus $V^D$ is a nonzero $G$-invariant subspace of the irreducible representation $V$, and consequently $V^D=V$.  Thus, every element of $D$ acts trivially on $V$. Elements of $G$ that differ by an element of $D$ act in the same way on $V$. Thus, the action is really an action of the quotient group
	\begin{equation}
		G^{\rm ab}=G/D.
	\end{equation}
	Consequently, $V$ is an irreducible representation of the abelian group $G^{\rm ab}$. Since all elements of an abelian group commute, an irreducible complex representation of such a group is one-dimensional. Next, if $V$ is one-dimensional, then $\rho(g),\rho(h)\in\mathbb C^\times$ commute, and therefore
	\begin{equation}
		\rho([g,h]) = \rho(g)\rho(h)\rho(g)^{-1}\rho(h)^{-1} =1.
	\end{equation}
	Therefore the entire commutator subgroup $D=[G,G]$ acts trivially: $D\subseteq\ker\rho$. Hence $V^D=V\neq0$. We can now take $D$-invariants of \cref{eq:HoweDecomposition}. Indeed, since $D$ acts only on the color factor
	\begin{equation}
		\begin{aligned}
			\cF^D&=\left(\widehat{\bigoplus}_{\lambda\in\Lambda}V^G_\lambda\otimes\Theta(V^G_\lambda)\right)^D\,,\\
			&=\widehat{\bigoplus}_{\lambda\in\Lambda}(V^G_\lambda)^D\otimes\Theta(V^G_\lambda)\,.
		\end{aligned}
	\end{equation}
	Above we have established that
	\begin{equation}
		\left(V_\lambda^G\right)^D = 
		\begin{cases} V_\lambda^G\simeq\mathbb C,& V_\lambda^G\text{ one-dimensional},\\[2mm] 
			0,&\text{otherwise}. 
		\end{cases}
	\end{equation}
	The surviving $V_\lambda^G$'s are precisely the one-dimensional representations of $G$, equivalently the characters of
	\begin{equation} 
		G^{\rm ab}=G/[G,G].
	\end{equation}
	Therefore, the singlet Hilbert space under the derived color group decomposes as
	\begin{equation}
		\cF^{[G,G]}=\widehat{\bigoplus}_{\chi\in\widehat{G^{\rm ab}}_{\rm occ}}\Theta(\chi)\ .
		\label{eq:AbelianizationSectorTheorem}
	\end{equation}
	Here $\widehat{G^{\rm ab}}$ is the character group of the abelianization, while the subscript ``occ'' reminds us that only characters which actually occur in the oscillator Fock space contribute.
	
	The group-theoretic part of this argument is elementary. The non-trivial part is that each isotypic component is an irreducible module of the master algebra, which is the content of the Howe duality. Our central point is the physical interpretation of \cref{eq:AbelianizationSectorTheorem} and its implementation in the collective description. The argument also works verbatim for any closed normal subgroup $H \subset G: V^{H} \neq 0$ if and only if $V$ factors through $G/H$. The only special feature of $H = [G,G]$ is that $G/[G,G]$ is abelian, so the surviving labels are one-dimensional characters.
	
	The group $G^{\rm ab}$ determines the \emph{allowed labels}. Bose/Fermi statistics and the number $K$ of one-particle modes determine which labels are actually realized.  The usual $G$-singlet Hilbert space is simply the trivial-character term in \cref{eq:AbelianizationSectorTheorem}.
	
	\subsection{The three classical color groups}
	For the groups considered in this paper,
	\begin{center}
		\begin{tabular}{lccc}
			\toprule
			$G$ & $[G,G]$ & $G^{\rm ab}$ & $\widehat{G^{\rm ab}}$\\
			\midrule
			$U(N)$ & $SU(N)$ & $U(1)$ & $\widehat{U(1)}=\mathbb Z$\\
			$O(N)$ & $SO(N)$ & $\mathbb Z_2$ & $\widehat{\mathbb Z_2}=\mathbb Z_2$\\
			$Sp(N)$ & $Sp(N)$ & $\sets*{e}$ & $\sets*{e}$\\
			\bottomrule
		\end{tabular}
	\end{center}
	The decomposition \cref{eq:AbelianizationSectorTheorem} therefore gives us
	\begin{align}
		\cF^{SO(N)}&=\cH_+\oplus\cH_-, & \cH_+&=\Theta(\mathbf 1), & \cH_-&=\Theta(\det),\label{eq:ONsectors}\\
		\cF^{SU(N)}&=\widehat{\bigoplus}_{n\in\mathbb Z_{\rm occ}}\cH_n, & \cH_n&=\Theta((\det)^n),\label{eq:UNsectors}\\
		\cF^{Sp(N)}&=\cH_0.\label{eq:SpNsector}
	\end{align}
	For $O(N)$ the two characters are $1$ and $\det$, and $\det^2=1$.  For $U(N)$ the integer $n$ is the determinant charge, $\chi_n(U)=(\det U)^n$.  For $Sp(N)$ there is no non-trivial character and hence no second Howe module.
	
	The calculations of \cref{sec:extremalstates,sec:constraintscentralchar} provide the explicit extremal vectors and their bilocal weights.  They also reveal a useful warning: distinct abelianization sectors need not be distinguished by the center of the bilocal enveloping algebra.  In particular, the two $O(N)$ bosonic sectors have the same $\mathfrak{sp}(2K)$ Casimir eigenvalues even though they are distinct Howe modules (\cref{subsec:sectorobservable}).  This makes the sector-resolved character a more informative object than the Casimir data alone.
	
	\subsection{$GL(K)$ content of the sectors}
	\label{subsec:modecontent}
	Besides its bilocal module structure, each sector can be organized under the mode group $GL(K)$, generated by the $B_{pq}$. For bosons, $GL(N)\times GL(K)$ duality gives
	\begin{equation}
		\mathrm{Sym}\big(\mathbb{C}^N\otimes\mathbb{C}^K\big)=\bigoplus_{\ell(\mu)\le\min(N,K)}S_\mu(\mathbb{C}^N)\otimes S_\mu(\mathbb{C}^K)\,,\label{eq:bosonONmodecontent}
	\end{equation}
	where $S_\mu$ is the Schur functor and $\ell(\mu)$
	the number of rows of the partition $\mu$.
	For fermions, the dual Cauchy decomposition instead gives
	\begin{equation}
		\bigwedge(\mathbb{C}^N \otimes \mathbb{C}^K) =\bigoplus_{\substack{\ell(\mu)\leq N\\ \mu_1\leq K}}
		S_\mu(\mathbb{C}^N)\otimes S_{\mu^T}(\mathbb{C}^K)\,,\label{eq:fermionONmodecontent} 
	\end{equation}
	where $\mu^T$ denotes the transpose partition.
	In particular, the fermionic restriction $\mu_1\leq K$
	replaces the bosonic restriction $\ell(\mu)\leq K$.

	The projection onto the two $O(N)$ characters then reduces to a condition on $\mu$. By the Cartan--Helgason theorem for the compact symmetric pair $(U(N),O(N))$, an irreducible representation of $U(N)$ with highest weight $\lambda\in\mathbb{Z}^N$ contains an $O(N)$-invariant vector if and only if every $\lambda_i$ is even, and the invariant is then unique~\cite{Helgason2000,GoodmanWallach}.
	Since $\det$ and $\det^{-1}$ coincide on $O(N)$, $S_\mu(\mathbb{C}^N)$ contains the character $\det$ exactly when $S_\mu\otimes\det^{-1}$, of highest weight $\mu-(1^N)$, contains an invariant, that is, when $\ell(\mu)=N$ and all parts of $\mu$ are odd. For bosons the two sectors of \cref{eq:ONsectors} therefore decompose under $GL(K)$ as
	\begin{equation}
		\mathcal{H}_+\simeq\bigoplus_{\substack{\mu_1\ge\cdots\ge\mu_N\ge0\\ \mu_i\ \mathrm{even}}}S_\mu(\mathbb{C}^K)\,,
		\qquad
		\mathcal{H}_-\simeq\bigoplus_{\substack{\mu_1\ge\cdots\ge\mu_N\ge1\\ \mu_i\ \mathrm{odd}}}S_\mu(\mathbb{C}^K)\,,
		\label{eq:ONmodecontent}
	\end{equation}
	and for fermions in the same way, with $S_\mu(\mathbb{C}^K)$ replaced by $S_{\mu^T}(\mathbb{C}^K)$.
	
	The lowest $GL(K)$ type in $\mathcal{H}_-$ is $\wedge^N\mathbb{C}^K$ for bosons, spanned by the epsilon states of \cref{subsubsec:SOBosons}, and $\mathrm{Sym}^N\mathbb{C}^K$ for fermions. 
	
	The $U(N)$ sectors decompose in the same way under the two mode groups $GL(K)\times GL(K)$ generated by $B_{pq}$ and $C_{pq}$. For bosons the $\det^n$ component of $S_\lambda(\mathbb C^N)\otimes S_{\mu}(\mathbb C^N)^\ast$ requires $\lambda = \mu + (n^N)$ for $n\geq 0$, so that
	\begin{equation}
		\mathcal H_n \simeq \bigoplus_{\ell(\mu)\leq N} S_{\mu+(n^{N})}(\mathbb C^K)\otimes S_{\mu}(\mathbb C^K),\qquad n\geq 0\,,\label{eq:UNmodecontent}
	\end{equation}
	with the two factors exchanged for $n<0$. For fermions the same holds with each $S_{\nu}(\mathbb C^K)$ replaced by $S_{\nu^{\scriptscriptstyle\mathrm{T}}}(\mathbb C^K)$. These decompositions are the $GL(K)$ content of the theta lifts of the characters of $O(N)$ and $U(N)$~\cite{Howe:1989nad,Kashiwara:1978kv}.
	
	The abelianization sector decomposition determines the possible sectors. Their actual occurrence in the oscillator Fock space follows from this mode-space content and is summarized in the last column of \cref{tab:sixmodels}. The extremal states in \cref{sec:extremalstates} realize it explicitly.
	
	For $O(N)$ bosons a determinant requires $N$ distinct one-particle modes, so the odd sector appears only for $K\geq N$.  For $O(N)$ fermions all $N$ colors can occupy one mode and the odd sector exists already at $K=1$.  For $U(N)$ bosons the same distinct-mode requirement implies that nonzero determinant charge exists only when $K\ge N$ and once a determinant exists, arbitrary bosonic powers give all $n\in\mathbb Z$.  For $U(N)$ fermions the Pauli principle instead gives the finite range $-K\le n\le K$.
	
	\subsection{Physical interpretation}\label{subsec:physicalinterpretationofsectors}
	
	After gauging only $[G,G]$, the abelianization $G^{\rm ab}$ survives as a global symmetry: a $\mathbb Z_2$ \emph{baryon parity} for $SO(N)$, and the \emph{baryon number} $U(1)$ for $SU(N)$. The sector decomposition can be seen as charge decomposition. The bilocal operators, which are the single-trace operators dual to the higher-spin fields, are $G^{\rm ab}$-neutral and act irreducibly on each $\mathcal H_\chi$. The sectors are therefore superselected with respect to the master algebra and are invisible to any expansion built on top of them. They are \emph{not} superselected with respect to the full algebra of $[G,G]$-invariant operators: that algebra contains the baryon operators $\mathcal O$ and $\mathcal U^{a,b}$ found in \cref{sec:extremalstates}, which determine the $G^{\rm ab}$ charge. Gauging the full group $G$ amounts to projecting to the neutral sector. In the bulk, the charged sectors correspond to heavy non-perturbative states. For the $SO(N)$ and $SU(N)$ baryons of AdS/CFT see~\cite{Witten:1998xy}.

	\section{Extremal states}\label{sec:extremalstates}
	The non-trivial abelianization sectors are realized by explicit extremal states, denoted by $\ketv\Omega$, whose oscillator representations determine the corresponding Howe-dual bilocal modules. In this section we construct these extremal states for every sector and record their weights. We use Cartan generators that are diagonal in the modes: $H_p = B^{\rm W}_{pp}$ for $\mathfrak{sp}(2K)$, $H_p = -B^{\rm W}_{pp}$ for $\mathfrak{so}(2K)$, and with suitable signs of the diagonals of the generator matrix in the unitary cases. We choose positive roots so that the Fock vacuum is annihilated by every positive root vector, with this convention all singlet modules are \emph{highest-weight} modules. Further details can be found in \cref{subsec:hwc}.
	
	\subsection{$O(N)$ models}
	The decomposition \cref{eq:ONsectors} contains two $SO(N)$ singlet sectors, paired with the trivial and determinant representations of $O(N)$. 
	\subsubsection{Bosons}\label{subsubsec:SOBosons}
	In the bosonic oscillator realization the determinant sector exists only for $K\ge N$.  The Levi--Civita epsilon state below is its extremal state. We use the Cartan generators $H_p = B^{\rm W}_{pp}$, so that weights are read off from occupation numbers, and the positive root vectors are: $A_{pq}$ ($p\leq q$) and $B^{\rm W}_{pq}$ ($p < q$). With this convention, the Fock vacuum $\ketv 0$ is a \emph{highest-weight} state, annihilated by all positive root vectors, and 
	\begin{equation}
		H_p\ketv 0 = \frac N2\ketv 0\,.
	\end{equation}
	Thus 
	\begin{equation}
		\lambda_{\ketv{0}} = \pts*{\frac N2,\frac N2,\ldots, \frac N2}\ .
	\end{equation}
	We define the new background states, $\mathcal O(p_1, p_2, \ldots, p_N)\ketv 0$ as follows,
	\begin{equation}
		\mathcal O(p_1, p_2, \ldots, p_N)\ketv 0 = \varepsilon_{i_1i_2\cdots i_N}a^\dagger_{i_1}(p_1)a^\dagger_{i_2}(p_2)\cdots a^\dagger_{i_N}(p_N)\ketv 0\ .
	\end{equation}
	Bosonic statistics paired with antisymmetry of the Levi--Civita symbol dictates that $p_1, p_2, \ldots, p_N$ must all be distinct for these states to be non-trivial. They span the lowest $GL(K)$ type $\wedge^N\mathbb C^K$ of $\mathcal H_-$ found in \cref{subsec:modecontent}. It can be easily shown that all positive root vectors of the type $A$ annihilate these states, \emph{i.e.}
	\begin{equation}
		A_{pq}\mathcal O(p_1,p_2,\ldots,p_N)\ketv 0 = 0\ ,\quad(p\leq q,\;\;\forall\; p_i)
	\end{equation}
	but not all positive root vectors of the type $B^{\rm W}$ annihilate these states, as we have
	\begin{equation}
		B_{pq}\mathcal O(p_1,p_2,\ldots,p_N)\ketv 0 = \sum_{l=1}^N\delta_{p_lq}\mathcal{O}(\ldots, p_{l-1},p,p_{l+1},\ldots)\ketv 0\ . \label{eq:Btypeaction}
	\end{equation}
	For the specific state, $\ketv\Omega = \mathcal O(1,2,\ldots, N)\ketv 0$, \cref{eq:Btypeaction} gives for all $p<q$,
	\begin{equation}
		B_{pq}\ketv \Omega = 0\ .
	\end{equation}
	It is clearly also a highest-weight state. In this case however, the weights are, 
	\begin{equation}
		\lambda_{\ketv \Omega} = \pts*{\underbrace{\frac N2 +1,\ldots, \frac N2 + 1}_{N\text{ times}}, \underbrace{\frac N2,\ldots,\frac N2}_{(K-N)\text{ times}}}\ .
	\end{equation}
	We can check that successive operation of two background operators gives, 
	\begin{equation}
		\mathcal O(p_1,\ldots,p_N)\mathcal O(q_1,\ldots,q_N)\ketv 0 = \sum_{\sigma \in S_N} \operatorname{sgn}(\sigma)\bar A_{p_1q_{\sigma(1)}} \bar A_{p_2q_{\sigma(2)}}\cdots \bar A_{p_Nq_{\sigma(N)}}\ketv{0}\ ,
	\end{equation}
	and we recover a state from the vacuum Fock space. This is the oscillator form of $\det^2=1$. It agrees with the observation of~\cite{Sundborg:2026cut} that in the $SO(N)$ vector model a product of two baryons reduces to products of bilinears, so that there is no independent two-baryon sector.
	
	\subsubsection{Fermions}\label{subsubsec:SOFermions}
	The same $\mathbb Z_2$ abelianization gives two sectors for fermions: the two chiral $\mathfrak{so}(2K)$ modules.  Here Fermi statistics allow all $N$ colors to occupy a single mode, so the determinant sector exists already for $K=1$.  We now use the Cartan generators $H_{p} = -B^{\rm W}_{pp}$ and the positive root vectors $A_{pq}$ ($p<q$) and $B^{\rm W}_{pq}$ ($p>q$). Again we see that the Fock vacuum is a highest-weight state with the weight, 
	\begin{equation}
		\lambda_{\ketv{0}} = \pts*{\frac N2,\frac N2,\ldots, \frac N2}\ .
	\end{equation}
	We define the new background states, $\mathcal O^f(p_1,p_2,\ldots,p_N)\ketv 0$, as follows
	\begin{equation}
		\mathcal O^f(p_1,p_2,\ldots,p_N)\ketv 0 = \varepsilon_{i_1i_2\cdots i_N}b^\dagger_{i_1}(p_1)b^\dagger_{i_2}(p_2)\cdots b^\dagger_{i_N}(p_N)\ketv 0\ .
	\end{equation}
	Fermionic statistics and the antisymmetry of the Levi--Civita symbol dictate that $p_1, p_2,\ldots p_N$ need not be distinct for these states to be non-trivial. They span the lowest $GL(K)$ type $\operatorname{Sym}^N\mathbb C^K$ of $\mathcal H_-$ found in \cref{subsec:modecontent}.
	It can be shown that all positive root vectors of the type $A$ annihilate these states, \emph{i.e.}
	\begin{equation}
		A_{pq}\mathcal O^f(p_1,p_2,\ldots,p_N)\ketv 0 = 0\ ,\quad(p<q,\;\;\forall\; p_i)
	\end{equation}
	yet again not all positive root vectors of the type $B^{\rm W}$ annihilate these states, as we now have
	\begin{equation}
		B_{pq}\mathcal O^f(p_1,p_2,\ldots,p_N)\ketv 0 = \sum_{l=1}^N\delta_{p_lq}\mathcal{O}^f(\ldots, p_{l-1},p,p_{l+1},\ldots)\ketv 0\,.
	\end{equation}
	However if we consider the specific state $\ketv\Omega = \mathcal O^f(K,K,\ldots,K)\ketv 0$, then we have, 
	\begin{equation}
		B^{\rm W}_{pq}\ketv\Omega = 0\,. \quad(p>q)
	\end{equation}
	Thus we have a highest-weight state with the following weight,
	\begin{equation}
		\lambda_{\ketv\Omega} = \pts*{\frac N2, \ldots,\frac N2, -\frac N2}.
	\end{equation}
	The two states generate inequivalent $\mathfrak{so}(2K)$ modules which are exchanged by the outer automorphism of $\mathfrak{so}(2K)$. None of the even-order Casimirs can distinguish them. The Pfaffian Casimir however, can distinguish the Fock vacuum from the non-trivial epsilon background. More details can be found in \cref{subsec:sectoridentification}.	
	
	We can check that successive operation of two background operators gives, 
	\begin{equation}
		\mathcal O^f(p_1,\ldots,p_N)\mathcal O^f(q_1,\ldots,q_N)\ketv 0 = (-1)^{N(N-1)/2}\sum_{\sigma \in S_N} \bar A_{p_1q_{\sigma(1)}} \bar A_{p_2q_{\sigma(2)}}\cdots \bar A_{p_Nq_{\sigma(N)}}\ketv{0}\,,
	\end{equation}
	and again we recover a state in the vacuum Fock space. The extra factor $(-1)^{N(N-1)/2}$ arises from reordering the $2N$ fermionic creation operators into the pairs $b^\dagger_i(p_k)b^\dagger_i(q_{\sigma(k)})$, which takes $\frac{N(N-1)}{2}$ transpositions. 
	
	\subsection{$U(N)$ models}\label{subsec:UNModels}
	For $U(N)$ the abelianization is continuous \emph{i.e.} $U(N)^{\rm ab} = U(1)$, and its character group is $\mathbb Z$. 
	\subsubsection{Bosons}
	The two determinant operators introduced below generate charges $n=+1$ and $n=-1$.  Bosonic statistics require $K\ge N$ for either determinant to exist, but once it exists, arbitrary powers generate an unbounded tower $n\in\mathbb Z$.  The explicit weights and generalized constraints below are therefore the oscillator realization of the sectors $\cH_n=\Theta((\det)^n)$. The bilocals in \cref{eq:UNBilocals1,eq:UNBilocals2} form the generator matrix
	\begin{equation}
		L = \begin{pmatrix}
			-D & A\\
			-\bar A & B
		\end{pmatrix}\,.
	\end{equation}
	$L$ is the generator of the non-compact $\mathfrak{u}(K,K)$ algebra. 
	\begin{equation}
		[L_{PQ}, L_{RS}] = L_{PS}\delta_{RQ} - \delta_{PS}L_{RQ}\ .
	\end{equation}
	We make the choice $\sets*{H_p \equiv - L_{pp}}_{p=1}^{K}\cup \sets*{H_p \equiv L_{pp}}_{p=K+1}^{2K}$ for the basis of the Cartan subalgebra and the positive root vectors are $A_{pq}$ ($\forall\;p,q$), $B_{pq}$ ($p<q$) and $D_{pq}$ ($p>q$). The Fock vacuum is a highest-weight state trivially. We find that its weight is,
	\begin{equation}
		\lambda_{\ketv{0}} = \pts*{N^K, 0^K}.
	\end{equation}
	In this case  we have two different kinds of background operators $\mathcal U^a$ and $\mathcal U^b$, defined as follows,
	\begin{gather}
		\mathcal U^a(p_1,p_2,\ldots, p_N) = \varepsilon_{i_1 i_2\cdots i_N}a^{\dagger\,i_1}(p_1)a^{\dagger\,i_2}(p_2)\cdots a^{\dagger\,i_N}(p_N)\ ,\\
		\mathcal U^b(p_1,p_2,\ldots, p_N) = \varepsilon^{i_1 i_2\cdots i_N}b^\dagger_{i_1}(p_1)b^\dagger_{i_2}(p_2)\cdots b^\dagger_{i_N}(p_N)\ .
	\end{gather}
	Again, bosonic statistics along with the antisymmetry of Levi--Civita symbol dictates that $p_1, p_2,\ldots p_N$ must be distinct. We consider a particular state, $\mathcal U^a(1,2,\ldots,N)\ketv 0$. It is annihilated by all positive root vectors and thus is a highest-weight state whose weight is,
	\begin{equation}
		\lambda_{\mathcal U^a\ketv 0} = \pts*{N^K, 1^N, 0^{K-N}}\ .
	\end{equation}
	And likewise we consider the state, $\mathcal U^b(1,2,\ldots, N)\ketv 0$, and find that it too is a highest-weight state, this time the weight is, 
	\begin{equation}
		\lambda_{\mathcal U^b\ketv 0} = \pts*{(N+1)^N, N^{K-N}, 0^K}\ .
	\end{equation}
	We can check the successive operation of two distinct background operators gives,
	\begin{equation}
		\mathcal U^a(p_1,\ldots,p_N)\mathcal U^b(q_1,\ldots,q_N)\ketv 0 = \sum_{\sigma \in S_N} \operatorname{sgn}(\sigma)\bar A_{p_1q_{\sigma(1)}} \bar A_{p_2q_{\sigma(2)}}\cdots \bar A_{p_Nq_{\sigma(N)}}\ketv{0}\ ,
	\end{equation}
	and we recover a state in the vacuum Fock space. However, successive operation of the \emph{same} kind of background operators may not necessarily reproduce a state in the vacuum Fock space. In general, 
	\begin{equation}
		\pts[\Big]{\mathcal U^a(1,2,\ldots, N)}^n\ketv 0,\qquad n>0\,,
	\end{equation}
	and 
	\begin{equation}
		\pts*{\mathcal U^b(1,2,\ldots, N)}^{\absv n}\ketv 0,\qquad n<0\,,
	\end{equation}
	are all highest-weight states, whose weights are 
	\begin{equation}
		\lambda_{\mathcal U^a\cdots \mathcal U^a\ketv 0} = \pts*{N^K, n^N, 0^{K-N}}\,,
	\end{equation}
	\begin{equation}
		\lambda_{\mathcal U^b\cdots \mathcal U^b\ketv 0} = \pts*{(N+ \absv n)^N, N^{K-N}, 0^K}\,.
	\end{equation}
	where $n$ is the charge under the linear Casimir $Q$ defined in \cref{eq:UNBlinearCasimir}, and $\absv n$ is the number of background operators used to construct them. This number is unbounded: $\pts{\mathcal U^a(1,\ldots,N)}^n\ketv 0$ is the $n$\textsuperscript{th} power of a determinant of commuting creation operators, which is non-zero for every $n$. 
	
	Products of $n$ operators $\mathcal U^a$ with arbitrary mode labels span the lowest $GL(K)$ type of $\mathcal H_n$, $S_{(n^{N})}(\mathbb C^K)$ on the particle modes, which is the term $\mu = 0$ of \cref{eq:UNmodecontent}. For $n<0$, products of $\absv n$ operators $\mathcal U^b$ span $S_{(\absv n^N)}(\mathbb C^K)$ on the antiparticle modes. Each of the $\absv n$ columns of this Young diagram contains $N$ distinct modes, which is why $K\geq N$ is required. 
	
	\subsubsection{Fermions}
	The group-theoretic labels are again the integers $n$, but fermionic statistics truncate the realized character set.  A determinant fills one complete mode, so at most $K$ determinants of either sign can be inserted.  The tower of highest weights constructed below is therefore the finite sequence $-K\le n\le K$. The bilocals \cref{eq:UNFilocals,eq:UNFilocals2}, form the generator matrix
	\begin{equation}
		L = \begin{pmatrix}
			D & A\\
			\bar A & B
		\end{pmatrix}\,.
	\end{equation}
	$L$ is the generator of the compact $\mathfrak{u}(2K)$ algebra. 
	\begin{equation}
		[L_{PQ}, L_{RS}] = L_{PS}\delta_{RQ} - \delta_{PS}L_{RQ}\ .
	\end{equation}
	For the basis of the Cartan subalgebra, we make the standard choice $\sets*{H_p \equiv L_{pp}}_{p=1}^{2K}$ and the positive root vectors are $A_{pq}$ ($\forall\;p,q$), $B_{pq}$ ($p<q$) and $D_{pq}$ ($p>q$). The vacuum state $\ket{0}$ is annihilated by all the positive root vectors and its weight is,
	\begin{equation}
		\lambda_{\ketv{0}} = \pts*{N^K, 0^K}\ .
	\end{equation}
	Again we see that we now have two different kinds of background operators $\mathcal U^b$ and $\mathcal U^d$, defined as follows,
	\begin{gather}
		\mathcal U^b(p_1,p_2,\ldots, p_N) = \varepsilon_{i_1 i_2\cdots i_N}b^{\dagger\,i_1}(p_1)b^{\dagger\,i_2}(p_2)\cdots b^{\dagger\,i_N}(p_N)\ ,\\
		\mathcal U^d(p_1,p_2,\ldots, p_N) = \varepsilon^{i_1 i_2\cdots i_N}d^\dagger_{i_1}(p_1)d^\dagger_{i_2}(p_2)\cdots d^\dagger_{i_N}(p_N)\ .
	\end{gather}
	Fermionic statistics combined with antisymmetry of the Levi--Civita symbol dictates that $p_1, p_2,\ldots, p_N$ need not be distinct. We consider a particular state $\mathcal{U}^b(1,1,\ldots, 1)\ketv{0}$, and one may check that it is annihilated by all the positive root vectors. We find its weight,
	\begin{equation}
		\lambda_{\mathcal U^b\ketv{0}} = \pts*{N^{K+1},0^{K-1}}\ .
	\end{equation}
	In fact we see that $\mathcal{U}^b(1,\ldots, 1)\mathcal{U}^b(2,\ldots, 2)\ketv{0}$ is also annihilated by all the positive root vectors, and its weight is
	\begin{equation}
		\lambda_{\mathcal U^b\mathcal U^b\ketv{0}} = \pts*{N^{K+2},0^{K-2}}\ .
	\end{equation}
	This pattern terminates at the fully $b^\dagger$-occupied state, $\mathcal U^b(1,\ldots)\mathcal U^b(2,\ldots)\cdots \mathcal U^b(K,\ldots)\ketv{0}$ with the weight,
	\begin{equation}
		\lambda_{\mathcal{U}^b\cdots \mathcal{U}^b\ketv 0} = \pts*{N^{2K}}\ .
	\end{equation}
	Likewise for the particular state constructed using the background operator $\mathcal U^d(1,1,\ldots,1)\ketv 0$, we find that it too is a highest-weight state with weight, 
	\begin{equation}
		\lambda_{\mathcal U^d\ketv{0}} = \pts*{N^{K-1},0^{K+1}}\ .
	\end{equation} 
	Proceeding in a similar fashion as earlier, we check that the fully $d^\dagger$-occupied state, \\$\mathcal U^d(1,\ldots)\mathcal U^d(2,\ldots)\cdots \mathcal U^d(K,\ldots)\ketv{0}$ is the last highest-weight state we can construct using the tower of $\mathcal U^d$ operators whose weight is,
	\begin{equation}
		\lambda_{\mathcal{U}^d\cdots \mathcal{U}^d\ketv 0} = \pts*{0^{2K}}\ .
	\end{equation}
	For $n>0$, products of $n$ operators $\mathcal U^b$ with arbitrary mode labels span the lowest $GL(K)$ type of $\mathcal H_n$, $S_{(N^n)}(\mathbb C^K)$ on the particles modes. And for $n<0$, products of $\absv n$ operators $\mathcal U^d$ span $S_{(N^{\absv n})}(\mathbb C^K)$ on the antiparticle modes. This is the term $\mu = 0$ of \cref{eq:UNmodecontent} with rows and columns exchanged. The Young diagram has $\absv n$ rows, so $S_{(N^{\absv n})}(\mathbb C^K)$ is non-zero only for $\absv n\leq K$, which is the Pauli truncation.
	
	We can check the successive operation of two distinct background operators gives,
	\begin{equation}
		\mathcal U^b(p_1,\ldots,p_N)\mathcal U^d(q_1,\ldots,q_N)\ketv 0 = (-1)^{N(N-1)/2}\sum_{\sigma \in S_N}\bar A_{p_1q_{\sigma(1)}} \bar A_{p_2q_{\sigma(2)}}\cdots \bar A_{p_Nq_{\sigma(N)}}\ketv{0}\ ,
	\end{equation}
	and we recover a state in the vacuum Fock space. As seen earlier for $O(N)$ fermions, the factor of $(-1)^{N(N-1)/2}$ arises yet again due to reordering of fermionic creation operators. 
	
	\subsection{$Sp(N)$ models}\label{subsec:SpNModels}
	The decomposition \cref{eq:SpNsector} contains no additional sector because $Sp(N)$ is its own derived subgroup. The Pfaffian/Hafnian identities below provide a direct oscillator verification: the apparent epsilon background is already a word in the ordinary bilocal creation operators and therefore remains inside the original Howe module.
	We consider $N$ to be strictly even for this section. There is a geometric identity, 
	\begin{equation}
		\varepsilon_{i_1i_2\cdots i_N} = \frac{1}{2^{\frac N2}(\frac N2)!}\sum_{\sigma\in S_N}\sgn(\sigma)\Omega_{i_{\sigma(1)}i_{\sigma(2)}}\Omega_{i_{\sigma(3)}i_{\sigma(4)}}\cdots \Omega_{i_{\sigma(N-1)}i_{\sigma(N)}}\ .\label{eq:epsilontensoridentity}
	\end{equation} 
	The above is equivalent to 
	\begin{equation}
		\epsilon = \frac{1}{(\frac N2)!}\underbrace{\omega\wedge \omega \wedge\cdots\wedge \omega}_{\frac N2 \text{ times}}\,,
	\end{equation}
	where $\Omega_{ij}$ is the $Sp(N)$-invariant symplectic form, normalized so that $\operatorname{Pf}\Omega = 1$, and $\omega$ is the corresponding two-form. Define the $N\times N$ matrix $M$ with entries $M_{ab} \equiv \bar A_{p_a p_b}$, where $\bar A_{pq}$ is the $Sp(N)$-invariant bilocal creation operator. For bosons, $M$ is antisymmetric and for fermions it is symmetric. Using \cref{eq:epsilontensoridentity}, we can prove that 
	\begin{equation}
		\varepsilon_{i_1 i_2\cdots i_N}a^{\dagger\,i_1}(p_1)\cdots a^{\dagger\,i_N}(p_N) = \operatorname{Pf}(M)\ ,\label{eq:newbSpboson}
	\end{equation}
	for bosons and 
	\begin{equation}
		\varepsilon_{i_1 i_2\cdots i_N}b^{\dagger\,i_1}(p_1)\cdots b^{\dagger\,i_N}(p_N) = \operatorname{Hf}(M)\ ,\label{eq:newbSpfermion}
	\end{equation}
	for fermions. The Pfaffian of an antisymmetric matrix is defined as,
	\begin{equation}
		\operatorname{Pf}(M) = \frac{1}{2^{\frac N2}(\frac N2)!}\sum_{\sigma \in S_N}\sgn(\sigma)\prod_{i=1}^{\frac N2}M_{\sigma(2i-1)\sigma(2i)}\ .
	\end{equation}
	The Hafnian of a symmetric matrix is defined as,
	\begin{equation}
		\operatorname{Hf}(M) = \frac{1}{2^{\frac N2}(\frac N2)!}\sum_{\sigma \in S_N}\prod_{i=1}^{\frac N2}M_{\sigma(2i-1)\sigma(2i)}\ .
	\end{equation}
	Using \cref{eq:newbSpboson,eq:newbSpfermion} we can show that the new background operators create words of length $\frac N2$ using $\bar A$ as letters. Therefore, the new background operators create states living in the original Fock space itself. This is an instance of the first fundamental theorem of invariant theory for $Sp(N)$: \textit{every $Sp(N)$-invariant polynomial in the creation operators is a polynomial in the bilinears $\bar{A}_{pq}$}~\cite{Howe:1989nad}.\\

	\begin{table}[t]
		\centering
		\renewcommand{\arraystretch}{1.3}
		\begin{tabular}{llll}
			\toprule
			model & sector & extremal state & highest weight\\
			\midrule
			$O(N)$ bosons & $+$ & $\ketv0$ & $\pts*{(\frac N2)^K}$\\
			& $-$ & $\mathcal O(1,\ldots,N)\ketv0$ & $\pts*{(\frac N2+1)^N,(\frac N2)^{K-N}}$\\
			$O(N)$ fermions & $+$ & $\ketv0$ & $\pts*{(\frac N2)^K}$\\
			& $-$ & $\mathcal O^f(K,\ldots,K)\ketv0$ & $\pts*{(\frac N2)^{K-1},-\frac N2}$\\
			$U(N)$ bosons & $n\ge0$ & $\pts*{\mathcal U^a(1,\ldots,N)}^n\ketv0$ & $\pts*{N^K,n^N,0^{K-N}}$\\
			& $n<0$ & $\pts*{\mathcal U^b(1,\ldots,N)}^{\absv{n}}\ketv0$ & $\pts*{(N+\absv{n})^N,N^{K-N},0^K}$\\
			$U(N)$ fermions & $-K\le n\le K$ & $\mathcal U^b$ or $\mathcal U^d$ towers & $\pts*{N^{K+n},0^{K-n}}$\\
			$Sp(N)$ & --- & $\ketv0$ & single module\\
			\bottomrule
		\end{tabular}
		\caption{Extremal states and highest weights of all abelianization sectors.}
		\label{tab:extremalweights}
	\end{table}
	
	\noindent We summarize the weights of the extremal states constructed in this section in \cref{tab:extremalweights}.

\section{Finite-$N$ constraints and Casimirs}\label{sec:constraintscentralchar}
The finite-$N$ relations among the bilocals follow from the way the color Lie algebra acts on each sector. This gives a uniform derivation of the constraints in every sector and of the identities satisfied by the extremal states.

\paragraph{Constraints.}
For $U(N)$ bosons, the $\mathfrak{u}(N)$ color generators are 
\begin{equation}
L^i_j = \sum_{p=1}^K\pts*{a^{\dagger\,i}(p)a_j(p)-b^\dagger_j(p)b^i(p)}\,.
\end{equation}
The $\mathcal H_n = \Theta\pts{\det^n}$ is characterized by $L^i_j\ketv\psi = n\delta^i_j\ketv\psi$ for every state $\ketv\psi \in \mathcal H_n$. Consider the operator expression $\pts*{\bar A A - B^2 + K B}_{pq}$, normal ordering in the oscillators gives us
\begin{equation}
\pts*{\bar A A - B^2 + K B}_{pq} = -\sum_{i,j}a^{\dagger\,i}(p)L^j_ia_j(q) = -\sum_{i,j} a^{\dagger\,i}(p)a_j(q)L^j_i + N B_{pq}\,.
\end{equation}
On an arbitrary state $\ketv\psi\in \mathcal H_n$, we therefore have
\begin{equation}
\bar A A + (K+n-N)B - B^2 = 0,\qquad \text{on all of $\mathcal H_n$}\,.
\end{equation}
In the same way, with $D_{pq} = C_{pq} + N\delta_{pq}$ we have,
\begin{equation}
\pts*{A \bar A - D^2 - KD}_{pq} = \sum_{i,j}b^i(p)L^j_{i}b^\dagger_j(q) = \pts*{n-N} D_{pq},\qquad \text{on all of $\mathcal H_n$}\,,
\end{equation}
which implies
\begin{equation}
A\bar A - (K+n-N)D -D^2 = 0,\qquad \text{on all of $\mathcal H_n$}\,.
\end{equation}
or equivalently,
\begin{equation}
A\bar A - (C + N\mathbf{1})(C + (K+n)\mathbf{1}) = 0,\qquad \text{on all of $\mathcal H_n$}\,.
\end{equation}
We extend this argument for $U(N)$ fermions, and find that on $\mathcal H_n$,
\begin{gather}
\bar A A - (K - n + N)B + B^2 = 0\,,\\
A\bar A - (N\mathbf 1 - C)(C + (K-n)\mathbf 1) = 0\,.
\end{gather}
For $O(N)$ bosons, the $\mathfrak{so}(N)$ generators $L_{ij} = \sum_p\pts*{a^\dagger_i(p)a_j(p) - a^\dagger_j(p)a_i(p)}$ annihilate \emph{both} sectors $\mathcal H_{\pm}$, because both are $SO(N)$ singlets. The same computation as above gives,
\begin{equation}
(\bar A A - B^2 + KB)_{pq} = \sum_{i,j}a^\dagger_i(p)L_{ij} a_j(q) = \sum_{i,j} a^\dagger_i(p)a_j(q)L_{ij} + (N-1)B_{pq},
\end{equation}
so that 
\begin{equation}
\bar A A  + (K+1 - N)B - B^2 = 0,\qquad \text{on all of $\mathcal F^{SO(N)} = \mathcal H_+\oplus \mathcal H_-$}\,.
\end{equation}
For $O(N)$ fermions we get, 
\begin{equation}
\bar A A  - (K-1 + N)B + B^2 = 0,\qquad \text{on all of $\mathcal F^{SO(N)} = \mathcal H_+\oplus \mathcal H_-$}\,.
\end{equation}

\paragraph{Relations on the lowest $GL(K)$ type.} Let $\mathcal H_{\chi}^{(0)}$ denote the lowest $GL(K)$ type of a non-trivial sector \emph{i.e.} the extremal states defined in \cref{sec:extremalstates}. These states are polynomials in a single species of creation operators, and transform under $GL(N)$ acting on the color indices as a power of the determinant: $\det$ for $O(N)$ epsilon states, and $\det^{\absv n}$ for the $U(N)$ determinant towers. The $\mathfrak{gl}(N)$ generators $E_{ij}$ of that particular species act on $\mathcal H_{\chi}^{(0)}$ as follows,
\begin{gather}
	E_{ij}\;\mathcal O\ketv 0 = \sum_{p}a^\dagger_i(p)a_j(p)\mathcal O\ketv 0 = \delta_{ij}\mathcal O\ketv 0\,,\qquad \text{($O(N)$ bosons)}\\
	E_{ij}\;\mathcal O^f\ketv 0 = \sum_{p}b^\dagger_i(p)b_j(p)\mathcal O^f\ketv 0 = \delta_{ij}\mathcal O^f\ketv 0\,,\qquad \text{($O(N)$ fermions)}\\
	\begin{split}
		E^i_{j}\;\mathcal U^a\cdots \mathcal U^a\ketv 0 &= \sum_{p}a^{\dagger\, i}(p)a_j(p)\;\mathcal U^a\cdots \mathcal U^a\ketv 0 = n \delta^i_{j}\;\mathcal U^a\cdots \mathcal U^a\ketv 0\,,\\
		E_i^{\;j}\mathcal U^b\cdots \mathcal U^b\ketv 0 &= \sum_{p}b^\dagger_i(p)b^j(p)\mathcal U^b\cdots \mathcal U^b\ketv 0 = \absv n\delta_{i}^{\,j}\mathcal U^b\cdots\mathcal U^b\ketv 0\,,\\
	\end{split}\quad \text{($U(N)$ bosons)}\\
	\begin{split}
		E^i_{j}\;\mathcal U^b\cdots \mathcal U^b\ketv 0 &= \sum_{p}b^{\dagger\, i}(p)b_j(p)\;\mathcal U^b\cdots \mathcal U^b\ketv 0 = n \delta^i_{j}\;\mathcal U^b\cdots \mathcal U^b\ketv 0\,,\\
		E_i^{\;j}\mathcal U^d\cdots \mathcal U^d\ketv 0 &= \sum_{p}d^\dagger_i(p)d^j(p)\mathcal U^d\cdots \mathcal U^d\ketv 0 = \absv n\delta_{i}^{\,j}\mathcal U^d\cdots\mathcal U^d\ketv 0\,,
	\end{split}\quad \text{($U(N)$ fermions)}
\end{gather}
where $n$ is the charge of the $U(N)$ determinant sectors. From above, we can derive the following identities: for $O(N)$ models,
\begin{equation}
\sum_r	\bar A_{rq} B_{pr}\ketv\psi = \xi\bar A_{pq}\ketv\psi\,,\label{eq:specialONB}
\end{equation}
and for $U(N)$ models, 
\begin{equation}
	\begin{gathered}
		\sum_r \bar A_{rq}B_{pr}\ketv\psi = \xi \absv n\bar A_{pq}\ketv\psi\,,\quad \text{(fundamental sector with charge $n$)}\\
		\sum_r\bar A_{pr}C_{rq}\ketv\psi = \xi \absv n\bar A_{pq}\ketv\psi\,,\quad \text{(antifundamental sector with charge $-n$)}
	\end{gathered}\label{eq:specialUNB}
\end{equation}
where $\ketv\psi\in \mathcal H_\chi^{(0)}$. These relations hold only on $\mathcal H_{\chi}^{(0)}$, on higher types (\emph{i.e.} bilocal excitations on extremal states), the $\mathfrak{gl}(N)$ generators no longer act as scalars. The full sector $\mathcal H_\chi$ is a quotient of the module induced from $\mathcal H^{(0)}_{\chi}$ on which the bilocal creation operators act freely. Even at the first excitation, the induced module contains states or $GL(K)$ types that are physically forbidden in $\mathcal H_\chi$: types with more than $N$ rows for bosonic color, and types with first row longer than $N$ for fermionic color. The identities \cref{eq:specialONB,eq:specialUNB} precisely quotient out these states. We shall see in \cref{sec:collectivenewbackground} that these are enforced by null operators.

\subsection{Characteristic equations and Casimirs to all orders}\label{subsec:Casimirs} 
We now derive formulae for Casimir eigenvalues of highest-weight states at all orders. The $N\to -N$ correspondence is also made explicit between relevant pairs of models.
\subsubsection{Orthogonal and symplectic}\label{subsubsec:orthsymp}
For the orthogonal and symplectic models we define the $2K\times 2K$ generator matrix,
\begin{equation}
	\mathcal M(\varepsilon,\xi) = \begin{pmatrix}
		B^{\rm W}_{pq} & \bar A_{pq}\\
		-\varepsilon A_{pq} & -B_{qp}^{\rm W}
	\end{pmatrix}\,,\label{eq:unifiedorthogonalsymplecticgenerator}
\end{equation}
where 
\begin{equation}
	B^{\rm W}_{pq} \equiv B_{pq} +\xi\frac N2\delta_{pq}\,.
\end{equation}
Its relation to the generators $\mathcal L$ of the master algebras $\mathfrak{sp}(2K)$ and $\mathfrak{so}(2K)$ is shown explicitly in \cref{subsec:generatormatricesON}.

We define the Casimirs as,
\begin{equation}
	C_n = \tr(\mathcal M^n)\,.\label{eq:CasimirNormalization}
\end{equation}
Consider the matrix $\mathcal M^2 - c_1\mathcal M + c_0\mathbf{1}$. With coefficients, $c_1 = K + \varepsilon$ and $c_0 = \xi\frac{N}{2}\pts*{K + \varepsilon - \xi\frac N2}$, every entry of this matrix is proportional to the color generators defined earlier, present on the rightmost end. Therefore, it vanishes on the entire $[G,G]$-singlet space. For $SO(N)$ bosons, for example, 
\begin{equation}
\mathcal M^2 - (K+1)\mathcal M + \frac N2\pts*{K+1-\frac N2}\mathbf 1 = \begin{pmatrix}
-\sum_{i,j} a^\dagger_i(p)a_j(q)L_{ij} & - \sum_{i,j} a^\dagger_i(p)a^\dagger_j(q) L_{ij}\\
\sum_{i,j}a_i(p)a_j(q)L_{ij} & -\sum_{i,j}a^\dagger_i(q)a_j(p)L_{ij}
\end{pmatrix}\,.
\end{equation}
Thus on the singlet, $\mathcal M$ satisfies the characteristic equation,
\begin{equation}
\mathcal M^2 - (K+\varepsilon)\mathcal M + \xi\frac N2\pts*{K + \varepsilon - \xi\frac N2}\mathbf 1 = 0\,.
\end{equation}
Tracing this relation, implies that we have the recursion relation for $C_n$'s
\begin{equation}
C_n - (K+\varepsilon)C_{n-1} + \xi\frac{N}{2}\pts*{K+\varepsilon - \xi\frac N2}C_{n-2} = 0\,,
\end{equation}
with 
\begin{equation}
C_{0} = \tr \mathbf 1 = 2K,\qquad C_1 = \tr\mathcal M = 0\,.
\end{equation}
This can be solved by standard methods, and we obtain\footnote{Equivalently, without the spurious pole at $\xi N = K + \varepsilon$,
\[ C_{n} = -\frac{\xi NK(2K + 2\varepsilon - \xi N)}{2}\sum_{j=0}^{n-2}\pts*{\frac{\xi N}{2}}^j\pts*{\frac{2K + 2\varepsilon - \xi N}{2}}^{n-2-j},\qquad n\geq 2\,. \]}
\begin{equation}
C_n = \frac{\xi KN\pts*{2K + 2\varepsilon - \xi N}}{2\pts*{\xi N - K - \varepsilon}}\pts*{\pts*{\frac{2K + 2\varepsilon - \xi N}{2}}^{n-1} - \pts*{\frac{\xi N}{2}}^{n-1}}\,.
\end{equation}
The $N\to -N$ correspondence is very explicit above for the given type of algebra ($\varepsilon$ fixed). From the initial conditions, we can also see easily that 
\begin{equation}
C_3 = (K+\varepsilon)C_2\,.\label{eq:C3C2relation}
\end{equation}
Using \cref{eq:C3C2relation} we can show that all odd Casimirs are linear combinations of even Casimirs. Explicitly we have, 
for $SO(N)$ bosons,
\begin{equation}
C_n = \frac{NK(2K + 2 - N)}{2(N-K-1)}\sets*{\pts*{\frac{2K+2-N}{2}}^{n-1} - \pts*{\frac{N}{2}}^{n-1}}\,.
\end{equation}
For $Sp(N)$ bosons,
\begin{equation}
C_n = \frac{NK(2K - 2 - N)}{2(N-K+1)}\sets*{\pts*{\frac{2K-2-N}{2}}^{n-1} - \pts*{\frac{N}{2}}^{n-1}}\,.
\end{equation} 
For $SO(N)$ fermions, 
\begin{equation}
C_n = \frac{NK(2K - 2 + N)}{2(N+K-1)}\sets*{\pts*{\frac{2K-2+N}{2}}^{n-1} - \pts*{-\frac{N}{2}}^{n-1}}\,.
\end{equation}
For $Sp(N)$ fermions, 
\begin{equation}
C_n = \frac{NK(2K + 2 + N)}{2(N+K+1)}\sets*{\pts*{\frac{2K+2+N}{2}}^{n-1} - \pts*{-\frac{N}{2}}^{n-1}}\,.
\end{equation}
For $\mathfrak{so}(2K)$ the center contains, in addition to the $C_n$ (for $n = 2, 4, \ldots, 2K-2$), the Pfaffian. Its values on the two $SO(N)$ fermionic sectors are of opposite parity, so the Pfaffian distinguishes them.

\subsubsection{Unitary}\label{subsubsec:unitary}
In the sector of charge $n$, the constraints are given above. The generator matrix is 
\begin{equation}
L = \begin{pmatrix}
-\xi D & A\\
-\xi \bar A & B
\end{pmatrix}\,.
\end{equation}
Like before, we find the characteristic equation
\begin{equation}
L^2 - (K + \xi n - \xi N)L = 0\,.
\end{equation}
And 
\begin{equation}
\tr L = N (n-\xi K)\,.
\end{equation}
This sets up a recursion relation whose solution is,
\begin{equation}
\tr L^m = N(n-\xi K)(K + \xi n - \xi N)^{m-1}\,.
\end{equation}
However, we need to subtract the trace and obtain the generator matrix for $\mathfrak{su}(K,K)$ (or $\mathfrak{su}(2K)$),
\begin{equation}
\hat L = L - \frac{\tr L}{2K}\mathbf 1\,.
\end{equation}
Define 
\begin{equation}
C_m = \frac12 \tr(\hat{L}^m)\,.
\end{equation}
where the factor $\frac12$ is a convention chosen so that $C_0 = K$.
The characteristic equation is now,
\begin{equation}
\hat L^2 - \pts*{K + \xi n - n\frac NK}\hat L + \frac{\xi N(2K-\xi N)(K^2-n^2)}{4K^2}\mathbf 1 = 0\,.
\end{equation}
This sets up a recursion relation with initial conditions,
\begin{equation}
C_0 = K\,,\quad C_1 = 0\,.
\end{equation}
Explicitly, for $SU(N)$ bosons we have
\begin{equation}
C_m = -\frac{N(2K-N)(K^2-n^2)}{4K(K+n-N)}\sets*{\pts*{\frac{(2K-N)(K+n)}{2K}}^{m-1} - \pts*{\frac{N(K-n)}{2K}}^{m-1}}\,.
\end{equation}
For $SU(N)$ fermions we have
\begin{equation}
C_m = \frac{N(2K+N)(K^2-n^2)}{4K(K-n+N)}\sets*{\pts*{\frac{(2K+N)(K-n)}{2K}}^{m-1} - \pts*{-\frac{N(K+n)}{2K}}^{m-1}}\,.
\end{equation}
Now we can see the $(N, n) \to (-N, -n)$ map explicitly between the two cases.

\subsection{Sector identification}\label{subsec:sectoridentification}
There are many competing definitions of what one calls a Casimir, when we use the term in this paper, we simply mean any element contained in the center of the universal enveloping algebra $U(G^\ast)$ of the master algebra $G^\ast$. Usually, if one is able to obtain the eigenvalues of all \emph{independent} Casimirs, then one can pinpoint the representation. However, for infinite-dimensional representations this is not necessarily true. 

Consider the Casimir eigenvalues we found for $SO(N)$ bosonic sectors. For the singlets, there is a common characteristic equation and thus all $C_n$ agree. However, one can also see the reason in terms of highest weights. With positive roots defined according to our convention, the Weyl vector is $\rho = (-1, -2, \ldots, -K)$. From \cref{tab:extremalweights} we have,
\begin{equation}
\lambda_{\ketv 0} + \rho = \pts*{\frac N2 - 1, \frac N2 - 2,\ldots, -\frac N2, -\frac N2 - 1,\ldots, \frac N2 - K}\,,
\end{equation}
and 
\begin{equation}
\lambda_{\ketv\Omega} + \rho = \pts*{\frac N2, \frac N2-1,\ldots, \frac N2 - (N-1), \frac N2 - (N+1), \frac N2 - (N+2),\ldots, \frac N2 - K}\,.
\end{equation}
The components of vectors $\lambda_{\ketv 0} + \rho$ and $\lambda_{\ketv\Omega} + \rho$ agree everywhere except for a single entry. As multisets, the two vectors differ only by $-\frac N2 \longleftrightarrow \frac N2$, so that they are related by a Weyl-group element,
\begin{equation}
\lambda_{\ketv\Omega} + \rho = w(\lambda_{\ketv{0}} + \rho), \quad w\in W(C_K) = S_K\ltimes \mathbb Z^K_2\,,
\end{equation}
and by the Harish-Chandra isomorphism~\cite{HarishChandra1951,Humphreys:1972} every element of the center of $U(\mathfrak{sp}(2K))$ \emph{i.e.} every Casimir, takes the same value on the two modules. We can see this explicitly for the quadratic Casimir: on a highest-weight state with weight $\lambda$, one finds
\begin{equation}
	\frac12 C_2 = \innerx{\lambda}{\lambda + 2\rho} = \innerx{\lambda+\rho}{\lambda+\rho} - \innerx{\rho}{\rho}\,,
\end{equation}
which depends on $\lambda$ only through the $W(C_K)$-invariant length $\lambda + \rho$. 

The Weyl group $W(C_K)$ contains all the sign changes of the components of $\lambda + \rho$, which is why $\mathfrak{sp}(2K)$ has no Pfaffian-type Casimir. This coincidence is possible only because $\cH_+$ and $\cH_-$ are infinite-dimensional. A finite-dimensional irreducible module has a dominant integral highest weight $\lambda$, and the Weyl-group orbit of $\lambda+\rho$ contains exactly one strictly dominant vector. Two inequivalent finite-dimensional irreducible modules therefore always differ in the value of some Casimir~\cite{Humphreys:1972}. Infinite dimensionality is necessary but not sufficient: the $U(N)$ bosonic sectors are also infinite-dimensional, yet the linear Casimir $Q$ separates them.

The $SO(N)$ fermionic sectors illustrate the converse. Their $\mathfrak{so}(2K)$ modules are finite-dimensional, so some Casimir must separate them, and since all $C_n$ agree it is the Pfaffian,
\begin{equation}
	C'_K = \sum^{2K}_{P_1,Q_1,\ldots,P_K,Q_K=1}\varepsilon_{\scriptscriptstyle P_1Q_1\cdots P_KQ_K}L_{P_1Q_1}L_{P_2Q_2}\cdots L_{P_KQ_K}\,,
\end{equation}
It is free of ordering ambiguities, since the factors carry disjoint indices and commute. On a highest-weight state of weight $\lambda$, it takes the value
\begin{equation}
	C'_K = (-1)^{\frac{K(K-1)}{2}}2^K K!\prod_{i=1}^K(\lambda_i + \rho_i)\,,
\end{equation}
with $\rho = (K-1, K-2, \ldots, 0)$, and it distinguishes the trivial vacuum from the non-trivial background as
\begin{equation}
	\prod_{i=1}^K(\lambda_i + \rho_i) = \pm\prod_{j=0}^{K-1}\pts*{\frac N2 + j}\qquad \text{on }\mathcal H_{\pm}\,.\label{eq:Pfaffianvalues}
\end{equation}

\subsection{Beyond the center of $U(\mathfrak{sp}(2K))$}\label{subsec:sectorobservable}
An observable that separates the two extremal states, such as the weight or the particle number, need not label the two entire modules. For example, in the case of orthogonal vector models the total particle number $\nu = \sum_{p}B_{pp}$ has spectrum $2\mathbb Z_{+}$ on $\mathcal H_+$ and $N + 2\mathbb Z_{+}$ on $\mathcal H_-$, which overlap for even $N$. 

We now construct an observable for $O(N)$ bosons, that takes the value $+1$ on all of $\mathcal H_+$ and $-1$ on all of $\mathcal H_-$, first in terms of oscillators and then in terms of the bilocals. Define the following operator $\mathcal R$,
\begin{equation}
\mathcal R=\exp\pts*{i\pi\sum_{p=1}^K a^\dagger_1(p)a_1(p)}\,,\label{eq:reflectioncharge}
\end{equation}
where 
\begin{equation}
	\begin{gathered}
		\mathcal R a_1(p) \mathcal R^{-1} = -a_1(p)\,,\\ 
		\mathcal R a_{i}(p) \mathcal R^{-1} = a_i(p), \quad(i\neq1)\,.
	\end{gathered}
\end{equation}
$\mathcal R$ implements the reflection $\operatorname{diag}(-1,1,\ldots,1)\in O(N)$ as it satisfies $\mathcal R^\dagger=\mathcal R$ and $\mathcal R^2=1$. Each bilocal contains color $1$ twice, so
\begin{equation}
\sq*{\mathcal R,A_{pq}}=\sq*{\mathcal R,\bar A_{pq}}=\sq*{\mathcal R,B_{pq}}=0\,,
\end{equation}
while each term of the epsilon tensor contains color $1$ exactly once, so $\mathcal R\,\mathcal O\,\mathcal R^{-1}=-\mathcal O$. Hence
\begin{equation}
\mathcal R\big|_{\cH_+}=+1\,,\qquad \mathcal R\big|_{\cH_-}=-1\,,
\end{equation}
on the \emph{entire} modules. On $\cF^{SO(N)}$ every orientation-reversing element of $O(N)$ acts in the same way, because any two such elements differ by an element of $SO(N)$. So $\mathcal R$ is canonically the generator of $O(N)/SO(N)$, the baryon parity of \cref{subsec:physicalinterpretationofsectors}. Restricted to $\cF^{SO(N)}$, $P_\pm=\frac12\pts*{\mathbf 1\pm\mathcal R}$ is therefore the projector onto $\cH_\pm$. It reappears as the sector projector of \cref{sec:sectorcharacters}, so the observable and the sector character are the same object. The particle-number parity $e^{i\pi\nu}$ implements $-\mathbf 1\in O(N)$, and $\det(-\mathbf 1)=(-1)^N$. For odd $N$, $-\mathbf 1$ is orientation reversing and $e^{i\pi\nu}=\mathcal R$ on $\cF^{SO(N)}$; for even $N$, $-\mathbf 1\in SO(N)$ and $e^{i\pi\nu}=1$ on both sectors. 

The same construction, with $b_1(p)$, applies to $O(N)$ fermions, where by \cref{eq:Pfaffianvalues} the Pfaffian is proportional to $\mathcal R$ on $\cF^{SO(N)}$ \emph{i.e.} there the reflection \emph{is} a central polynomial.

The operator $\mathcal R$ is written in terms of the color oscillators. To express it through the bilocals alone we use the mode-space content \cref{eq:ONmodecontent}: under the $\mathfrak{u}(K)$ subalgebra generated by the normal-ordered $B_{pq}$, the $GL(K)$ types of $\cH_+$ have all rows of even length, while those of $\cH_-$ have exactly $N$ rows, all of odd length. Let $P_\mu$ be the spectral projector onto the $GL(K)$ type $S_\mu(\mathbb C^K)$; it is a function of the center of $U(\mathfrak u(K))$, hence of the $B_{pq}$ alone. Then
\begin{equation}
\Pi_{\rm biloc}\equiv\sum_\mu(-1)^{\mu_1}P_\mu=\mathcal R\qquad\text{on }\cF^{SO(N)}\,.\label{eq:Pibiloc}
\end{equation}
The sum converges in the strong operator topology and defines a bounded involution; it is not a polynomial in the bilocals. The projectors can be made explicit with the Capelli generating function,
\begin{equation}
D_K(u)=\operatorname{cdet}\sq*{B_{pq}+(u+K-p)\delta_{pq}}\,,\qquad D_K(u)\big|_{S_\mu(\mathbb C^K)}=\prod_{i=1}^K\pts*{u+\ell_i}\,,\qquad \ell_i=\mu_i+K-i\,,
\end{equation}
where $\operatorname{cdet}$ denotes the column determinant of a matrix $X$ with non-commuting entries,
\begin{equation}
\operatorname{cdet}X=\sum_{\sigma\in S_K}\operatorname{sgn}(\sigma)\,X_{\sigma(1)1}X_{\sigma(2)2}\cdots X_{\sigma(K)K}\,,
\end{equation}
in which the entries of the first column stand leftmost, those of the second column next, and so on. For commuting entries it reduces to the ordinary determinant. The diagonal shifts $K-p$ make $D_K(u)$ commute with every $B_{pq}$. The coefficients of $D_K(u)$ generate the center of $U(\mathfrak u(K))$. The sector is read off from the largest root, $(-1)^{\mu_1}=(-1)^{\ell_1-K+1}$.

\paragraph{A worked out example.} Let $N=K=2$. With $\nu=B_{11}+B_{22}$, 
\begin{equation}
	\begin{gathered}
		J_z=\frac12\pts*{B_{11}-B_{22}}\,,\\ 
		J_+=B_{12},\quad  J_-=B_{21}\,.
	\end{gathered}
\end{equation}
 and $\mathbf J^2=J_z^2+\frac12\pts*{J_+J_-+J_-J_+}$, a $GL(2)$ type $(\mu_1,\mu_2)$ has $\nu=\mu_1+\mu_2$ and $\mathbf J^2=j(j+1)$ with $j=\frac12(\mu_1-\mu_2)$. Hence
\begin{equation}
\Pi_2=\exp\sq*{i\pi\pts*{\frac\nu2-\frac{\sqrt{1+4\mathbf J^2}-1}{2}}}=(-1)^{\mu_2}
\end{equation}
equals $+1$ on $\cH_+$, where $(\mu_1,\mu_2)=(2m_1,2m_2)$, and $-1$ on $\cH_-$, where $(\mu_1,\mu_2)=(2m_1+1,2m_2+1)$. The states $\bar A_{12}\ketv0$ and $\varepsilon_{ij}a^\dagger_i(1)a^\dagger_j(2)\ketv0$ have identical Cartan weights, $B_{11}=B_{22}=1$, but $j=1$ and $j=0$ respectively, so $\Pi_2=+1$ and $-1$. A single central element of $U(\mathfrak u(2))$ does not suffice. For example, $D_2(0)=(B_{11}+1)B_{22}-B_{21}B_{12}$ takes the value $(\mu_1+1)\mu_2=6$ both on the type $(2,2)\subset\cH_+$ and on the type $(5,1)\subset\cH_-$. The sector is determined only by the joint values of both $\mathfrak u(2)$ Casimirs, combined non-polynomially as in $\Pi_2$.\\

\noindent The observables $\mathcal R$ and $\Pi_{\rm biloc}$ are necessarily non-polynomial. Let $T\in U(\mathfrak{sp}(2K))$ act as a scalar $t_\pm$ on $\cH_\pm$. Since $U(\mathfrak g)$ is locally finite under the adjoint action, $T$ lies in a finite-dimensional, completely reducible $\operatorname{ad}\mathfrak g$-module $W$. Its image in $\operatorname{End}(\cH_+\oplus\cH_-)$ is $\mathfrak g$-invariant, hence equals the image of the invariant part $W^{\mathfrak g}\subset Z(U(\mathfrak g))$. Because every Casimir takes the same value on $\cH_+$ and $\cH_-$, $t_+=t_-$. 

The argument does not exclude polynomial observables whose spectra merely differ on the two sectors, such as joint spectra of $\mathfrak u(K)$ Casimirs. Therefore we have the same eigenvalues of all $\mathfrak{sp}(2K)$ Casimirs but different mode-space content, hence distinguished by a non-polynomial $\mathbb Z_2$ observable. In particular, the sector label is polynomial in the bilocals exactly when some Casimir takes different values on the two modules, as for $U(N)$ models and for $O(N)$ fermions.

\section{Sector-resolved characters and partition functions}
\label{sec:sectorcharacters}

The decomposition \cref{eq:AbelianizationSectorTheorem} suggests a natural refinement of the usual singlet partition function.  Instead of first summing all $[G,G]$ singlets, one should retain the character $\chi\in\widehat{G^{\rm ab}}$ and compute the contribution of the corresponding Howe module separately.

\subsection{The universal sector projector}
Let $U(g)$ denote the action of the color group on the microscopic Fock space.  For a one-dimensional character $\chi$ of $G^{\rm ab}$, pulled back to $G$, define
\begin{equation}
P_\chi=\int_G dg\,\overline{\chi(g)}\,U(g)\ .\label{eq:sectorprojector}
\end{equation}
With normalized Haar measure this projects onto the $\chi$-isotypic color subspace.  Introduce bilocal Cartan fugacities collectively as $\mathbf z^{\mathbf H}$.  The sector-resolved character is
\begin{align}
Z_\chi(\mathbf z)&=\Tr_{\cF}\left(P_\chi\,\mathbf z^{\mathbf H}\right)\nonumber\\
&=\int_Gdg\,\overline{\chi(g)}\Tr_{\cF}\left(U(g)\mathbf z^{\mathbf H}\right).
\label{eq:sectorpartitiongeneral}
\end{align}
Using the Howe decomposition \cref{eq:HoweDecomposition} and orthogonality of $G$ characters, only the color representation $V^G=\chi$ survives.  Consequently
\begin{equation}
Z_\chi(\mathbf z)={\rm ch}_{\Theta(\chi)}(\mathbf z)\ .\label{eq:sectorcharacterHowe}
\end{equation}
This is the character-level form of \cref{eq:AbelianizationSectorTheorem}.  Summing over the characters which actually occur gives
\begin{equation}
Z_{[G,G]}(\mathbf z)=\sum_{\chi\in\widehat{G^{\rm ab}}_{\rm occ}}Z_\chi(\mathbf z)\ .\label{eq:derivedpartition}
\end{equation}
Thus the microscopic group integral and the bilocal representation character are two descriptions of the same quantity.  This equality gives a direct way to test the proposed sector assignments. For $O(N)$ restricted to $\mathcal F^{SO(N)}$ the projectors onto $\mathcal H_{\pm}$ are $P_{\pm} = \frac12\pts*{\mathbf 1 \pm \mathcal R}$ with $\mathcal R$ as in \cref{eq:reflectioncharge}. The sector observable of \cref{subsec:sectorobservable} and the sector character are therefore two aspects of the same object.

\subsection{\texorpdfstring{$O(N)$}{O(N)}: even and odd characters}
For $O(N)$,
\begin{equation}
\chi_+(g)=1,\qquad \chi_-(g)=\det g.
\end{equation}
For $K$ equal-frequency bosonic modes, with $x=e^{-\beta\omega}$,
\begin{equation}
{\cal Z}^{O,b}_{\rm Fock}(g;x)=\frac{1}{\det(1-xg)^K},
\end{equation}
and hence
\begin{equation}
Z_\pm^{O,b}(x)=\int_{O(N)}dg\,\chi_\pm(g)\frac{1}{\det(1-xg)^K}\ .\label{eq:ONbosonsectorZ}
\end{equation}
For $K<N$ the odd integral vanishes because the determinant representation does not occur in the bosonic Fock space.  For $K\ge N$ both sectors occur.

For fermions,
\begin{equation}
{\cal Z}^{O,f}_{\rm Fock}(g;x)=\det(1+xg)^K,
\end{equation}
and
\begin{equation}
Z_\pm^{O,f}(x)=\int_{O(N)}dg\,\chi_\pm(g)\det(1+xg)^K\ .\label{eq:ONfermionsectorZ}
\end{equation}
Both sectors are realized.  In either statistics
\begin{equation}
Z_{SO(N)}=Z_++Z_-\ ,
\end{equation}
while the full $O(N)$-singlet partition function is $Z_+$. 

We introduce mode fugacities $y_p$ for the $B_{pp}$, keep $x=e^{-\beta\omega}$, and drop the zero-point energies (restore $x^{NK/2}$ for Weyl ordering). From the mode-space content \cref{eq:fermionONmodecontent,eq:ONmodecontent} one obtains, for all $N$ and $K$,
\begin{align}
Z^{O,b}_+(x;y)&=\sum_{\substack{\mu\ \mathrm{even}\\ \ell(\mu)\le\min(N,K)}}x^{\absv{\mu}}s_\mu(y)\,,\qquad
Z^{O,b}_-(x;y)=x^N\sum_{\substack{\nu\ \mathrm{even}\\ \ell(\nu)\le N}}x^{\absv{\nu}}s_{\nu+(1^N)}(y)\,,\label{eq:ONbosonSchur}\\
Z^{O,f}_\pm(x;y)&=\sum_{\mu\in\mathcal P_\pm,\ \mu_1\le K}x^{\absv{\mu}}s_{\mu^T}(y)\,,\label{eq:ONfermionSchur}
\end{align}
where $s_\mu$ are Schur polynomials, $\mathcal P_+$ is the set of partitions with all parts even and $\mathcal P_-$ the set of partitions with exactly $N$ parts, all odd; $Z^{O,b}_-=0$ for $K<N$. Several closed forms follow:
\begin{itemize}
\item For $N\ge K$, Littlewood's identity~\cite{Macdonald:1995} gives $Z^{O,b}_+(x;y)=\prod_{p\le q}\pts*{1-x^2y_py_q}^{-1}$.
\item For $K=N$, $s_{\nu+(1^N)}(y)=(y_1\cdots y_N)\,s_\nu(y)$, so that
\begin{equation}
Z^{O,b}_-(x;y)=x^N\,y_1\cdots y_N\,Z^{O,b}_+(x;y)=\frac{x^N\,y_1\cdots y_N}{\prod_{p\le q}\pts*{1-x^2y_py_q}}\,.\label{eq:ZminusKequalsN}
\end{equation}
At $K=N$ the odd sector is isomorphic to the even sector tensored with $\det$ as a $GL(K)$-module, but not as an $\mathfrak{sp}(2K)$-module.
\item For odd $N$, $-\mathbf 1\in O(N)$ has determinant $-1$ and acts as $x\to-x$, so $Z^{O,b}_-(x)=\frac12\sq*{I(x)-I(-x)}$, with $I$ the integral over $SO(N)$: the odd sector is the odd-degree part.
\item For $K>N$ and $y_p=1$, for example,
\begin{equation}
\begin{aligned}
	O(2),\ K=3:&\quad Z_+=\frac{1-x^6}{(1-x^2)^6}\,,&\qquad Z_-&=\frac{3x^2}{(1-x^2)^5}\,,\\
	O(2),\ K=4:&\quad Z_+=\frac{1+3x^2+6x^4}{(1-x^2)^7}\,,&\qquad Z_-&=\frac{x^2(6+3x^2+x^4)}{(1-x^2)^7}\,,\\
	O(3),\ K=4:&\quad Z_+=\frac{1-x^8}{(1-x^2)^{10}}\,,&\qquad Z_-&=\frac{4x^3}{(1-x^2)^{9}}\,.
\end{aligned}
\end{equation}
The factors $1-x^{2(N+1)}$ are the rank conditions $\det_{(N+1)\times(N+1)}\bar A=0$. In these examples both sectors have the same Gelfand--Kirillov dimension, $NK-\frac{N(N-1)}{2}$, and the same leading coefficient, so they have the same leading high-temperature behavior and differ only at subleading order.
\item For fermions, particle--hole conjugation maps the Fock vacuum to the completely filled state, which transforms as $(\det g)^K$; hence $Z_{\chi\cdot\det^K}(x)=x^{NK}Z_\chi(1/x)$. For $K$ odd this exchanges $Z^{O,f}_+$ and $Z^{O,f}_-$, while for $K$ even, each is self-conjugate. In both cases $\dim\cH_+=\dim\cH_-$. However self-conjugacy alone does not give equal dimensions, it follows from the outer automorphism of $\mathfrak{so}(2K)$ exchanging the two modules. 
\end{itemize}
The characters \cref{eq:ONbosonSchur} are those of the unitary positive-energy $\mathfrak{sp}(2K,\mathbb R)$ modules with the extremal weights of \cref{tab:extremalweights}~\cite{Kashiwara:1978kv}; Enright and Willenbring give resolutions and Hilbert series for exactly these modules~\cite{Enright:2004ew}. For fermions $Z^{O,f}_\pm$ is the Weyl character of the $\mathfrak{so}(2K)$ representation of highest weight $\pts*{(\frac N2)^{K-1},\pm\frac N2}$. All these formulas agree with the group integrals \cref{eq:ONbosonsectorZ,eq:ONfermionsectorZ}.

\subsection{\texorpdfstring{$U(N)$}{U(N)}: determinant-charge-resolved partition functions}
The residual determinant charge is measured by the linear Casimir $Q$ defined in \cref{eq:UNBlinearCasimir}. 
On the sector $\cH_n$, $Q|\Psi_n\rangle=n|\Psi_n\rangle$.  Since every bilocal contains equal total fundamental and antifundamental color charge, the bilocal algebra preserves $n$.

For $U(N)$, $\chi_n(U)=(\det U)^n$, and the exact sector projector gives
\begin{equation}
Z_n(x)=\int_{U(N)}dU\,(\det U)^{-n}{\cal Z}_{\rm Fock}(U;x)\ .\label{eq:UNsectorprojector}
\end{equation}
For the bosonic theory,
\begin{equation}
{\cal Z}^{U,b}_{\rm Fock}(U;x)=\frac{1}{\det(1-xU)^K\det(1-xU^{-1})^K},
\end{equation}
so
\begin{equation}
Z_n^{U,b}(x)=\int_{U(N)}dU\,(\det U)^{-n}\frac{1}{\det(1-xU)^K\det(1-xU^{-1})^K}\ .\label{eq:UNbosonZq}
\end{equation}
Using the Cauchy identity
\begin{equation}
\frac{1}{\det(1-xU)^K}=\sum_\lambda x^{|\lambda|}s_\lambda(U)s_\lambda(1^K),
\end{equation}
and the corresponding formula for $U^{-1}$, character orthogonality implies for $n\ge0$ that $\lambda=\mu+(n^N)$.  Therefore
\begin{equation}
Z_n^{U,b}(x)=x^{Nn}\sum_\mu x^{2|\mu|}s_\mu(1^K)s_{\mu+(n^N)}(1^K),\qquad n\ge0\ .
\label{eq:UNbosonSchurZq}
\end{equation}
For equal fundamental and antifundamental spectra, $Z_{-n}^{U,b}=Z_n^{U,b}$.  Since $s_{\mu+(|n|^N)}(1^K)$ vanishes when its Young diagram has more than $K$ rows,
\begin{equation}
\mathbb Z_{\rm occ} =\begin{cases}\{0\},&K<N,\\ \mathbb Z,&K\ge N.\end{cases}\label{eq:UNbosonqrestriction}
\end{equation}
Thus the restriction inferred from the determinant operator is reproduced directly by the partition function. With particle and antiparticle mode fugacities $y_p$ and $\bar y_p$, \cref{eq:UNbosonSchurZq} refines to $Z^{U,b}_n=x^{N\absv{n}}\sum_\mu x^{2\absv{\mu}}s_{\mu+(\absv{n}^N)}(y)\,s_\mu(\bar y)$ for $n\geq 0$ and with $y\leftrightarrow \bar y$ for $n<0$, in agreement with \cref{eq:UNmodecontent}. For $K=N$ this gives $Z^{U,b}_n=(x^Ny_1\cdots y_N)^n\,Z^{U,b}_0$ for $n\ge0$.

For fermions,
\begin{equation}
{\cal Z}^{U,f}_{\rm Fock}(U;x)=\det(1+xU)^K\det(1+xU^{-1})^K,
\end{equation}
so
\begin{equation}
Z_n^{U,f}(x)=\int_{U(N)}dU\,(\det U)^{-n}\det(1+xU)^K\det(1+xU^{-1})^K\ .\label{eq:UNfermionZq}
\end{equation}
Using the dual Cauchy identity, one finds that the bilocal $U(2K)$ representation in the charge-$n$ sector has highest weight
\begin{equation}
\lambda_n=\left(\underbrace{N,\ldots,N}_{K+n},\underbrace{0,\ldots,0}_{K-n}\right),\qquad -K\le n\le K\ .\label{eq:UNfermionlambdaq}
\end{equation}
Equivalently, $\cH_n=V^{U(2K)}_{(N^{K+n})}$.  For equal oscillator frequency,
\begin{equation}
Z_n^{U,f}(x)=x^{NK}s_{(N^{K+n})}\left(\underbrace{x,\ldots,x}_{K},\underbrace{x^{-1},\ldots,x^{-1}}_{K}\right),\qquad -K\le n\le K\ .
\label{eq:UNfermionSchurZq}
\end{equation}
The finite range is the character-theoretic form of the Pauli truncation displayed directly in \cref{subsec:UNModels}.  Finally,
\begin{align}
Z_{SU(N)}^{b}(x)&=\sum_{n\in\mathbb Z_{\rm occ}}Z_n^{U,b}(x),\\
Z_{SU(N)}^{f}(x)&=\sum_{n=-K}^{K}Z_n^{U,f}(x).
\end{align}

\subsection{\texorpdfstring{$Sp(N)$}{Sp(N)}: no sector refinement}
Since $[Sp(N),Sp(N)]=Sp(N)$, there is no non-trivial abelianization character and hence no sector index.  The only projector is
\begin{equation}
Z^{Sp}(x)=\int_{Sp(N)}dg\,{\cal Z}_{\rm Fock}^{Sp}(g;x)\ .
\end{equation}
For bosons and fermions,
\begin{equation}
{\cal Z}_{\rm Fock}^{Sp,b}(g;x)=\frac{1}{\det(1-xg)^K},\qquad
{\cal Z}_{\rm Fock}^{Sp,f}(g;x)=\det(1+xg)^K.
\end{equation}
The Pfaffian and Hafnian identities in \cref{subsec:SpNModels} explain microscopically why an epsilon insertion does not produce a second character: it is already a polynomial in the ordinary bilocal creation operators.  

For fermions the single module is the finite-dimensional $\mathfrak{sp}(2K)$ representation of highest weight $\pts*{\frac N2,\ldots, \frac N2}$, so that with fugacities $z_p$ for the Cartan generators
\begin{equation}
\chi(z) = sp_{(\frac{N}{2})^K}(z) = \frac{\det\sq*{z_j^{\frac N2 + K - i + 1} - z_j^{-\pts*{\frac N2 + K - i + 1}}}^{K}_{i,j=1}}{\det\sq*{z^{K - i + 1}_j - z^{-(K-i+1)}_{j}}_{i,j=1}^K}\,,
\end{equation}
where $sp_\lambda$ is the symplectic Schur polynomial.

\section{Collective realization around the new backgrounds}
\label{sec:collectivenewbackground}

We now solve the constraints of \cref{sec:constraintscentralchar} around each new background in terms of oscillators. As in the usual collective description~\cite{Jevicki:1979mb,Das:2003vw,deMelloKoch:2010wdf,deMelloKoch:2014vnt,Jevicki:2014mfa} we rescale $A\to A/\sqrt N$ and $\bar A\to\bar A/\sqrt N$, and introduce bosonic oscillators $\alpha_{pq}$ for the bilocal fluctuations. The new ingredient is a set of auxiliary oscillators, of the statistics opposite to that of the color oscillators, which carry the background. The realizations below reproduce the bilocal algebra \emph{exactly}; they are of Dyson--Maleev type~\cite{Dyson:1956zz,Maleev:1958}, i.e.\ non-unitary, $\bar\alpha_{pq}\neq\alpha_{pq}^\dagger$, as for the Holstein--Primakoff construction~\cite{Holstein:1940zp} in its Dyson form. The expansion in $1/N$ enters only when the auxiliary oscillators are expressed in terms of the bilocals and when the adjoint is computed (\cref{appx:collective}). Since the bosonic epsilon sectors require $K\ge N$, terms such as $(K+1)/N$ are not small there, and the exactness of the realization is essential. 

\subsection{$O(N)$ bosons}\label{subsec:collONboson}
As seen in \cref{sec:constraintscentralchar} the quadratic constraint holds on both sectors,
\begin{equation}
\bar A A + (K + 1 - N)B - B^2 = 0\,.
\end{equation}
We rescale the operators, $A \to \frac{1}{\sqrt N} A$ and $\bar A \to \frac{1}{\sqrt N} \bar A$, and thus we have,
\begin{equation}
\bar A_{pr} A_{rq} = B_{pq} + \frac1N\pts[\Big]{ B_{pr}B_{rq} - (K+1)B_{pq}}\,,\label{eq:SOnB-modifiedconstraint}
\end{equation}
and 
\begin{equation}
[A_{pq}, \bar A_{rs}] = \delta_{sp}\delta_{rq} + \delta_{rp}\delta_{sq} + \frac1N\pts[\Big]{\delta_{sp}B_{rq} + B_{sp}\delta_{rq} + \delta_{rp}B_{sq} + B_{rp}\delta_{sq}}\,.\label{eq:SOnB-modifiedcommutation}
\end{equation}
We introduce $\alpha_{pq}, \bar \alpha_{pq}$ and $\beta_p, \bar\beta_{p}$,
\begin{equation}
[\alpha_{pq}, \bar\alpha_{rs}] = \delta_{pr}\delta_{sq} + \delta_{ps}\delta_{rq}\,,
\end{equation}
\begin{equation}
\sets{\beta_p,\bar \beta_q} = \delta_{pq}\,.
\end{equation}
We have the ansatz for $\bar A, B$, 
\begin{equation}
\bar A_{pq} = \bar\alpha_{pq},\qquad B_{pq} = \bar\alpha_{pr}\alpha_{rq} + \bar\beta_{p}\beta_q\,.\label{eq:SOnB-ansatz}
\end{equation}
The auxiliary fermions carry a fixed fermion number $\sum_p\bar\beta_p\beta_p=N$. On the epsilon background they form a filled Fermi sea, while the oscillators $\alpha_{pq}$ describe bilocal excitations on top of it. This gives an algebraic origin for the Fermi-gas description of the baryon spectrum in~\cite{Sundborg:2026cut}; the field-theory version is discussed in \cref{subsec:oddground}.

From the action of $A$, it is natural to consider the ansatz 
\begin{equation}
A_{pq} = \alpha_{pq} + \frac{1}{N}\pts[\Big]{\bar\alpha_{rs}\alpha_{pr}\alpha_{sq} + \alpha_{pr}\bar\beta_{r}\beta_{q} + \alpha_{qr}\bar\beta_{r}\beta_{p}}\,.\label{eq:SOnB-solutionA}
\end{equation}
It is easy to check that the ansatz reproduces the bilocal algebra exactly. Substituting everything back into the rescaled constraint gives,
\begin{equation}
\begin{aligned}
&\bar A_{pr} A_{rq} - B_{pq} +\frac 1N\pts[\Big]{(K+1)B_{pq} - B_{pr}B_{rq}} \\
&= \frac 1N\pts[\Big]{\bar\alpha_{pr}\alpha_{qs}\bar\beta_{s}\beta_{r} - \bar\alpha_{rs}\alpha_{sq}\bar\beta_{p}\beta_{r}} -\bar\beta_p\beta_q + \frac 1N\bar\beta_r\beta_r\bar\beta_p\beta_q\,,
\end{aligned}\label{eq:constraint-remainder}
\end{equation}
In terms of the operator
\begin{equation}
\mathcal N_{pq} = \bar\alpha_{rq}\bar\beta_{p}\beta_r - \bar\alpha_{rp}\bar\beta_{q}\beta_r\,,\label{eq:SOnBnulloperator}
\end{equation}
the right-hand side of \cref{eq:constraint-remainder} takes the compact form
\begin{equation}
\bar A_{pr} A_{rq} - B_{pq} +\frac 1N\pts[\Big]{(K+1)B_{pq} - B_{pr}B_{rq}}=-\frac1N\sum_s\mathcal N_{ps}\alpha_{sq}+\frac{\sum_r\bar\beta_r\beta_r-N}{N}\,\bar\beta_p\beta_q\,.\label{eq:constraintcompact}
\end{equation}
The fermion number $\sum_r\bar\beta_r\beta_r$ commutes with the whole realization and labels its superselection sectors. The second term vanishes only for $\sum_r\bar\beta_r\beta_r=0$ and for $\sum_r\bar\beta_r\beta_r=N$: \emph{the constraint itself selects the two abelianization sectors}. The value $0$ gives $\cH_+$, and the value $N$, a filled Fermi sea, gives $\cH_-$.

Moreover,
\begin{equation}
[\bar A_{pq}, \mathcal N_{rs}] = 0\,.
\end{equation}
\begin{equation}
[B_{pq}, \mathcal N_{rs}] = \mathcal N_{ps}\delta_{qr} - \mathcal N_{pr}\delta_{qs}\,.
\end{equation}
and
\begin{gather}
\sq*{A_{pq}, \mathcal N_{rs}} = \frac{1}{N}\sq[\Big]{\delta_{ps}\mathcal N_{rt}\alpha_{tq} + \delta_{qs}\mathcal N_{rt}\alpha_{tp} - \delta_{qr}\mathcal N_{st}\alpha_{tp} - \delta_{pr}\mathcal N_{st}\alpha_{tq}}\,,\label{eq:ACcommutator}\\
\sq*{\mathcal N_{pq}, \mathcal N_{rs}} = \bar A_{qr}\mathcal N_{ps} - \bar A_{qs}\mathcal N_{pr} - \bar A_{pr}\mathcal N_{qs} + \bar A_{ps}\mathcal N_{qr}\,.
\end{gather}
The derivations, together with those for $\bar{\mathcal N}_{pq}=\alpha_{pr}\bar\beta_r\beta_q-\alpha_{qr}\bar\beta_r\beta_p$, are given in \cref{appx:collective}. Hence the image of $\mathcal N$ is invariant under the bilocal algebra: $\bar A$ commutes with $\mathcal N$, $B$ maps it to itself, and $A$ maps it to $\mathcal N\alpha$. It is a submodule, and the residual \cref{eq:constraintcompact} lies in it. On the quotient by this submodule the constraint holds exactly. Microscopically, the quotient implements the identity \cref{eq:specialONB}.

\subsection{$O(N)$ fermions}\label{subsec:collONFermions}
The quadratic constraint, valid on both sectors as seen in \cref{sec:constraintscentralchar}, is
\begin{equation}
\bar A A - (K+N-1)B + B^2  = 0\,.
\end{equation}
We rescale the operators to obtain, 
\begin{equation}
\bar A_{pr}A_{rq} = B_{pq} + \frac{1}{N}\pts*{(K-1)B_{pq} - B_{pr}B_{rq}}\,.
\end{equation}
The only commutator that changes is
\begin{equation}
\sq{A_{pq}, \bar A_{rs}} = \delta_{sp}\delta_{rq} - \delta_{sq}\delta_{rp} - \frac{1}{N}\pts*{B_{sp}\delta_{rq} + B_{rq}\delta_{sp} - B_{sq}\delta_{rp} - B_{rp}\delta_{sq}}\,.
\end{equation}
We introduce $\alpha_{pq}, \bar\alpha_{pq}, \bar\beta_p$ and $\beta_p$ to solve the constraints. Now $\beta, \bar\beta$ are bosonic oscillators with fixed boson number $N$ or 
\begin{equation}
\sum_{p}\bar\beta_p\beta_p = N\,.
\end{equation}
The commutation relations are
\begin{equation}
\sq{\alpha_{pq}, \bar\alpha_{rs}} = \delta_{ps}\delta_{rq} - \delta_{pr}\delta_{qs},\qquad \sq{\beta_p, \bar\beta_q} = \delta_{pq}\,.
\end{equation}
We use the ansatz,
\begin{equation}
\begin{gathered}
\bar A_{pq} = \bar\alpha_{pq},\qquad B_{pq} = \bar\alpha_{pr}\alpha_{rq} + \bar\beta_p\beta_q\,,\\
A_{pq} = \alpha_{pq} - \frac{1}{N}\pts*{\bar\alpha_{rs}\alpha_{pr}\alpha_{sq} + \alpha_{pr}\bar\beta_{r}\beta_{q} - \alpha_{qr}\bar\beta_{r}\beta_p}\,.
\end{gathered}
\end{equation}
When we substitute the above into the constraint (along with the fixed boson number condition), we get
\begin{equation}
\bar A_{pr}A_{rq} - B_{pq} -\frac{1}{N}\pts*{(K-1)B_{pq} - B_{pr}B_{rq}} = \frac{1}{N}\mathcal N_{pr}\alpha_{rq}\,,
\end{equation}
where 
\begin{equation}
\mathcal N_{pq} = \bar\beta_{p}\beta_r\bar\alpha_{rq} + \bar\beta_{q}\beta_{r}\bar\alpha_{rp}\,.
\end{equation}
$\mathcal N$ is an operator that generates the null states and must be quotiented out for the physically relevant sector. Microscopically this is equivalent to enforcing the identity \cref{eq:specialONB}.
\subsection{$U(N)$ bosons}\label{subsec:collUNboson}
Recall, the quadratic constraints in the sector of charge $n$ are,
\begin{gather}
\bar A A + (K+n -N)B - B^2 = 0\,,\\
A\bar A - (C+N\mathbb I)(C+(K+n)\mathbb I) = 0\,,
\end{gather}
We rescale the operators, $A \to \frac{1}{\sqrt N} A$ and $\bar A \to \frac{1}{\sqrt N}\bar A$, to obtain
\begin{gather}
\bar A_{pr} A_{rq} = B_{pq} + \frac{1}{N}\pts[\Big]{B_{pr}B_{rq} - (K+n)B_{pq}}\,,\\
A_{pr}\bar A_{rq} = C_{pq} + (K+n)\delta_{pq} + \frac{1}{N}\pts[\Big]{C_{pr}C_{rq} + (K+n)C_{pq}}\,,
\end{gather}
and 
\begin{equation}
[A_{pq}, \bar A_{rs}] = \delta_{ps}\delta_{rq} + \frac1N\pts*{C_{ps}\delta_{rq} + B_{rq}\delta_{ps}}\,.
\end{equation}
All other commutators remain unchanged. We introduce $\alpha_{pq}, \bar\alpha_{pq}, \beta_p, \bar\beta_p$ and  $\gamma_p,\bar \gamma_p$ to solve our constraints
\begin{gather}
\sq{\alpha_{pq}, \bar\alpha_{rs}} = \delta_{ps}\delta_{rq}\,,\\
\sets{\beta^k_p, \bar\beta^l_q} = \delta^{kl}\delta_{pq},\quad (k, l = 1,\ldots, n_\beta)\,,\\
\sets{\gamma^k_p, \bar\gamma^l_q} = \delta^{kl}\delta_{pq},\quad (k, l = 1,\ldots, n_\gamma)\,.
\end{gather}
The superscripts $k,l$ denote the flavor. Each flavor $k$ labels a single determinant: $\bar\beta^k_{p_1}\cdots\bar\beta^k_{p_N}$ for $\mathcal U^a(p_1,\ldots,p_N)$ and $\bar\gamma^k_{p_1}\cdots\bar\gamma^k_{p_N}$ for $\mathcal U^b(p_1,\ldots,p_N)$ respectively. They only appear summed, and we deal with the following objects
\begin{equation}
\bar\beta_p\beta_q = \sum_{k=1}^{n_\beta} \bar\beta^k_p\beta^k_q,\qquad \bar\gamma_p\gamma_q = \sum_{k=1}^{n_\gamma} \bar\gamma^k_p\gamma^k_q.
\end{equation}
They satisfy the following relations,
\begin{equation}
\sq{\bar\beta_p\beta_q, \bar\beta_r\beta_s} = \delta_{rq}\bar\beta_p\beta_s - \delta_{ps}\bar\beta_r\beta_q,\qquad \sq{\bar\gamma_p\gamma_q, \bar\gamma_r\gamma_s} = \delta_{rq}\bar\gamma_p\gamma_s - \delta_{ps}\bar\gamma_r\gamma_q\,.
\end{equation}
We also impose the restriction
\begin{equation}
\sum_{p}\bar\beta^k_p\beta^l_p = N\delta^{kl},\qquad \sum_{p}\bar\gamma^k_p\gamma^l_p = N\delta^{kl}.\label{eq:UNBflavorconstraint}
\end{equation}
The diagonal part of \cref{eq:UNBflavorconstraint} states that each flavor carries exactly $N$ quanta, and the off-diagonal part requires the auxiliary state to transform in a one-dimensional representation of $U(n_\beta)$ (respectively $U(n_\gamma)$); together they select $\det^N$. The constraint is therefore the Gauss law of an emergent $U(n_\beta)$ (respectively $U(n_\gamma)$) gauge symmetry rotating the auxiliary flavors. Since the bilocals depend on $\beta$ and $\gamma$ only through the flavor-summed combinations above, this rotation is a redundancy of the auxiliary description rather than a symmetry of the model. A natural ansatz follows,
\begin{equation}
\begin{gathered}
\bar A_{pq} = \bar\alpha_{pq},\quad B_{pq} = \bar\alpha_{pr}\alpha_{rq} + \bar\beta_p\beta_q,\quad C_{pq} = \bar\alpha_{rq}\alpha_{pr} + \bar\gamma_{q}\gamma_{p}\,,\\
A_{pq} = \alpha_{pq} + \frac{1}{N}\pts*{\bar\alpha_{rs}\alpha_{pr}\alpha_{sq} + \alpha_{pr}\bar\beta_r\beta_q + \alpha_{rq}\bar\gamma_r\gamma_p}\,.
\end{gathered}
\end{equation}
This ansatz reproduces the algebra exactly. The charge $n = n_\beta - n_\gamma$ of the operator $Q$ (\cref{eq:UNBlinearCasimir}) labels the module. Now we substitute the ansatz and $n$ in terms of $n_\beta, n_\gamma$ and note,
\begin{gather}
\bar A_{pr}A_{rq} - B_{pq} - \frac{1}{N}\pts*{B_{pr}B_{rq} - (K+n_\beta - n_\gamma)B_{pq}} = -\frac{1}{N}\mathcal N_{pr}\alpha_{rq} - \frac{n_\gamma}{N}\bar\beta_p \beta_q\,,\label{eq:UNBasymptoticconstraint1}\\
A_{pr}\bar A_{rq} - (K+n_\beta-n_\gamma)\delta_{pq} - C_{pq} - \frac{1}{N}\pts*{C_{pr}C_{rq} + (K+n_\beta-n_\gamma)C_{pq}} = \frac{1}{N}\mathcal N_{rq}\alpha_{pr} - \frac{n_\beta}{N}\bar\gamma_q \gamma_p\,,\label{eq:UNBasymptoticconstraint2}
\end{gather}
with 
\begin{equation}
\mathcal N_{pq} = \bar\beta_p\beta_r\bar\alpha_{rq} - \bar\gamma_q\gamma_r\bar\alpha_{pr} - (n_\beta - n_\gamma)\bar\alpha_{pq}\,.
\end{equation}
On the RHS of \cref{eq:UNBasymptoticconstraint1,eq:UNBasymptoticconstraint2} the terms containing $\mathcal N$ lie in its image, which is a submodule. Hence they vanish on the quotient. For the sake of argument, assume that we have the new background state constructed from both $\beta$ and $\gamma$ fermions and assume that we are at the lowest level ($\alpha$ annihilates the state), the first constraint gives
\begin{equation}
-\frac{1}{N}n_\gamma\bar\beta_p\beta_q\ketv\Psi\,,
\end{equation}
and the second gives
\begin{equation}
-\frac{1}{N}n_\beta\bar\gamma_q\gamma_p\ketv\Psi\,.
\end{equation}
Take the trace and we get $-n_\beta n_\gamma\ketv \Psi$ for both. For the constraints to simultaneously hold,  $n_\beta n_\gamma = 0$ or in other words, only one kind of auxiliary fermion is allowed to exist at a time.

\paragraph{$\bm{n>0}$ sector.} We now set $n_\gamma = 0$ and accordingly $n_\beta = n$. The constraints and the null operator are as follows,
\begin{align}
\bar A_{pr}A_{rq} - B_{pq} - \frac{1}{N}\pts*{B_{pr}B_{rq} - (K+n)B_{pq}} &= -\frac{1}{N}\mathcal N_{pr}\alpha_{rq}\,,\\
A_{pr}\bar A_{rq} - (K+n)\delta_{pq} - C_{pq} - \frac{1}{N}\pts*{C_{pr}C_{rq} + (K+n)C_{pq}} &= \frac{1}{N}\mathcal N_{rq}\alpha_{pr}\,,
\end{align}
\begin{equation}
\mathcal N_{pq} = \bar\beta_p\beta_r\bar\alpha_{rq} - n\bar\alpha_{pq}\,.
\end{equation}
It is obvious that $\sq{\bar\alpha, \mathcal N} = 0$ and therefore the constraints map every state into the subspace generated by $\mathcal N$ on the background state. When we quotient out those states, the constraints hold exactly. This is equivalent to the microscopic statement \cref{eq:specialUNB}
for every background $\ketv\Omega$ of type $\mathcal U^a$. Thus we can see that $\mathcal N$ is a null state generator. 

\paragraph{$\bm{n < 0}$ sector.} We now analogously set $n_\beta = 0$ and $n_\gamma = \absv n$. The logic of quotienting out the null states holds here as well, due to the analogous microscopic statement \cref{eq:specialUNB}
for every background $\ketv\Omega$ of type $\mathcal U^b$.
\subsection{$U(N)$ fermions}\label{subsec:collUNfermion}
After rescaling the operators we obtain
\begin{gather}
\bar A_{pr}A_{rq} = B_{pq} -\frac1N\pts*{B_{pr}B_{rq} - (K-n)B_{pq}}\,,\\
A_{pr}\bar A_{rq} = C_{pq} + (K-n)\delta_{pq} - \frac{1}{N}\pts*{C_{pr}C_{rq} + (K-n)C_{pq}}\,,
\end{gather}
and 
\begin{equation}
\sq{A_{pq}, \bar A_{rs}} = \delta_{ps}\delta_{rq} - \frac{1}{N}\pts*{C_{ps}\delta_{rq} + B_{rq}\delta_{ps}}\,.
\end{equation}
We introduce oscillators $\alpha_{pq}, \bar\alpha_{pq}, \beta_p, \bar\beta_p, \gamma_p$ and $\bar\gamma_p$ to solve our constraints. The auxiliary fields are now Schwinger bosons. 
\begin{gather}
\sq{\alpha_{pq}, \bar\alpha_{rs}} = \delta_{ps}\delta_{rq}\,,\\
\sq{\beta^k_p, \bar\beta^l_q} = \delta^{kl}\delta_{pq},\quad (k, l = 1,\ldots, n_\beta)\,,\\
\sq{\gamma^k_p, \bar\gamma^l_q} = \delta^{kl}\delta_{pq},\quad (k, l = 1,\ldots, n_\gamma)\,.
\end{gather}
Now $n_\beta, n_\gamma \leq K$, because each $\mathcal U^b$ or $\mathcal U^d$ fills one mode with all $N$ colors, and at most $K$ such modes fit. We once again deal with the following objects
\begin{equation}
\bar\beta_p\beta_q = \sum_{k=1}^{n_\beta} \bar\beta^k_p\beta^k_q,\qquad \bar\gamma_p\gamma_q = \sum_{k=1}^{n_\gamma} \bar\gamma^k_p\gamma^k_q.
\end{equation}
They satisfy the following relations,
\begin{equation}
\sq{\bar\beta_p\beta_q, \bar\beta_r\beta_s} = \delta_{rq}\bar\beta_p\beta_s - \delta_{ps}\bar\beta_r\beta_q,\qquad \sq{\bar\gamma_p\gamma_q, \bar\gamma_r\gamma_s} = \delta_{rq}\bar\gamma_p\gamma_s - \delta_{ps}\bar\gamma_r\gamma_q\,.
\end{equation}
We also impose the restriction
\begin{equation}
\sum_{p}\bar\beta^k_p\beta^l_p = N\delta^{kl},\qquad \sum_{p}\bar\gamma^k_p\gamma^l_p = N\delta^{kl}.\label{eq:UNFflavorconstraint}
\end{equation}
This implies that each flavor carries exactly $N$ quanta. The ansatz we use is,
\begin{equation}
\begin{gathered}
\bar A_{pq} = \bar\alpha_{pq},\quad B_{pq} = \bar\alpha_{pr}\alpha_{rq} + \bar\beta_p\beta_q,\quad C_{pq} = \bar\alpha_{rq}\alpha_{pr} + \bar\gamma_{q}\gamma_{p}\,,\\
A_{pq} = \alpha_{pq} - \frac{1}{N}\pts*{\bar\alpha_{rs}\alpha_{pr}\alpha_{sq} + \alpha_{pr}\bar\beta_r\beta_q + \alpha_{rq}\bar\gamma_r\gamma_p}\,.
\end{gathered}
\end{equation}
Substituting the above back into the constraints gives
\begin{gather}
\bar A_{pr}A_{rq} - B_{pq} + \frac{1}{N}\pts*{B_{pr}B_{rq} - (K-n_\beta + n_\gamma)B_{pq}} = \frac{1}{N}\mathcal N_{pr}\alpha_{rq} - \frac{n_\gamma}{N}\bar\beta_p \beta_q\,,\label{eq:UNFasymptoticconstraint1}\\
A_{pr}\bar A_{rq} - (K-n_\beta+n_\gamma)\delta_{pq} - C_{pq} + \frac{1}{N}\pts*{C_{pr}C_{rq} + (K-n_\beta+n_\gamma)C_{pq}} = -\frac{1}{N}\mathcal N_{rq}\alpha_{pr} - \frac{n_\beta}{N}\bar\gamma_q \gamma_p\,,\label{eq:UNFasymptoticconstraint2}
\end{gather}
with 
\begin{equation}
\mathcal N_{pq} = \bar\beta_p\beta_r\bar\alpha_{rq} - \bar\gamma_q\gamma_r\bar\alpha_{pr} + (n_\beta - n_\gamma)\bar\alpha_{pq}\,.
\end{equation}
For similar reasons, we can see explicitly that $\beta$ and $\gamma$ bosons can't coexist. And again we quotient out the submodule generated by $\mathcal N$ to obtain physical states. 

\subsection{Null operators and the selection of physical states}\label{subsec:nullquotient}
In every case the residual of the constraints is of the form (null operator)$\times\alpha$ plus a term that fixes the allowed auxiliary charges. In each case the null operator satisfies $\sq{\bar A,\mathcal N}=0$, $\sq{B,\mathcal N}\propto\mathcal N$ and $\sq{A,\mathcal N}\propto\mathcal N\alpha$. Its image is therefore a submodule, and the physical module is the quotient. This quotient is the oscillator form of the relations \cref{eq:specialONB,eq:specialUNB} on the lowest $GL(K)$ type.

For $O(N)$ bosons the compact form \cref{eq:constraintcompact} shows that the auxiliary fermion number can only be $0$ or $N$. Likewise, for $U(N)$ we saw that $n_\beta n_\gamma=0$. In both cases the abelianization sectors are not an input of the collective description but the only consistent values of the auxiliary charges.

The charge-$n$ sector of a $U(N)$ vector model is realized by $\absv{n}$ flavors of auxiliary oscillators of the \emph{opposite} statistics. The flavor constraint \cref{eq:UNBflavorconstraint} is the Gauss law of an emergent $U(\absv{n})$ gauge symmetry: its off-diagonal part requires the auxiliary state to transform in a one-dimensional representation of $U(\absv{n})$, and its diagonal part, which fixes $N$ quanta per flavor, selects $\det^N$. The auxiliary oscillators thus carry a dual pair $(U(\absv{n}),GL(K))$ in which the roles of the color rank $N$ and the charge $\absv{n}$ are exchanged with respect to the original one. Some facts support this reading:
\begin{enumerate}
\item The lowest $GL(K)$ type is $(\absv{n}^N)$ for bosonic color and $(N^{\absv{n}})$ for fermionic color, the transposes of one another. These are exactly the $\det^N$ sectors of $(U(\absv{n}),GL(K))$ skew and ordinary Howe duality on the auxiliary oscillators respectively.
\item Bosonic color needs $N$ auxiliary fermions per flavor in $K$ modes, so $K\ge N$. Fermionic color needs a $\det^N$ state of $U(\absv{n})$ bosons in $K$ modes, so $K\ge\absv{n}$.
\end{enumerate}
The $O(N)$ realizations of \cref{subsec:collONboson,subsec:collONFermions} fit the same pattern with a single flavor: a single Fermi sea for bosons and a single condensate for fermions. With one flavor the flavor constraint has no off-diagonal part, and the number of auxiliary quanta is superselected rather than imposed; the constraint itself restricts it to $0$ or $N$.

Together with the $N\to-N$ map in \cref{subsubsec:orthsymp} and the $(N,n)\to (-N,-n)$ map in \cref{subsubsec:unitary} for Casimir eigenvalues, this ties the bosonic and fermionic descriptions together.
\subsection{Fluctuation counting and phase space}\label{subsec:fluctuations}
Null vectors are generic at finite $N$. In the bosonic models they appear whenever $K>N$, in the vacuum sector at level $N+1$ (the rank conditions $\det_{(N+1)\times(N+1)}\bar A=0$) and in the epsilon sector already at level one: $\mathrm{Sym}^2\mathbb C^K\otimes\wedge^N\mathbb C^K=S_{(3,1^{N-1})}(\mathbb C^K)\oplus S_{(2,1^N)}(\mathbb C^K)$, and the second summand is null. Only for $K=N$ are both bosonic sectors free Fock spaces of $K(K+1)/2$ oscillators. The invariant notion of ``number of free oscillators'' is the Gelfand--Kirillov dimension, which is the same in every sector at fixed $(N,K)$~\cite{Enright:2004ew}: $NK-\frac{N(N-1)}2$ for $O(N)$ and $N(2K-N)$ for $U(N)$ when $K\ge N$. Heuristically this counts the $\bar\alpha_{pq}$ with at least one index inside the Fermi sea, $p\le N$; around the epsilon background the null relation identifies the remaining ones with particle--hole excitations of the auxiliary fermions. 

For fermionic color the modules are finite-dimensional, and the counting is as follows. For the sake of argument, let $n>0$. Then the extremal state is $\ketv\Omega=\sq*{\det_{1\le k,p\le n}\bar\beta^k_p}^N\ketv0$, which satisfies \cref{eq:UNFflavorconstraint} and has the modes $1,\ldots, n$ fully occupied, $\bar\beta_p\beta_p = N$ for $p\leq n$. (The state $\prod_k(\bar\beta^k_k)^N\ketv0$ has the same occupations but violates the off-diagonal part of \cref{eq:UNFflavorconstraint}.) On these modes $\ketv\Omega$ transforms as the $N$\textsuperscript{th} power of the determinant under the $GL(n)$ generated by $\bar\beta_p\beta_r$ with $p,r\le n$, while $\beta_r\ketv\Omega=0$ for $r>n$. Hence $\bar\beta_p\beta_r\ketv\Omega=N\delta_{pr}\ketv\Omega$ for $p\le n$ and any $r$, and with $n_\beta=n$, $n_\gamma=0$,
\begin{equation}
	\mathcal N_{pq}\ketv\Omega = \bar\beta_p\beta_r\bar\alpha_{rq}\ketv\Omega+n\bar\alpha_{pq}\ketv\Omega=(N+n)\bar\alpha_{pq}\ketv\Omega,\quad (p\leq n, \text{but for any }q)\,.
\end{equation}
This implies that the $nK$ states $\bar\alpha_{pq}\ketv\Omega$ with $p\leq n$ are null. Microscopically, $\bar A_{pq}\ketv\Omega=0$ identically for $p\le n$: the mode $p$ is Pauli-blocked.

The remaining free oscillators are the pairs $\bar\alpha_{pq}$ with $p>n$ and $q$ free ($K(K-n)$ of them in total) and $\frac{\bar\beta_{p}\beta_q}{\sqrt N}$ with $q \leq n$ and $p > n$ ($n(K-n)$ in total)\footnote{They are the Holstein--Primakoff fluctuations~\cite{Holstein:1940zp} of the auxiliary condensate with 
\[ \frac{1}{N}\sq{\bar\beta_q\beta_p, \bar\beta_p\beta_q} = \frac{1}{N}\pts*{\bar\beta_q\beta_q - \bar\beta_p\beta_p} \simeq 1\;\;(\text{on the background}) \]}. They constitute $(K+n)(K-n)$ free oscillators in total, which perfectly agrees with the dimension of the classical phase space through the background 
\begin{equation}
\operatorname{Gr}(K+n, 2K|\mathbb C) = \frac{U(2K)}{U(K+n)\times U(K-n)}\,.
\end{equation} 
This is consistent with what one would expect from the Pauli exclusion principle, and the fact that one has a compact phase space for fermions. Indeed, $\operatorname{Gr}(K+n,2K|\mathbb C)$ is the coadjoint orbit of $U(2K)$ through the highest weight $\pts*{N^{K+n},0^{K-n}}$, and its quantization at level $N$ reproduces the module.

\section{Abelianization sectors in the free vector model on \texorpdfstring{$\mathbb{R}\times S^2$}{R x S2}}
\label{sec: field theory}

In the preceding sections the oscillator label $p$ ran over $K$ modes of equal frequency. We now take the modes to be the single-particle states of the free vector model in three dimensions, quantized on $\mathbb{R}\times S^2$, so that energies are scaling dimensions of local operators. Since the number of modes is now infinite, the occurrence conditions of \cref{sec:abelianizationtheorem}, namely $K\ge N$ for the bosonic determinant sectors and $|n|\le K$ for the $U(N)$ fermions, are always satisfied. What replaces them is an energy cost: each non-trivial abelianization sector is separated from the vacuum by a gap that grows with $N$. In this section we determine this gap, the degeneracy and spin of the lowest states in each sector, and the lowest energy at fixed angular momentum.

The leading behavior of these energies is known: the lightest baryons of the free scalar and fermion theories have dimensions $\simeq\frac23N^{3/2}$ and $N$~\cite{Shenker:2011zf,Aharony:2015mjs}, and for closed shells the scalar result is exact~\cite{Sundborg:2026cut}. Here we complement these results with the exact finite-$N$ structure implied by the mode content of the sectors: the energy and degeneracy for arbitrary $N$, the spin content of the ground states, the $U(N)$ towers, and the dependence on angular momentum.

\subsection{Single-particle spectrum}
\label{subsec:spectrum}

We consider $N$ free conformally coupled scalars on $\mathbb{R}\times S^2$ with unit radius. The curvature coupling $\xi R\,\phi_i\phi_i/2$ with $\xi=1/8$ gives an effective mass squared $1/4$ on the sphere, so that
\begin{equation}
\omega_l=\sqrt{l(l+1)+\tfrac14}=l+\tfrac12\,,\qquad l=0,1,2,\ldots\,,
\label{eq:scalarspectrum}
\end{equation}
each with degeneracy $2l+1$. We label the modes by $p=(l,m)$ with $|m|\le l$; under the state--operator map they correspond to $\phi$ and its traceless descendants, with $\Delta_\phi=\tfrac12$. For the Majorana fermion the positive-frequency modes are labelled by $p=(j,m)$ with $j\in\mathbb{Z}_{\ge0}+\tfrac12$ and have $\omega_j=j+\tfrac12$, consistent with $\Delta_\psi=1$. The single-particle partition functions are
\begin{equation}
z_B(x)=\frac{x^{1/2}(1+x)}{(1-x)^2}\,,\qquad z_F(x)=\frac{2x}{(1-x)^2}\,,
\end{equation}
and in the $U(N)$ theories particles and antiparticles each carry this spectrum. Since the zero-point energies vanish in zeta-function regularization,\footnote{For the scalar, $\tfrac12\sum_l(2l+1)(l+\tfrac12)=\zeta(-2,\tfrac12)=0$; for the fermion, $\sum_n n^2=\zeta(-2)=0$.} the Weyl-ordering constants of \cref{sec:bilocalhowedualityfiniteN} drop out and $\ketv{0}$ has zero energy.

Both the Hamiltonian and the rotation generator are diagonal bilocals,
\begin{equation}
H=\sum_p\omega_p\,B_{pp}\,,\qquad J_z=\sum_p m_p\,B_{pp}\,,
\label{eq:HJcartan}
\end{equation}
and hence lie in the Cartan subalgebra of the bilocal algebra. Ordering the modes by energy, we can therefore read off the energy and spin of each extremal state of \cref{sec:extremalstates} directly from its weight. Moreover, the conformal generators are themselves color-singlet bilinears, so $\mathfrak{so}(3,2)$ preserves every abelianization sector and the lowest states in each sector are conformal primaries.

The lowest states of a sector are most easily found from its decomposition under the mode group $GL(K)$, \cref{subsec:modecontent}. To evaluate energies we use the fact that a weight vector of $S_\lambda(\mathbb C^K)$ corresponds to a semistandard tableau $T$ of shape $\lambda$ with entries in the mode labels, and by \cref{eq:HJcartan} carries
\begin{equation}
E[T]=\sum_{\square\in\lambda}\omega_{T(\square)}\,,\qquad J_z[T]=\sum_{\square\in\lambda}m_{T(\square)}\,.
\label{eq:tableaucharges}
\end{equation}
Every box thus contributes positive energy, and the entries in a given column are distinct modes.

\subsection{Lowest states of the odd sector}
\label{subsec:oddground}

The lowest states of the odd sector now follow from \cref{eq:tableaucharges}. By \cref{eq:ONmodecontent}, every state in the bosonic odd sector lies in some $S_\mu(\mathbb{C}^K)$ with $\ell(\mu)=N$. The first column of a tableau of this shape contains $N$ distinct modes, and every further box costs positive energy, so the energy is minimized by $\mu=(1^N)$ with the first column occupied by the $N$ lowest modes. To describe these modes we write
\begin{equation}
L\equiv\lfloor\sqrt N\rfloor\,,\qquad n\equiv N-L^2\in\{0,1,\ldots,2L\}\,,
\end{equation}
so that they fill the shells $l<L$ completely and occupy $n$ of the $2L+1$ modes of the shell $l=L$. The ground states of the bosonic odd sector are therefore the epsilon states built on the shells $l<L$ together with $n$ distinct modes of the shell $l=L$, with energy
\begin{equation}
E_N=\sum_{l=0}^{L-1}(2l+1)\Big(l+\frac12\Big)+n\Big(L+\frac12\Big)
=\frac{2L+1}{2}\left[N-\frac{L(L+1)}{3}\right].
\label{eq:ENexact}
\end{equation}
The only remaining freedom is the choice of which $n$ of the $2L+1$ modes of the open shell are occupied, so the ground space is $\wedge^nV_L$, where $V_L$ has spin $L$, and its dimension is
\begin{equation}
d_N=\binom{2L+1}{N-L^2}\,.
\label{eq:dN}
\end{equation}

For closed shells, $N=L^2$, \cref{eq:ENexact} reduces to $L(4L^2-1)/6$, in agreement with~\cite{Sundborg:2026cut}, where it was also noted that the ground state is degenerate for general $N$; \cref{eq:dN} makes this degeneracy explicit. In terms of the auxiliary fermions introduced in \cref{subsec:collONboson}, the bosonic ground states are filled Fermi seas, and the degeneracy \cref{eq:dN} is that of a partially filled shell. The energy \cref{eq:ENexact} is continuous across shell closures, since setting $n=2L+1$ reproduces the closed-shell value for $L+1$.\footnote{One may equally use the convention $L^2<N\le(L+1)^2$, in which the shell $l=L$ holds $1\le n\le 2L+1$ quanta; \cref{eq:ENexact} is unchanged.} Writing $\sqrt N=L+t$, where $t\in[0,1)$ is the fractional part of $\sqrt N$ and $n=t(2L+t)$, it takes the exact form
\begin{equation}
E_N=\frac23N^{3/2}+\Big(t(1-t)-\frac16\Big)\sqrt N+\frac{t(1-t)(1-2t)}{6}\,.
\label{eq:ENasymptotic}
\end{equation}
The leading term is the energy of a Fermi sea in the two-dimensional space of modes with linear dispersion.\footnote{On $\mathbb{R}\times S^{d-1}$ the scalar has $\omega_l=l+\frac{d-2}{2}$ with degeneracy approaching $2l^{d-2}/(d-2)!$ at large $l$, and the same argument gives $E_N\simeq\frac{d-1}{d}\big(\frac{(d-1)!}{2}\big)^{1/(d-1)}N^{d/(d-1)}$, for instance $E_N\simeq1.08\,N^{4/3}$ in four dimensions; the scaling $N^{d/(d-1)}$ was found in~\cite{Shenker:2011zf}. The fermionic gap is $\frac{d-1}{2}N$ in any $d$. The case $d=2$ requires separate treatment because of the scalar zero mode.} The second term is a shell correction, whose coefficient oscillates between $-\tfrac16$ at closed shells, where $E_{L^2}=\frac23L^3-\frac L6$, and $\tfrac1{12}$ at half filling, $t=\tfrac12$. The last term never exceeds $\sqrt3/108$ in magnitude.

The $SO(3)$ content of the ground space is that of $n$ equivalent electrons in an atomic shell, here without spin. Its character follows from the generating function of $n$-element subsets of $\{-L,\ldots,L\}$,
\begin{equation}
\mathrm{Tr}_{\wedge^nV_L}\,y^{J_z}=y^{\,n(n-1)/2-nL}\begin{bmatrix}2L+1\\ n\end{bmatrix}_y\,,
\qquad
\begin{bmatrix}M\\ n\end{bmatrix}_y=\prod_{k=1}^n\frac{1-y^{M-n+k}}{1-y^k}\,,
\label{eq:wedgecharacter}
\end{equation}
and the multiplicity of spin $J$ is the coefficient of $y^J$ minus that of $y^{J+1}$. The ground state is a rotational singlet for closed shells and a single spin-$L$ multiplet for $n=1$, and by the particle--hole symmetry $\wedge^nV_L\simeq\wedge^{2L+1-n}V_L$ also for $n=2L$. In general the spins extend up to $J_{\max}=n(2L+1-n)/2$; for example $\wedge^2V_2=V_3\oplus V_1$ and $\wedge^3V_3=V_6\oplus V_4\oplus V_3\oplus V_2\oplus V_0$.

Since the lowest states of a sector are primaries, these ground states are the lightest primaries in the odd sector, that is, the lightest baryons of the $SO(N)$ theory. \Cref{tab:smallN} lists them for small $N$. For $N=2$ the ground state is the $SO(2)$ current, which is indeed odd under the reflections in $O(2)$. Since $SO(2)\simeq U(1)$, the odd sector of the $N=2$ theory is the \emph{charge-conjugation}-odd sector of a complex scalar.

\begin{table}[t]
\centering
\begin{tabular}{cccccl}
\toprule
$N$ & $(L,n)$ & $E_N$ & $d_N$ & $SO(3)$ & primary operator\\
\midrule
1 & $(1,0)$ & $1/2$ & 1 & $V_0$ & $\phi^1$\\
2 & $(1,1)$ & $2$ & 3 & $V_1$ & $\varepsilon_{ij}\phi^i\partial_\mu\phi^j$\\
3 & $(1,2)$ & $7/2$ & 3 & $V_1$ & $\varepsilon_{ijk}\phi^i\partial_\mu\phi^j\partial_\nu\phi^k$\\
4 & $(2,0)$ & $5$ & 1 & $V_0$ & $\varepsilon_{ijkl}\,\varepsilon^{\mu\nu\rho}\phi^i\partial_\mu\phi^j\partial_\nu\phi^k\partial_\rho\phi^l$\\
5 & $(2,1)$ & $15/2$ & 5 & $V_2$ & \\
9 & $(3,0)$ & $35/2$ & 1 & $V_0$ & \\
\bottomrule
\end{tabular}
\caption{Lowest states of the bosonic odd sector for small $N$, with energy $E_N$, degeneracy $d_N$, $SO(3)$ content and, where simple, the corresponding primary operator.}
\label{tab:smallN}
\end{table}

For fermions the same argument applies with rows and columns exchanged. Every $GL(K)$ type in the fermionic odd sector has a first row of length $N$, each of whose boxes costs at least $\omega_{1/2}=1$, and the minimum is attained by the shape $(N)$ with all entries in the two modes of the shell $j=\frac12$. Hence
\begin{equation}
E^F_N=N\,,\qquad d^F_N=N+1\,,
\label{eq:EFN}
\end{equation}
and the ground space is $\mathrm{Sym}^NV_{1/2}\simeq V_{N/2}$. Because all colors can occupy a single mode, the gap grows only linearly in $N$~\cite{Aharony:2015mjs}, and the ground states $\varepsilon_{i_1\cdots i_N}\psi^{i_1}_{(\alpha_1}\cdots\psi^{i_N}_{\alpha_N)}$ form a single multiplet of maximal spin $N/2$. In terms of the auxiliary bosons of \cref{subsec:collONFermions}, this ground space is a condensate in the two-dimensional lowest shell, and $V_{N/2}$ consists of its rotations. The bosonic ground state at a closed shell, by contrast, is a rotational singlet with energy of order $N^{3/2}$. For $N=2$ the fermionic ground state is again the $SO(2)$ current.

\subsection{Determinant charge in the \texorpdfstring{$U(N)$}{U(N)} theories}
\label{subsec:UNcharge}

The same argument applies to the sectors $\mathcal{H}_n$ of the $U(N)$ theories, in which the color representation is $\det^n$, with Young diagram $(|n|^N)$. For $n>0$ the lowest states involve only particle oscillators, and for $n<0$ only antiparticle oscillators. In the bosonic theory the mode representation has the same shape, and each of its $|n|$ columns contains $N$ distinct modes, so that
\begin{equation}
E^{B}_n=|n|\,E_N\simeq\frac23\,|n|\,N^{3/2}\,.
\end{equation}
In the ground space every closed-shell mode is occupied $|n|$ times, while the open shell carries the $GL(2L+1)$ representation $S_{(|n|^{N-L^2})}$; the degeneracy $\dim S_{(|n|^{N-L^2})}(\mathbb{C}^{2L+1})$ reduces to \cref{eq:dN} for $|n|=1$.

In the fermionic theory the mode representation is the transpose $(N^{|n|})$, so the lowest states fill $|n|$ modes with all $N$ colors. The particle shells have $\omega=1,2,\ldots$ with $2,4,\ldots$ modes, and writing $|n|=M(M+1)+r$ with $0\le r\le 2M+1$ we find
\begin{equation}
E^{F}_n=N\left[\frac{M(M+1)(2M+1)}{3}+r(M+1)\right]\simeq\frac23\,N\,|n|^{3/2}\,,
\end{equation}
with degeneracy $\dim S_{(N^r)}(\mathbb{C}^{2M+2})$. The Pauli bound $|n|\le K$ of \cref{subsec:UNModels} is thus replaced by an energy cost. The two asymptotic forms are exchanged under $N\leftrightarrow|n|$, as expected from the exchange $\mu\leftrightarrow\mu^T$ between the bosonic and fermionic Howe decompositions, \cref{eq:bosonONmodecontent,eq:fermionONmodecontent}. This is the field theoretic counterpart of the statistics and rank flip of \cref{subsec:nullquotient}.

\subsection{Angular momentum and the yrast line}
\label{subsec:yrast}

We now ask how the lowest energy in the bosonic odd sector depends on the spin. Let $E_{\rm yr}(J)$ be the lowest energy among states of spin $J$ in $\mathcal{H}_-$. As $E_{\rm yr}$ is non-decreasing, it coincides with the lowest energy at fixed $J_z=J$. Minimizing at fixed $J_z$ is awkward because the constraint couples all occupied modes, so we consider instead the Legendre-transformed problem of minimizing $H-\Omega J_z$ for $0\le\Omega<1$. The energy of each box of a tableau is then replaced by
\begin{equation}
\omega_p-\Omega\,m_p=l-\Omega m+\tfrac12\ \ge\ (1-\Omega)\,l+\tfrac12\,,
\end{equation}
which is still positive, so by the argument of \cref{subsec:oddground} the minimizer is again an epsilon state built on $N$ distinct modes. Physically, it is the ground state in a frame rotating with angular velocity $\Omega$, and $\Omega<1$ is the condition that the equator not move faster than light.

Let $(E_*,J_*)$ denote the charges of the minimizer. Since it minimizes $E-\Omega J_z$ over all states,
\begin{equation}
E\ \ge\ E_*+\Omega\,(J_z-J_*)
\label{eq:supportingline}
\end{equation}
for every state in $\mathcal{H}_-$. For $J_z=J_*$ this gives $E\ge E_*$, so $(E_*,J_*)$ lies on the yrast line, and each value of $\Omega$ yields one point on it. As $\Omega$ increases from $0$ to $1$, $J_*$ grows from $0$ to $N(N-1)/2$, so varying $\Omega$ sweeps out the yrast line. Moreover, \cref{eq:supportingline} states that all states lie on or above the straight line $E=E_*+\Omega(J-J_*)$ in the $(J,E)$ plane, and this line touches the yrast line at $(J_*,E_*)$. It is therefore tangent to the yrast line, and $\Omega=dE_{\rm yr}/dJ$ at $J=J_*$.

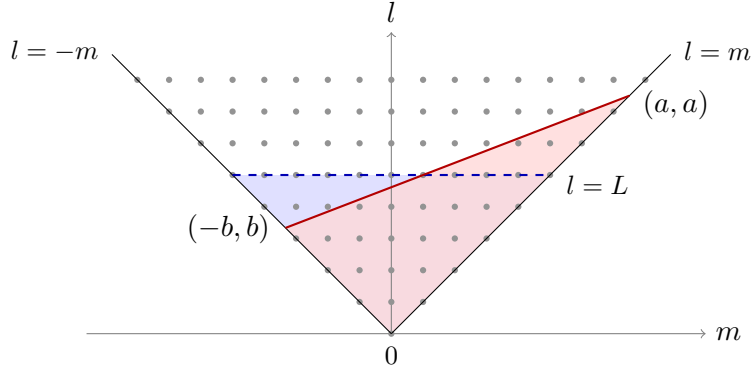
\begin{figure}[t]
\centering
\begin{tikzpicture}[scale=0.42]
\fill[blue!12] (0,0) -- (5,5) -- (-5,5) -- cycle;
\fill[red!15, opacity=0.8] (0,0) -- (7.5,7.5) -- (-3.333,3.333) -- cycle;
\draw[->, gray] (-9.6,0) -- (9.9,0) node[right, black] {$m$};
\draw[->, gray] (0,0) -- (0,9.5) node[above, black] {$l$};
\foreach \l in {0,...,8}{
	\pgfmathtruncatemacro{\mmin}{-\l}
	\foreach \m in {\mmin,...,\l}{\fill[gray!85] (\m,\l) circle (2.8pt);}
}
\draw[thin] (0,0) -- (8.8,8.8);
\draw[thin] (0,0) -- (-8.8,8.8);
\draw[dashed, thick, blue!70!black] (-5,5) -- (5,5);
\draw[thick, red!70!black] (-3.333,3.333) -- (7.5,7.5);
\node[below] at (0,-0.1) {\small $0$};
\node[right] at (7.6,7.2) {$(a,a)$};
\node[left] at (-3.5,3.3) {$(-b,b)$};
\node[right] at (5.2,4.7) {\small $l=L$};
\node[right] at (8.9,8.9) {\small $l=m$};
\node[left] at (-8.9,8.9) {\small $l=-m$};
\end{tikzpicture}
\caption{Modes of the conformally coupled scalar in the $(m,l)$ plane, one for each lattice point in the wedge $|m|\le l$. The shaded triangles have area $N=25$: the closed-shell region below $l=L=5$ (blue), and the region below the tilted line $l=\mu+\Omega m$ for $\Omega=5/13$ (red), whose vertices on the edges of the wedge are $(a,a)$ and $(-b,b)$. Up to boundary effects, the occupied modes are the lattice points inside these regions.}
\label{fig:wedge}
\end{figure}

To find the minimizer explicitly, note that on an epsilon state $H-\Omega J_z$ is a sum of independent contributions of the occupied modes, each mode contributing $l-\Omega m+\frac12$. The minimum over all choices of $N$ distinct modes is therefore obtained by occupying the $N$ modes with the smallest values of $l-\Omega m$, in the same way that fermions fill the lowest single-particle levels. Denoting by $\mu$ the largest occupied value, which plays the role of a Fermi level, the occupied modes are precisely those with
\begin{equation}
l\le\mu+\Omega m\,,
\end{equation}
that is, the lattice points of the wedge lying on or below the straight line $l=\mu+\Omega m$, whose slope in the $(m,l)$ plane is $\Omega$. The value of $\mu$ is fixed by requiring that $N$ modes lie below the line. For $\Omega=0$ the line is horizontal and the occupied modes form complete shells, as in \cref{subsec:oddground}. For $\Omega>0$ the line is tilted, so that modes with $m>0$ are favored and the state acquires angular momentum (\cref{fig:wedge}).

At large $N$ the sums over occupied modes can be replaced by integrals. The modes are the points of the integer lattice inside the wedge $|m|\le l$, so each mode occupies unit area in the $(m,l)$ plane, and to leading order
\begin{equation}
\sum_{\rm occupied}f(l,m)\simeq\int_T f(l,m)\,dm\,dl\,,
\end{equation}
where $T$ is the part of the wedge below the line. This region is a triangle. One vertex is the origin, and the other two are the intersections of $l=\mu+\Omega m$ with the edges $l=m$ and $l=-m$ of the wedge,
\begin{equation}
(m,l)=(a,a)\,,\quad (m,l)=(-b,b)\,,\qquad a=\frac{\mu}{1-\Omega}\,,\quad b=\frac{\mu}{1+\Omega}\,.
\end{equation}
Taking $f=1$, $f=l$ and $f=m$, and neglecting the zero-point contribution $\frac12$ per mode, gives $N$, $E$ and $J$ respectively. The area of the triangle is $\frac12\,\big|a\cdot b-(-b)\cdot a\big|=ab$. For a function linear in $(m,l)$, the integral over a triangle equals its area times the value of the function at the centroid, which is the average of the three vertices, $\big(\frac{a-b}{3},\frac{a+b}{3}\big)$. Hence
\begin{equation}
N\simeq ab\,,\qquad E\simeq ab\cdot\frac{a+b}{3}=\frac{N(a+b)}{3}\,,\qquad J\simeq ab\cdot\frac{a-b}{3}=\frac{N(a-b)}{3}\,.
\label{eq:triangleEJ}
\end{equation}
The first relation fixes the Fermi level, $\mu=\sqrt{N(1-\Omega^2)}$, and the other two then give
\begin{equation}
E\simeq\frac23\,\frac{N^{3/2}}{\sqrt{1-\Omega^2}}\,,\qquad J\simeq\frac23\,\frac{N^{3/2}\,\Omega}{\sqrt{1-\Omega^2}}=\Omega\,E\,.
\label{eq:EJofOmega}
\end{equation}
At $\Omega=0$ this reproduces the leading term of \cref{eq:ENasymptotic}. Eliminating $\Omega$ gives the yrast line,
\begin{equation}
E_{\rm yr}(J)^2-J^2\simeq\frac49N^3\simeq E_N^2\,,
\label{eq:yrasthyperbola}
\end{equation}
and one checks that $dE_{\rm yr}/dJ=J/E=\Omega$, as required by the Legendre construction. The same rotating ensemble $\mathrm{Tr}\,e^{-\beta(H-\Omega J_z)}$ was studied recently in the singlet sector at finite temperature~\cite{Kumar:2026mag}, where the Gross--Witten--Wadia transition of the free theory occurs at $T_c=\frac{\sqrt3}{\pi}\sqrt{N(1-\Omega^2)}$. This is the same combination as the Fermi level $\mu=\sqrt{N(1-\Omega^2)}$ found above, which extends to $\Omega\neq0$ the observation of~\cite{Sundborg:2026cut} that the deconfinement scale is set by the Fermi energy.

The hyperbolic form has a geometric origin. In the light-cone coordinates $u=l+m$ and $v=l-m$, the tilted triangle is the image of the closed-shell triangle under $(u,v)\mapsto(e^\eta u,e^{-\eta}v)$ with $\tanh\eta=\Omega$. This map preserves the wedge and the density of modes, as well as $E^2-J^2=(\sum u)(\sum v)$.\footnote{The argument relies on the uniform density of modes in the wedge, which is specific to three dimensions.} Correspondingly, $E\simeq E_N\cosh\eta$ and $J\simeq E_N\sinh\eta$.

The hyperbola \cref{eq:yrasthyperbola} rests on the continuum approximation, which requires the occupied triangle to be wide compared with the lattice spacing in every direction. To see when this fails, it is useful to organize the modes by the light-cone coordinate $k\equiv v=l-m\ge0$. For fixed $k$ the modes $(l,m)=(l,l-k)$ form a single-particle Regge trajectory: a quantum on it contributes $l+\frac12$ to $E$ and $l-k$ to $J$, and hence a fixed amount $k+\frac12$ to the twist $\tau\equiv E-J$. The leading trajectory $k=0$ consists of the modes $\partial_z^l\phi$, of twist $\Delta_\phi=\frac12$. In terms of these trajectories,
\begin{equation}
E-J_z=\sum_p\Big(k_p+\frac12\Big)B_{pp}\,,
\label{eq:twistsum}
\end{equation}
and the tilted triangle of \cref{fig:wedge} extends over the trajectories $0\le k\lesssim2b$. Since $ab=N$ and $a-b=3J/N$ by \cref{eq:triangleEJ}, at large spin ($J\gg E_N$) one has
$a\gg b$ and hence $b\simeq N/a\simeq N^2/(3J)$. Only about $2N^2/(3J)$
trajectories are therefore occupied, and the continuum approximation
breaks down once $J$ is of order $N^2$.

In this regime the yrast line can be found directly. Let $n_k$ be the number of quanta on the trajectory $k$. Since the modes on a trajectory have distinct $l$, the smallest angular momentum carried by $n_k$ quanta is $n_k^2/2$, up to terms of order $n_kk$, and is obtained by occupying the lowest modes; any larger value can be reached by moving quanta up the trajectory, at no cost in twist. The lowest twist at spin $J$ is therefore obtained by minimizing \cref{eq:twistsum} over the $n_k$, subject to $\sum_kn_k=N$ and $\sum_kn_k^2/2\le J$. Writing $n_k=N\nu_k$ and $J=jN^2$, the minimum is attained by occupations that decrease linearly along $k$, the discrete counterpart of the triangle:
\begin{equation}
\nu_k=\frac{(\beta-k)_+}{\sum_{k'\ge0}(\beta-k')_+}\,,
\qquad
j=\frac12\sum_k\nu_k^2\,,
\qquad
F(j)\equiv\sum_k k\,\nu_k\,,
\label{eq:scalingform}
\end{equation}
where $x_+=\max(x,0)$ and $\beta\ge1$ parametrizes the curve. At large $N$ with $j$ fixed, the yrast line then takes the scaling form
\begin{equation}
E_{\rm yr}(J)=J+N\Big[\frac12+F\big(J/N^2\big)\Big]+o(N)\,.
\label{eq:yrastscaling}
\end{equation}
For $j\to0$ many trajectories are occupied, $\beta\simeq2/(3j)$, and $\frac12+F(j)\simeq2/(9j)$, which reproduces the large-$J$ limit $E-J\simeq E_N^2/(2J)$ of \cref{eq:yrasthyperbola}. As $j$ increases the outer trajectories empty one by one; the third trajectory, for instance, is occupied only for $j<5/18$. For $1\le\beta\le2$ only $k=0,1$ remain, and
\begin{equation}
j=\frac{\beta^2+(\beta-1)^2}{2(2\beta-1)^2}\,,\qquad F=\frac{\beta-1}{2\beta-1}\,,
\end{equation}
so that $F\simeq\frac12-j$ as $j\to\frac12$. At $j=\frac12$ the last subleading trajectory empties and $F$ vanishes.

The endpoint of this process can be determined exactly at finite $N$. Since every state in the odd sector contains at least $N$ quanta, \cref{eq:twistsum} implies
\begin{equation}
E-J_z\ \ge\ \frac12\sum_pB_{pp}\ \ge\ \frac N2\,.
\end{equation}
Equality requires exactly $N$ quanta on the leading trajectory. These have distinct $l$, so that $J\ge0+1+\cdots+(N-1)$, and correspond to the primaries $\varepsilon_{i_1\cdots i_N}\prod_k\partial_z^{l_k}\phi^{i_k}$. Consequently
\begin{equation}
E_{\rm yr}(J)=J+\frac N2\,,\qquad J\ge J_c\equiv\frac{N(N-1)}{2}\,.
\label{eq:yrastlinear}
\end{equation}
Just below $J_c$ the yrast states have $N-x$ quanta on the leading trajectory and $x$ on the first subleading one. This gives
\begin{equation}
E_{\rm yr}(J)=J+\frac N2+x(J)\,,
\label{eq:nearthreshold}
\end{equation}
where $x(J)$ is the smallest non-negative integer with $J_c-Nx+x^2\le J$. The twist therefore grows by one each time $J$ decreases by about $N$ below $J_c$, and is constant above $J_c$. In the variables of \cref{eq:yrastscaling} this is the kink of $F$ at $j=\frac12$, where its slope jumps from $-1$ to $0$. The point $J_c$ is thus a sharp threshold of the yrast line rather than a crossover. In condensed-matter language it resembles a Lifshitz transition, at which the last occupied subband empties.

\begin{figure}[t]
\centering
\includegraphics[width=\textwidth]{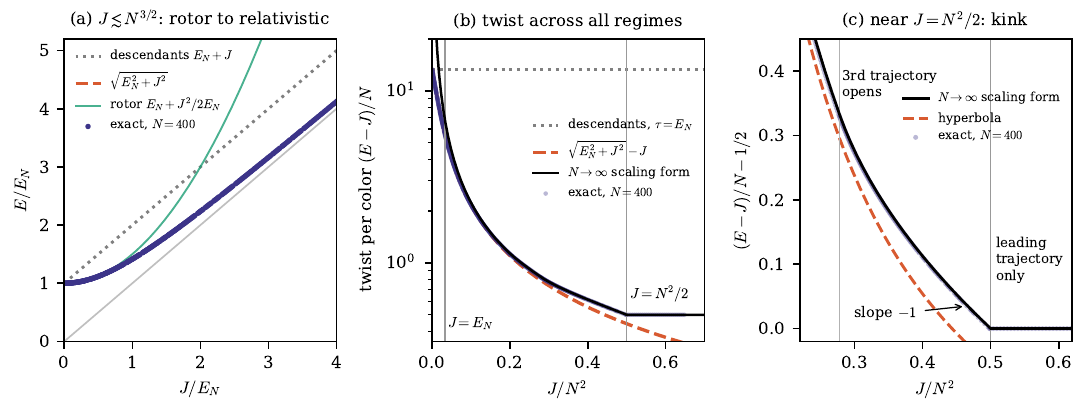}
\caption{Yrast line of the bosonic odd sector. Points are an exact minimization for $N=400$ ($E_N=5330$). (a) Small spin: the yrast line follows the hyperbola \cref{eq:yrasthyperbola}, which interpolates between the rotor $E_N+J^2/(2E_N)$ and the asymptote $E=J$, and lies below the descendants $E_N+J$ of the ground state. (b) Twist per color over the full range of $J$, together with the hyperbola, the large-$N$ scaling form \cref{eq:yrastscaling}, and the twist $E_N$ of the descendants. (c) Vicinity of $J=N^2/2$: the scaling form has a kink at the threshold, beyond which all quanta lie on the leading trajectory. The vertical line at $J/N^2=5/18$ marks where a third trajectory starts to be occupied.}
\label{fig:yrast}
\end{figure}

\Cref{tab:yrast} and \cref{fig:yrast} compare these results with an exact minimization for $N=400$. The hyperbola describes the yrast line accurately up to $J/N^2\approx0.2$, and the scaling form \cref{eq:yrastscaling} beyond; at $J_c$ the hyperbola gives $E-J\simeq4N/9$, compared with the exact value $N/2$. For open shells the yrast line is flat at $E_N$ over the spins contained in $\wedge^nV_L$, which extend up to $J_{\max}=O(N)$. This shell effect is of order $\sqrt N$ in energy and is not captured by the continuum approximation.

\begin{table}[t]
\centering
\begin{tabular}{cccc}
\toprule
$\Omega$ & $J$ & $E_{\rm yr}(J)$ & $\sqrt{E_N^2+J^2}$\\
\midrule
0 & 0 & 5330 & 5330\\
0.2 & 1080 & 5440 & 5438.3\\
0.5 & 3070 & 6153 & 6150.9\\
0.8 & 7098 & 8879 & 8876.4\\
0.99 & 36333 & 36734 & 36721.9\\
$\to1$ & 79800 & 80000 & 79977.8\\
\bottomrule
\end{tabular}
\caption{Yrast energies for $N=400$, where $E_N=5330$, obtained by minimizing $H-\Omega J_z$ over all choices of $N$ distinct modes, compared with \cref{eq:yrasthyperbola}. The last row is $J=J_c$, where \cref{eq:yrastlinear} holds exactly.}
\label{tab:yrast}
\end{table}

The yrast line therefore passes through three regimes. For $J\ll E_N$ it is that of a rotor with moment of inertia $E_N$, $E\simeq E_N+J^2/(2E_N)$. For $E_N\ll J\ll N^2$ it is that of an ultrarelativistic particle of mass $E_N$, $E\simeq J+2N^3/(9J)$; the crossover between these two regimes at $J\sim E_N\sim N^{3/2}$ is smooth. For $J\sim N^2$ the Fermi sea occupies only a few single-particle trajectories, and the yrast line follows \cref{eq:yrastscaling} until it reaches the leading trajectory at the threshold $J_c$.

The rotation along the yrast line is internal rather than rigid. The generator $P_+$ maps $(l,m)$ to $(l+1,m+1)$, and so raises $E$ and $J$ by one each while keeping every $k_p$ fixed. Acting on the ground state, it moves the Fermi sea rigidly along the trajectories, and the resulting descendants lie on $E=E_N+J$. The yrast states lie below this line because they reduce the twist by moving quanta towards the leading trajectory. Since $P_\mu$ raises $E$ by one and $J$ by at most one, it never lowers the twist. The yrast state of spin $J$ therefore belongs to the conformal multiplet of a primary with twist at most $\tau_{\rm yr}(J)<E_N$, and not to that of the ground state. As $J$ increases, primaries of progressively lower twist appear, until the twist reaches $N/2=N\Delta_\phi$, the value for $N$ free quanta on the leading trajectory. In the even sector, by contrast, the leading trajectory consists of the conserved higher-spin currents $\phi^i\partial_z^s\phi^i$, which contain only two quanta and have twist one.

For fermions $\omega_j-m=j-m+\tfrac12\ge\tfrac12$, so the same bound $E-J\ge N/2$ holds. It is already saturated by the ground states \cref{eq:EFN}, and at larger spin by placing all quanta in top modes $(j,m=j)$, with repetitions allowed. Hence
\begin{equation}
E^F_{\rm yr}(J)=J+\frac N2\,,\qquad J\in\frac N2+\mathbb{Z}_{\ge0}\,,
\end{equation}
and the fermionic yrast line is linear from the outset.

\subsection{Sector-resolved partition functions on \texorpdfstring{$S^1\times S^2$}{S1 x S2}}
\label{subsec:S1S2}

With the physical spectrum, the equal-frequency Fock-space traces of \cref{sec:sectorcharacters} are replaced by plethystic exponentials,
\begin{align}
\mathcal{Z}^{B}_{\rm Fock}(g;x)&=\exp\sum_{k\ge1}\frac{z_B(x^k)}{k}\,\mathrm{tr}\,g^k\,,\\
\mathcal{Z}^{F}_{\rm Fock}(g;x)&=\exp\sum_{k\ge1}\frac{(-1)^{k+1}z_F(x^k)}{k}\,\mathrm{tr}\,g^k\,.
\end{align}
Splitting $O(N)$ into its two components, each with normalized Haar measure, the sector partition functions become
\begin{equation}
Z_\pm(x)=\frac12\left[\int_{SO(N)}\!dg\;\mathcal{Z}_{\rm Fock}(g;x)\pm\int_{O^-(N)}\!dg\;\mathcal{Z}_{\rm Fock}(g;x)\right],
\end{equation}
and $Z_{SO(N)}=Z_++Z_-$. The results of \cref{subsec:oddground} fix the leading terms,
\begin{equation}
Z^B_-(x)=\binom{2L+1}{N-L^2}\,x^{E_N}\big[1+O(x)\big]\,,
\qquad
Z^F_-(x)=(N+1)\,x^{N}\big[1+O(x)\big]\,,\label{eq:Zleading}
\end{equation}
while $Z_+(x)=1+O(x)$. The corrections are of order $x$ because the next states in each sector cost at least one unit of energy, either by moving a quantum to a higher shell or by acting with $P_\mu$. A finite-$N$ evaluation of these integrals therefore provides a direct check of the sector assignments, and of \cref{eq:ENexact} and \cref{eq:dN}. Equivalently, $Z_\pm(x)=\Tr_{\cF^{SO(N)}}\sq*{\frac12\pts*{1\pm\mathcal R}x^{H}}$ with the reflection charge \cref{eq:reflectioncharge}. Evaluating the plethystic integrals exactly for small $N$ gives
\begin{center}
\renewcommand{\arraystretch}{1.25}
\small
\begin{tabular}{lll}
\toprule
model on $S^1\times S^2$ & $Z_+(x)$ & $Z_-(x)$\\
\midrule
$O(2)$ scalars & $1+x+4x^2+15x^3+43x^4+\cdots$ & $3x^2+11x^3+39x^4+112x^5+\cdots$\\
$O(3)$ scalars & $1+x+4x^2+15x^3+43x^4+\cdots$ & $3x^{7/2}+19x^{9/2}+73x^{11/2}+\cdots$\\
$O(4)$ scalars & $1+x+4x^2+15x^3+43x^4+\cdots$ & $x^5+16x^6+75x^7+\cdots$\\
$O(2)$ Majorana & $1+x^2+8x^3+19x^4+\cdots$ & $3x^2+8x^3+22x^4+48x^5+\cdots$\\
$O(3)$ Majorana & $1+x^2+8x^3+19x^4+\cdots$ & $4x^3+12x^4+38x^5+\cdots$\\
$O(4)$ Majorana & $1+x^2+8x^3+19x^4+\cdots$ & $5x^4+16x^5+54x^6+\cdots$\\
\bottomrule
\end{tabular}
\end{center}
The leading terms reproduce $\binom{2L+1}{N-L^2}x^{E_N}$ for the scalars ($3x^2$, $3x^{7/2}$, $x^5$) and $(N+1)x^N$ for the fermions, as stated in \cref{eq:Zleading}.

In large-$N$ analyses of the singlet sector the integral over $O^-(N)$ is usually replaced by that over $SO(N)$, since the two agree perturbatively in $1/N$ (see e.g.~\cite{Kumar:2026mag}). Their difference is precisely $2Z_-$, which is of order $x^{E_N}$. The thermodynamics of single-baryon states was studied in~\cite{Sundborg:2026cut} in a grand-canonical approximation, leading to a baryon condensation temperature slightly below the deconfinement temperature of~\cite{Shenker:2011zf}. The projector above gives instead the exact canonical partition function of the whole odd sector, including the dressing of baryons by bilinears and all finite-$N$ relations. Writing $I_\pm$ for the integrals over $SO(N)$ and $O^-(N)$, the ratio $Z_-/Z_+=(I_+-I_-)/(I_++I_-)$ provides the canonical counterpart of the condensation criterion of~\cite{Sundborg:2026cut}.

\section{Discussion}\label{sec:discussion}
In refs.~\cite{deMelloKoch:2026dfo,Ahn:2026ohj} the finite-$N$ Hilbert space of bilocal holography was identified with a single irreducible module of the master algebra, namely the one paired by Howe duality with the trivial representation of the color group. In this paper we have studied what happens when the singlet condition is imposed only for the derived subgroup $[G,G]$, as is natural for the color groups $SO(N)$ and $SU(N)$. The Hilbert space is then a direct sum of modules labelled by the characters of the abelianization $G^{\rm ab}=G/[G,G]$ that occur in the oscillator Fock space: an even and an odd sector for $O(N)$, a tower labelled by an integer determinant charge $n$ for $U(N)$, and no additional sector for $Sp(N)$. Bose/Fermi statistics decide which of these sectors are realized. Each new sector is built on an epsilon or determinant background with its own extremal weight, and its $GL(K)$ content follows from $GL(N)\times GL(K)$ duality. For $U(N)$ the quadratic constraints and the Casimirs depend on the charge $n$. For $O(N)$ the quadratic constraints are common to both sectors, which differ instead in relations that hold only on the lowest $GL(K)$ type. The character of each sector follows from a group-integral projector. The center of the bilocal enveloping algebra, however, does not always separate the sectors, a point to which we return below. The resulting picture clarifies the scope of bilocal holography at finite $N$. Although bilocal operators generate excitations within each irreducible sector, they cannot connect sectors carrying different abelianization charges. The full $SO(N)$ or $SU(N)$ singlet Hilbert space therefore requires distinct backgrounds, rather than excitations built on a single Fock vacuum.

The same structure governs the collective description. Around each new background the finite-$N$ constraints are solved by bilocal oscillators together with auxiliary oscillators of the opposite statistics, which carry the background. For $U(N)$ the charge-$n$ sector requires $\absv n$ flavors of auxiliary oscillators, whose flavor constraint is the Gauss law of an emergent $U(\absv n)$ gauge symmetry, so that color rank and sector charge trade places, $N\leftrightarrow \absv n$. In every case, the residual of the constraints is generated by a null operator whose image is a submodule, and the quotient by it is the oscillator form of the relations on the lowest $GL(K)$ type. The abelianization sectors are not an input of this description but the only consistent values of the auxiliary charges.

In the free vector model on $\mathbb{R}\times S^2$ the new sectors consist of baryonic primaries, whose energies, degeneracies and spin content we determined exactly at finite $N$; the exchange $N\leftrightarrow\absv n$ reappears in the energies of the charge-$n$ sectors. For $O(N)$ bosons the lowest energy at fixed angular momentum behaves, for $J\ll N^2$, as that of a relativistic particle of mass $E_N$. At $J\sim N^2$ it follows a scaling form, and it reaches a Regge trajectory of twist $N/2$ at $J=N(N-1)/2$. In this sense the extremal states, the constraints, the characters, the collective fields and the baryon spectrum are different expressions of the same representation-theoretic structure.

The $O(N)$ bosonic sectors also clarify what the finite-$N$ constraints can and cannot determine. In ref.~\cite{Ahn:2026ohj} the trace relations were identified with Casimir identities of the selected bilocal module, so that the quadratic constraints fix its central character. For $U(N)$ this information suffices: the determinant charge is itself central, through the linear Casimir $Q$, and the constraints of different sectors differ by the shift $K\to K\pm n$. For $O(N)$ fermions the even Casimirs agree, but the modules are finite-dimensional and the Pfaffian separates them. For $O(N)$ bosons, on the other hand, the two sectors obey the same quadratic constraint and every Casimir takes the same value on them, since their shifted highest weights $\lambda+\rho$ are related by an element of the Weyl group; this is possible only because the modules are infinite-dimensional. The finite-$N$ constraints therefore fix the central character of the bilocal Hilbert space, but not the Hilbert space itself. What distinguishes the two sectors is the baryon parity $\mathcal R$, which is a bounded function of the bilocals but not a polynomial in them, or equivalently the sector-resolved characters $Z_\pm$. In a collective description built on the quadratic constraints alone, the sector label must therefore be supplied as additional data, much like a superselection label.

A recurring theme of our analysis is a flip of statistics between the color oscillators and the auxiliary oscillators that carry the new backgrounds. For bosonic color the epsilon tensor antisymmetrizes the mode labels, so the background is a filled Fermi sea of auxiliary fermions, and the condition $K\ge N$ for the odd sector is the Pauli principle for these fermions. For fermionic color the mode labels are symmetrized, the auxiliary oscillators are bosons, and the background is a condensate in which all colors occupy the same mode; its fluctuations are of Holstein--Primakoff type and its phase space is compact. For the charge-$n$ sector of $U(N)$ fermions the bound $K\ge\absv n$ is then the condition that the $\absv n$ auxiliary flavors condense into distinct modes. On $\mathbb R\times S^2$ the Fermi surface becomes sharp, at the shell $l=L\simeq\sqrt N$, and the two pictures lead to different yrast lines. The Fermi sea must deform to carry angular momentum, which produces the relativistic dispersion described above, whereas the condensate simply moves to a single mode of high angular momentum, giving $E=J+N/2$ from the outset. Together with the exchange $N\leftrightarrow\absv n$ and the map $N\to-N$ of the Casimirs, this flip suggests a common correspondence between bosonic and fermionic descriptions. It would be interesting to understand its relation to the duality between bosonic and fermionic finite-$N$ Hilbert spaces with different numbers of modes discussed in ref.~\cite{deMelloKoch:2026dfo}. An important open question is whether these new collective backgrounds admit a direct semiclassical interpretation in the higher-spin bulk theory. In particular, the auxiliary Fermi seas arising for bosonic vector models may provide a useful starting point for identifying the bulk description of these non-perturbative sectors.

Another direction concerns Chern--Simons-matter theories, in which the color group is gauged with a Chern--Simons term at level $k$ and the choice between $U(N)$ and $SU(N)$, or between $O(N)$ and $SO(N)$, is part of the definition of the theory. In the $SU(N)$ theories the determinant sectors are the baryons, while in the $U(N)$ theories their role is played by monopole operators, and the two are exchanged under the bosonization dualities~\cite{Aharony:2015mjs}. Our results describe these sectors in the free limit $k\to\infty$. It would be interesting to follow the determinant towers to finite $k$, and to see whether the exchange $N\leftrightarrow\absv n$ between bosonic and fermionic charge-$n$ sectors has a counterpart in the map between baryons and monopoles. We hope to return to these questions in future work.

\acknowledgments
We would like to acknowledge discussions with Euihun Joung and Seok Kim. The work of A.J. is supported by the U.S. Department of Energy under contract DE-SC0010010. The work of RdMK was supported by a start up research fund of Huzhou University, a Zhejiang Province talent award and by a Changjiang Scholar award. J.Y. was supported by the National Research Foundation of Korea (NRF) grant funded by the Korean government (MSIT) (RS-2022-NR069038). J.Y. was supported by Brain Pool program funded by the Ministry of Science and ICT through the National Research Foundation of Korea (RS-2025-25457100).
B. A. was supported by Basic Science Research Program through the National Research Foundation of Korea (NRF) funded by the Korea government (MSIT) (RS-2026-25488427).

\appendix
\section{Conventions}\label{appx:conventions}
\subsection{Bilocal Algebras}\label{subsec:bilocalalgebras}
\paragraph{Orthogonal and symplectic algebras.}
Recall that we may treat the $O(N)$ and $Sp(N)$ vector models of both kinds (bosonic and fermionic) simultaneously with the help of two parameters: $\varepsilon, \xi = \pm 1$. We choose $\varepsilon = 1$ for $\mathfrak{sp}(2K)$ algebras and $\varepsilon = -1$ for $\mathfrak{so}(2K)$ algebras, and $\xi = 1$ for bosonic models and $\xi = -1$ for fermionic models. 
It follows,
\begin{gather}
A_{pq} = \varepsilon A_{qp},\qquad \bar A_{pq} = \varepsilon \bar A_{qp}\,,\\
B^{\rm W}_{pq} = B_{pq} + \xi\frac{N}{2}\delta_{pq}\,,
\end{gather}
and 
\begin{equation}
(A_{pq})^\dagger = \bar A_{qp}\,,
\end{equation}
where above $\bar A, A$ and $B$ are the pair-creation, pair-annihilation and number-preserving bilocals respectively. They close the algebra
\begin{gather}
[A_{pq}, \bar A_{rs}] = B^{\rm W}_{rp}\delta_{sq} + B^{\rm W}_{sq}\delta_{rp} + \varepsilon B^{\rm W}_{rq}\delta_{sp} + \varepsilon B^{\rm W}_{sp}\delta_{rq}\,,\\
[A_{pq}, A_{rs}] = 0 = [\bar A_{pq}, \bar A_{rs}]\,,\\
[B^{\rm W}_{pq}, B^{\rm W}_{rs}] = B^{\rm W}_{ps}\delta_{rq} - B_{rq}^{\rm W}\delta_{ps}\,,\\
[A_{pq}, B^{\rm W}_{rs}] = A_{ps}\delta_{rq} + \varepsilon A_{qs}\delta_{rp}\,,\\
[B^{\rm W}_{pq}, \bar A_{rs}] = \bar A_{ps}\delta_{rq} + \varepsilon \bar A_{pr}\delta_{sq}\,.
\end{gather}

\paragraph{Unitary algebras.} For $U(N)$ vector models, one has both fundamental and antifundamental oscillators. The algebra is closed by the pair-creation, pair annihilation bilocals $\bar A, A$ respectively along with \emph{two} number-preserving bilocals $B$ and $C$,
\begin{gather}
[A_{pq}, \bar A_{rs}] = N\delta_{ps}\delta_{rq} + \xi C_{ps}\delta_{rq} + \xi B_{rq}\delta_{ps}\,,\\
[A_{pq}, A_{rs}] = [\bar A_{pq}, \bar A_{rs}] = [B_{pq}, C_{rs}] = 0\,,\\
[B_{pq}, B_{rs}] = B_{ps}\delta_{rq} - B_{rq}\delta_{ps}\,,\\
[C_{pq}, C_{rs}] = -C_{ps}\delta_{rq} + C_{rq}\delta_{ps}\,,\\
[A_{pq}, B_{rs}] = A_{ps}\delta_{rq}\,,\\
[A_{pq}, C_{rs}] = A_{rq}\delta_{ps}\,,\\
[B_{pq}, \bar A_{rs}] = \bar A_{ps}\delta_{rq}\,,\\
[C_{pq}, \bar A_{rs}] = \bar A_{rq}\delta_{ps}\,.
\end{gather}
Above, we again used $\xi=1$ for the bosonic model and $\xi = -1$ for the fermionic model. 

\subsection{Generators for $O(N)$ master algebras}\label{subsec:generatormatricesON}
\paragraph{Bosons.}
We have the usual bosonic oscillators,
\begin{equation}
[a_i(p), a^\dagger_j(q)] = \delta_{ij}\delta_{pq}\,.
\end{equation}
We introduce the quadratures, 
\begin{equation}
\begin{aligned}
\chi^1_j(p) &= \frac1{\sqrt2}(a_j(p) + a^\dagger_j(p))\,,\\
\chi^2_j(p) &= \frac i{\sqrt2}(a_j(p) - a^\dagger_j(p))\,.
\end{aligned}
\end{equation}
Therefore,
\begin{equation}
[\chi^a_j(p),\chi^b_k(q)] = -i\varepsilon^{ab}\delta_{jk}\delta_{pq},\quad \text{with $\varepsilon^{12} = 1$}\,.
\end{equation}
Using the above, we define an operator symmetric in mode indices,
\begin{equation}
L_{ab}(p,q) \equiv \sum_{j=1}^N\frac{1}{2}\sets{\chi^a_j(p),\chi^b_j(q)}\,.
\end{equation}
One may check that we obtain,
\begin{equation}	
\begin{aligned}
	i[L_{ab}(p,q),L_{cd}(r,s)] &= \varepsilon^{bc}\delta_{qr}L_{ad}(p,s) + \varepsilon^{ac}\delta_{pr}L_{bd}(q,s)\\ 
	&\qquad + \varepsilon^{bd}\delta_{qs}L_{ac}(p,r) + \varepsilon^{ad}\delta_{ps}L_{bc}(q,r)\,,
\end{aligned}
\end{equation}
We may use the collective indices $(a,p)\mapsto P\equiv (a-1)K + p$ and define $\mathcal J_{PQ} = \varepsilon^{ab}\delta_{pq}$ and obtain
\begin{equation}
[L_{PQ}, L_{RS}] = -iL_{PS}\mathcal J_{QR} - iL_{QS}\mathcal J_{PR} -iL_{PR}\mathcal J_{QS} - iL_{QR}\mathcal J_{PS}\,.
\end{equation}
And we finally obtain the $\mathfrak{sp}(2K,\mathbb R)$ generator $\mathcal L_{PQ}$,
\begin{equation}
\mathcal L_{PQ} = -\mathcal J_{PR}L_{RQ}\,.
\end{equation}
Explicitly in terms of the bilocals we have,
\begin{align}
\mathcal L_{pq} &= \frac i2\pts*{\bar A_{pq} - A_{pq}  + B^{\rm W}_{pq} - B^{\rm W}_{qp}}\,,\\[.5\baselineskip]
\mathcal L_{p(K+q)} &= \frac12\pts*{A_{pq} + \bar A_{pq} - B^{\rm W}_{pq} - B^{\rm W}_{qp}}\,,\\[.5\baselineskip]
\mathcal L_{(K+p)q} &= \frac12\pts*{A_{pq} +\bar  A_{pq} + B^{\rm W}_{pq} + B^{\rm W}_{qp}}\,,\\[.5\baselineskip]
\mathcal L_{(K+p)(K+q)} &= \frac i2\pts*{A_{pq} - \bar A_{pq} + B^{\rm W}_{pq} - B^{\rm W}_{qp}}\,.
\end{align}
The quadratic Casimir in terms of $\mathcal L$ is,
\begin{equation}
C_2 = -\Tr(\mathcal L^2) = 2\Tr((B^{\rm W})^2) - \Tr(\bar A A + A \bar A)\,.
\end{equation}

\paragraph{Fermions.} We have the usual fermionic oscillators,
\begin{equation}
\sets{b_i(p), b^\dagger_j(q)} = \delta_{ij}\delta_{pq}\,.
\end{equation}
We introduce the Majorana operators, 
\begin{equation}
\begin{aligned}
\chi^1_j(p) &= \frac1{\sqrt 2}(b_j(p) + b^\dagger_j(p))\,,\\
\chi^2_j(p) &= \frac i{\sqrt 2}(b_j(p) - b^\dagger_j(p))\,.
\end{aligned}
\end{equation}
It follows,
\begin{equation}
\sets{\chi^a_i(p), \chi^b_j(q)} = \delta_{ij}\delta_{pq}\delta^{ab}\,.
\end{equation}
We define the $\mathfrak{so}(2K)$ generators,
\begin{equation}
\mathcal L_{ab}(p,q) = \frac i2\sum_{j=1}^N\sq{\chi^a_j(p),\chi^b_j(q)}\,.
\end{equation}
Again we use the collective indices $(a,p)\mapsto P \equiv (a-1)K + p$, following which one may check,
\begin{equation}
[\mathcal L_{PQ}, \mathcal L_{RS}] = i\mathcal L_{PS}\delta_{QR} + i\mathcal L_{QR}\delta_{PS} -i\mathcal L_{PR}\delta_{QS} - i\mathcal L_{QS}\delta_{PR}\,.
\end{equation}
Explicitly in terms of the bilocals we have,
\begin{align}
\mathcal L_{pq} &= \frac i2\pts*{\bar A_{pq} + A_{pq} + B^{\rm W}_{pq} - B^{\rm W}_{qp}}\,,\\[.5\baselineskip]
\mathcal L_{p(K+q)} &= \frac12\pts*{\bar  A_{pq} - A_{pq} - B^{\rm W}_{pq} - B^{\rm W}_{qp}}\,,\\[.5\baselineskip]
\mathcal L_{(K+p)q} &= \frac12\pts*{\bar A_{pq} - A_{pq} + B^{\rm W}_{pq} + B^{\rm W}_{qp}}\,,\\[.5\baselineskip]
\mathcal L_{(K+p)(K+q)} &=- \frac i2\pts*{\bar A_{pq} + A_{pq} - B^{\rm W}_{pq} + B^{\rm W}_{qp}} \,.
\end{align}
The quadratic Casimir in terms of $\mathcal L$ is,
\begin{equation}
C_2 = -\Tr(\mathcal L^2) = 2\Tr((B^{\rm W})^2) + \Tr(\bar A A + A \bar A)\,.
\end{equation}

\paragraph{Relation to the generator matrix $\bm{\mathcal M}$.} 
The generator matrix $\mathcal M(\varepsilon,\xi)$ (\cref{eq:unifiedorthogonalsymplecticgenerator}) is related to the matrix of our original generators $\mathcal L$ defined above as follows,
\begin{equation}
	\mathcal L = iUV^{-1}\mathcal M V U^{-1}\,,
\end{equation}
where 
\begin{equation}
	U = \frac{1}{\sqrt 2}\begin{pmatrix}
		\mathbb{I}_{K\times K} & \mathbb{I}_{K\times K}\\
		-i\mathbb{I}_{K\times K} & i\mathbb{I}_{K\times K}
	\end{pmatrix}	\,,\qquad V = \begin{pmatrix}
		\mathbb{I}_{K\times K} & 0\\
		0 & -\xi \mathbb{I}_{K\times K}
	\end{pmatrix}\,.
\end{equation}
Consequently
\begin{equation}
	\tr(\mathcal L^n) = i^n 	\tr(\mathcal M^n)\,.
\end{equation}

\subsection{Highest-weight conventions}\label{subsec:hwc}
In this work, we adopt a convention for positive roots to deal with \emph{highest-weight} modules. We elaborate upon our choice using $O(N)$ bosons as an example.

\paragraph{$\bm{O(N)}$ bosons.}
We make the standard choice $\sets*{H_p \equiv \frac12\pts*{\mathcal L_{(K+p)p} - \mathcal L_{p(K+p)}} = B^{\rm W}_{pp}}_{p=1}^K$ for the basis for the Cartan subalgebra. Roots live in the space dual to the Cartan subalgebra, where we define an orthonormal basis $\sets*{e_p}^{K}_{p=1}$ such that 
\begin{equation}
\innerx{e_p}{e_q} = \delta_{pq},\qquad e_p(H_q) = \delta_{pq}\,.
\end{equation}
All the remaining bilocals are the root vectors,
\begin{align}
[H_p, B^{\rm W}_{qr}] &= \pts*{\delta_{pq} - \delta_{pr}}B^{\rm W}_{qr}\,,\\
[H_p, \bar A_{qr}] &= \pts*{\delta_{pq}+\delta_{pr}}\bar A_{qr}\,,\\
[H_p, A_{qr}] &= -\pts*{\delta_{pq}+\delta_{pr}}A_{qr}\,.
\end{align}
The normalization of the root vectors is as follows,
\begin{equation}
\begin{gathered}
E_{-e_p - e_q} = \frac{i}{\sqrt{1+\delta_{pq}}}A_{pq},\qquad E_{e_p + e_q} = \frac{i}{\sqrt{1 + \delta_{pq}}}\bar A_{pq}\quad (p \leq q)\,,\\
E_{e_p - e_q} = iB^{\rm W}_{pq},\qquad E_{-e_p + e_q} = -iB^{\rm W}_{qp}\quad (p<q)\,.
\end{gathered}
\end{equation}
With these normalizations we have,
\begin{equation}
[E_{\alpha}, E_{-\alpha}] = H_{\alpha} \,,
\end{equation}
where $H_{\alpha} = \sum_{p}\alpha(H_p)H_p$. To ensure that we deal with \emph{highest-weight} states, we consider $\alpha = -e_p - e_q$ ($p\leq q$) and $\alpha = e_p - e_q$ for $p<q$ as \emph{positive} roots and therefore 
\begin{equation}
\begin{aligned}
\frac{i}{\sqrt{1+\delta_{pq}}}A_{pq}\quad (p\leq q),\qquad iB^{\rm W}_{pq}\quad (p<q) \hskip0.02\textwidth &:\text{(positive root vectors)},\\
\frac{i}{\sqrt{1+\delta_{pq}}}\bar A_{pq}\quad (p\leq q),\qquad -iB^{\rm W}_{pq}\quad (p>q) \hskip0.02\textwidth &:\text{(negative root vectors)}.
\end{aligned}
\end{equation}
Making this choice affects the Weyl vector, given that it is defined as 
\begin{equation}
\rho = \frac12\sum_{\alpha\text{ is positive}} \alpha.
\end{equation}
Summing our positive roots gives,
\begin{equation}
\rho = \sum_{p=1}^K -p\, e_p\,,
\end{equation}
or 
\begin{equation}
\rho = (-1, -2, \ldots, -K)\,.
\end{equation}
This differs from the standard Weyl vector for the $\mathfrak{sp}(2K)$ algebra, and one must be cautious while trying to check the argument used in \cref{subsec:sectoridentification}. 

\paragraph{$\bm{O(N)}$ fermions.} We use $\sets*{H_p\equiv \mathcal L_{p(K+p)} = -B^{\rm W}_{pp}}_{p=1}^K$ as the basis of the Cartan subalgebra. For the same reason as earlier, we make the choice (up to some normalization)
\begin{equation}
\begin{aligned}
A_{pq}\quad (p< q),\qquad B^{\rm W}_{pq}\quad (p>q) \hskip0.02\textwidth &:\text{(positive root vectors)},\\
\bar A_{pq}\quad (p< q),\qquad B^{\rm W}_{pq}\quad (p<q) \hskip0.02\textwidth &:\text{(negative root vectors)}.
\end{aligned}\label{eq:rootgenerators}
\end{equation}

\paragraph{$\bm{U(N)}$ bosons.} We use $\sets*{H_p \equiv - L_{pp}}_{p=1}^{K}\cup \sets*{H_p \equiv L_{pp}}_{p=K+1}^{2K}$ as the basis of the Cartan subalgebra and choose, 
\begin{equation}
\begin{aligned}
A_{pq}\quad (\forall\;\; p,q),\quad B_{pq}\quad (p<q), \quad D_{pq} = C_{pq}\quad (p>q) \hskip0.01\textwidth &:\text{(positive root generators)},\\
\bar A_{pq}\quad (\forall\;\; p,q),\quad B_{pq}\quad (p>q), \quad D_{pq} = C_{pq}\quad (p<q) \hskip0.01\textwidth &:\text{(negative root generators)}.
\end{aligned}
\end{equation}

\paragraph{$\bm{U(N)}$ fermions.} We use  $\sets*{H_p \equiv L_{pp}}_{p=1}^{2K}$ as the basis for the Cartan subalgebra and choose, 
\begin{equation}
\begin{aligned}
A_{pq}\quad (\forall\;\; p,q),\quad B_{pq}\quad (p<q), \quad D_{pq} = -C_{pq}\quad (p>q) \hskip0.01\textwidth &:\text{(positive root generators)},\\
\bar A_{pq}\quad (\forall\;\; p,q),\quad B_{pq}\quad (p>q), \quad D_{pq} = -C_{pq}\quad (p<q) \hskip0.01\textwidth &:\text{(negative root generators)}.
\end{aligned}
\end{equation}
Since the positive roots in the first block are $D_{pq}$ with $p>q$, we list the weights of the first block of length $K$ in the order $(H_K,\ldots, H_1)$, which is the dominant order.

\section{Collective realization details}\label{appx:collective}

This appendix collects the computations behind \cref{sec:collectivenewbackground}. We treat the $O(N)$ bosons in detail and list the results directly for the other models. Throughout, $A$ and $\bar A$ denote the rescaled bilocals of \cref{sec:collectivenewbackground}, and repeated mode indices are summed.

\subsection{\texorpdfstring{$O(N)$}{O(N)} bosons}\label{appx:collONboson}

The ansatz
\begin{equation}
	\bar A_{pq} = \bar\alpha_{pq},\qquad B_{pq} = \bar\alpha_{pr}\alpha_{rq} + \bar\beta_{p}\beta_q\,,\label{eq:appxbarABansatzON}
\end{equation}
\begin{equation}
	A_{pq} = \alpha_{pq} + \frac{1}{N}\pts[\Big]{\bar\alpha_{rs}\alpha_{pr}\alpha_{sq} + \alpha_{pr}\bar\beta_{r}\beta_{q} + \alpha_{qr}\bar\beta_{r}\beta_{p}}\,,\label{eq:appxsolutionA}
\end{equation}
reproduces the rescaled bilocal algebra exactly, for any value of the fermion number $\bar\beta_t\beta_t$. Since $\bar\alpha_{pr}\alpha_{rq}$ and $\bar\beta_p\beta_q$ are commuting $\mathfrak{gl}(K)$ generators,
\begin{equation}
	[B_{pq}, B_{rs}] = B_{ps}\delta_{rq} - \delta_{ps}B_{rq}\,,\qquad
	[B_{pq}, \bar A_{rs}] = \bar A_{ps}\delta_{rq} + \bar A_{pr}\delta_{sq}\,,
\end{equation}
and since every term of \cref{eq:appxsolutionA} is a symmetric $GL(K)$ tensor of the same type as $\alpha_{pq}$,
\begin{equation}
	[A_{pq}, B_{rs}] = A_{ps}\delta_{rq} + A_{qs}\delta_{rp}\,.
\end{equation}
For $[A,\bar A]$ the term of order one is $[\alpha_{pq},\bar\alpha_{rs}]$, and at order $\frac1N$
\begin{equation}
	\sq*{\bar\alpha_{tu}\alpha_{pt}\alpha_{uq} + \alpha_{pt}\bar\beta_t\beta_q + \alpha_{qt}\bar\beta_t\beta_p\,,\, \bar\alpha_{rs}} = \delta_{pr}B_{sq} + \delta_{ps}B_{rq} + \delta_{qr}B_{sp} + \delta_{qs}B_{rp}\,,
\end{equation}
which reproduces \cref{eq:SOnB-modifiedcommutation}. A longer computation of the same kind gives $[A_{pq}, A_{rs}] = 0$.

\paragraph{Residual of the constraint.} With the ansatz, we find
\begin{align}
	\bar A_{pr}A_{rq} &= \bar\alpha_{pr}\alpha_{rq} + \frac1N\pts*{\bar\alpha_{pr}\bar\alpha_{st}\alpha_{rs}\alpha_{tq} + \bar\alpha_{pr}\alpha_{rs}\bar\beta_s\beta_q + \bar\alpha_{pr}\alpha_{qs}\bar\beta_s\beta_r}\,,\\
	B_{pr}B_{rq} &= \bar\alpha_{ps}\bar\alpha_{rt}\alpha_{sr}\alpha_{tq} + (K+1)\bar\alpha_{ps}\alpha_{sq} + \bar\alpha_{ps}\alpha_{sr}\bar\beta_r\beta_q + \bar\beta_p\beta_r\bar\alpha_{rt}\alpha_{tq} + \pts*{K+1-\bar\beta_t\beta_t}\bar\beta_p\beta_q\,,
\end{align}
where we used $[\alpha_{sr},\bar\alpha_{rt}] = (K+1)\delta_{st}$ and $\bar\beta_p\beta_r\bar\beta_r\beta_q = \pts*{K+1-\bar\beta_t\beta_t}\bar\beta_p\beta_q$. In the rescaled constraint the terms quartic in $\alpha$, the terms $(K+1)\bar\alpha_{ps}\alpha_{sq}$ and the terms $\bar\alpha_{pr}\alpha_{rs}\bar\beta_s\beta_q$ cancel, which leaves \cref{eq:constraint-remainder}. 
\begin{equation}
	\frac1N\pts*{\bar\alpha_{pr}\alpha_{qs}\bar\beta_s\beta_r - \bar\alpha_{rs}\alpha_{sq}\bar\beta_p\beta_r} = -\frac1N\mathcal N_{ps}\alpha_{sq}\,,
\end{equation}
serves as the defining relation for the null operator $\mathcal N_{ps}$. The remaining terms combine into $\frac1N\pts*{\bar\beta_t\beta_t - N}\bar\beta_p\beta_q$ and finally we obtain \cref{eq:constraintcompact}.

\paragraph{Algebra of the null operator.} The operator $\mathcal N_{pq}$  contains only $\bar\alpha$ and $\bar\beta\beta$, so that $[\bar A_{pq},\mathcal N_{rs}]=0$ and
\begin{equation}
	[B_{pq}, \mathcal N_{rs}] = \mathcal N_{ps}\delta_{qr} - \mathcal N_{pr}\delta_{qs}\,.
\end{equation}
For $A$ we expand
\begin{equation}
	[A_{pq}, \mathcal N_{rs}] = [\alpha_{pq}, \bar\alpha_{ts}\bar\beta_r\beta_t] + \frac1N[\bar\alpha_{uv}\alpha_{pu}\alpha_{vq} + \alpha_{pu}\bar\beta_{u}\beta_{q} + \alpha_{qu}\bar\beta_{u}\beta_{p}, \bar\alpha_{ts}\bar\beta_r\beta_t] - (r\leftrightarrow s)\,.
\end{equation}
The first commutator gives $\delta_{ps}\bar\beta_r\beta_q + \delta_{qs}\bar\beta_r\beta_p - (r\leftrightarrow s)$. In the terms of order $\frac1N$, the commutators $[\bar\beta_u\beta_q,\bar\beta_r\beta_t]$ produce the same combination multiplied by $-\frac1N\bar\beta_t\beta_t$, and everything else assembles into products $\mathcal N\alpha$. The result is
\begin{equation}
	\begin{split}
		[A_{pq}, \mathcal N_{rs}] &= \frac{1}{N}\sq[\Big]{\delta_{ps}\mathcal N_{rt}\alpha_{tq} + \delta_{qs}\mathcal N_{rt}\alpha_{tp} - \delta_{qr}\mathcal N_{st}\alpha_{tp} - \delta_{pr}\mathcal N_{st}\alpha_{tq}}\\
		&\quad+\pts*{1-\frac{\bar\beta_t\beta_t}{N}}\pts*{\delta_{ps}\bar\beta_r\beta_q + \delta_{qs}\bar\beta_r\beta_p - \delta_{pr}\bar\beta_s\beta_q - \delta_{qr}\bar\beta_s\beta_p}\,.
	\end{split}\label{eq:appB-ANgeneral}
\end{equation}
The second line vanishes for $\bar\beta_t\beta_t=N$, and also for $\bar\beta_t\beta_t=0$, where $\bar\beta_r\beta_q$ annihilates the state. In both sectors \cref{eq:appB-ANgeneral} therefore reduces to \cref{eq:ACcommutator}. For two null operators,
\begin{align}
	[\mathcal N_{pq}, \mathcal N_{rs}] &= \pts[\Big]{\bar\alpha_{tq}\bar\alpha_{us}[\bar\beta_p\beta_t, \bar\beta_r\beta_u] - (r\leftrightarrow s)} - (p\leftrightarrow q)\,,\\
	&= \pts[\Big]{\bar\alpha_{rq}\bar\alpha_{ts}\bar\beta_p\beta_t - \bar\alpha_{tq}\bar\alpha_{ps}\bar\beta_r\beta_t - \bar\alpha_{sq}\bar\alpha_{tr}\bar\beta_p\beta_t + \bar\alpha_{tq}\bar\alpha_{pr}\bar\beta_s\beta_t} - (p\leftrightarrow q)\,,\\
	&= \bar A_{qr}\mathcal N_{ps} - \bar A_{qs}\mathcal N_{pr} - \bar A_{pr}\mathcal N_{qs} + \bar A_{ps}\mathcal N_{qr}\,.
\end{align}
The operator $\bar{\mathcal N}_{pq}=\alpha_{pr}\bar\beta_r\beta_q-\alpha_{qr}\bar\beta_r\beta_p$, which appears in the adjoint rules below, satisfies
\begin{align}
	[\bar A_{pq}, \bar{\mathcal N}_{rs}] &= -\delta_{pr}\bar\beta_q\beta_s - \delta_{qr}\bar\beta_p\beta_s + \delta_{ps}\bar\beta_q\beta_r + \delta_{qs}\bar\beta_p\beta_r\,,\\
	[B_{pq}, \bar{\mathcal N}_{rs}] &= \delta_{ps}\bar{\mathcal N}_{qr} - \delta_{pr}\bar{\mathcal N}_{qs}\,.
\end{align}

\paragraph{Null states at the first level.} Let $\ketv\psi$ be a state of the epsilon sector annihilated by every $\alpha_{pq}$, \emph{i.e.}\ a state of $N$ auxiliary fermions without bilocal quanta. On such a state $B_{pr}\ketv\psi=\bar\beta_p\beta_r\ketv\psi$ and $\bar A_{pq}\ketv\psi=\bar\alpha_{pq}\ketv\psi$, so that
\begin{equation}
	\mathcal N_{pq}\ketv\psi = \pts*{\bar A_{rq}B_{pr} - \bar A_{pq}}\ketv\psi - (p\leftrightarrow q)\,.\label{eq:appB-levelone}
\end{equation}
The null states at the first level are thus the antisymmetrized differences of the two sides of \cref{eq:specialONB}. At the first level the auxiliary states $\bar\alpha_{pq}\ketv\psi$ form $\mathrm{Sym}^2\mathbb C^K\otimes\wedge^N\mathbb C^K$, and for $K>N$ the null states span a non-zero $GL(K)$-submodule of it, contained in the summand $S_{(2,1^N)}(\mathbb C^K)$ that is absent from $\cH_-$ (\cref{subsec:fluctuations}). Since that summand is irreducible, the two coincide: at the first level the quotient by the image of $\mathcal N$ removes exactly the states forbidden by $GL(N)\times GL(K)$ duality.

\paragraph{Inversion and adjoint at order $\bm{1/N}$.} Because the realization is non-unitary, the adjoint of $\alpha_{pq}$ is not $\bar \alpha_{qp}$. We express the auxiliary oscillators through the bilocals and compute the adjoint order by order in $\frac1N$. From \cref{eq:appxsolutionA}, with $\bar\beta_r\beta_q=B_{rq}-\bar\alpha_{rs}\alpha_{sq}$ and $[\alpha_{pr}, B_{rq}]=(K+1)\alpha_{pq}$,
\begin{align}
	\alpha_{pq} &= A_{pq} - \frac1N\pts*{\bar\alpha_{rs}\alpha_{pr}\alpha_{sq} + \alpha_{pr}\bar\beta_r\beta_q + \alpha_{qr}\bar\beta_r\beta_p}\,,\\
	&= A_{pq} + \frac1N\pts*{\bar \alpha_{rs}\alpha_{pr}\alpha_{sq} - B_{rq}\alpha_{pr} - B_{rp}\alpha_{qr}}\,,
\end{align}
and replacing $\alpha$ by $A$ on the RHS,
\begin{equation}
	\alpha_{pq} = A_{pq} + \frac{1}{N}\pts*{\bar A_{rs}A_{pr}A_{sq} - B_{rq}A_{pr} - B_{rp}A_{qr}} + \mathcal O\pts*{\frac{1}{N^2}}\,.\label{eq:appB-alphainverse}
\end{equation}
Then $\bar\beta_p\beta_q = B_{pq} - \bar A_{pr}\alpha_{rq}$ gives
\begin{equation}
	\bar\beta_{p}\beta_{q} = B_{pq} - \bar A_{pr}A_{rq} - \frac1N\pts*{\bar A_{pr}\bar A_{st}A_{rs}A_{tq} - \bar A_{pr}B_{sq}A_{rs} - \bar A_{pr}B_{sr}A_{qs}} + \mathcal O\pts*{\frac{1}{N^2}}\,.\label{eq:appB-betainverse}
\end{equation}
The adjoint is fixed by the microscopic theory, $\pts*{A_{pq}}^\dagger=\bar A_{pq}$ and $\pts*{B_{pq}}^\dagger=B_{qp}$. Hence $\pts*{\bar\alpha_{pq}}^\dagger = A_{pq}$ is given by \cref{eq:appxbarABansatzON}, and from \cref{eq:appB-alphainverse},
\begin{align}
	\pts*{\alpha_{pq}}^\dagger &= \bar A_{pq} + \frac1N\pts*{\bar A_{pr}\bar A_{sq}A_{rs} - \bar A_{pr}B_{qr} - \bar A_{qr}B_{pr}} + \mathcal O\pts*{\frac{1}{N^2}}\,,\\
	&= \bar\alpha_{pq} - \frac1N\pts*{\bar\alpha_{pr}\bar\alpha_{sq}\alpha_{rs} + \bar\alpha_{pr}\bar\beta_q\beta_r + \bar\alpha_{qr}\bar\beta_p\beta_r} + \mathcal{O}\pts*{\frac1{N^2}}\,.\label{eq:appB-alphaadjoint}
\end{align}
In the same way, the adjoint of \cref{eq:appB-betainverse} is
\begin{equation}
	\pts*{\bar\beta_p\beta_q}^\dagger = \bar\beta_q\beta_p - \frac1N\pts*{\bar\alpha_{qr}\alpha_{rs}\bar\beta_s\beta_p - \bar\alpha_{rs}\bar\beta_q\beta_s\alpha_{pr}} + \mathcal{O}\pts*{\frac1{N^2}}\,.
\end{equation}
Since $\bar\alpha_{rs}\bar\beta_q\beta_s = \bar\alpha_{sq}\bar\beta_r\beta_s + \mathcal N_{qr}$ by \cref{eq:SOnBnulloperator}, this is the same as
\begin{equation}
	\pts*{\bar\beta_p\beta_q}^\dagger = \bar\beta_q\beta_p -\frac1N\bar\alpha_{qr}\bar{\mathcal N}_{rp} + \frac1N\mathcal N_{qr}\alpha_{rp} + \mathcal O\pts*{\frac1{N^2}}\,.\label{eq:appB-betaadjoint}
\end{equation}
The term $\mathcal N_{qr}\alpha_{rp}$ maps every state into the image of $\mathcal N$ and drops on the quotient, while $\bar{\mathcal N}$ annihilates the background, where $\alpha$ annihilates the state. Thus $\bar\beta\beta$ is self-adjoint up to terms of order $\frac1N$ that vanish on the background.

In the vacuum sector the auxiliary fermions are absent, and the ansatz can be resummed exactly,
\begin{equation}
	A_{pq} = \alpha_{pq}+\frac1N \bar{\alpha}_{rs}\alpha_{pr} \alpha_{sq} = \alpha_{pr}\sq*{\pts*{1-\frac{K+1}{N}}\delta_{rq} + \frac1N B_{rq}}\,.
\end{equation}
Taking the adjoint and inverting the matrix in brackets wherever it is invertible,
\begin{equation}
	\pts*{\alpha_{pq}}^\dagger=\pts*{1-\frac{K+1}{N}}^{-1}\sq*{\mathbb I+\frac{1}{N-K-1}B}^{-1}_{pr}\bar\alpha_{rq}\,.
\end{equation}
Expanded to first order in $\frac1N$, this reproduces \cref{eq:appB-alphaadjoint} at $\bar\beta\beta=0$, because $B_{pr}\bar\alpha_{rq}=\bar\alpha_{pr}\bar\alpha_{sq}\alpha_{rs}+(K+1)\bar\alpha_{pq}$.

\subsection{\texorpdfstring{$O(N)$}{O(N)} fermions}\label{appx:collONFermions}
The realization of \cref{subsec:collONFermions} works in the same way, with $\alpha_{pq}=-\alpha_{qp}$ and with bosonic auxiliary oscillators $\beta_p$. The ansatz reproduces the rescaled algebra quoted there, together with
\begin{equation}
	[B_{pq}, \bar A_{rs}] = \bar A_{ps}\delta_{rq} - \bar A_{pr}\delta_{sq}\,,\qquad [A_{pq}, B_{rs}] = A_{ps}\delta_{rq} - A_{qs}\delta_{rp}\,,\qquad [A_{pq}, A_{rs}]=0\,.
\end{equation}
Without imposing the boson number, the residual of the rescaled constraint is
\begin{equation}
	\bar A_{pr}A_{rq} - B_{pq} -\frac{1}{N}\pts*{(K-1)B_{pq} - B_{pr}B_{rq}} = \frac{1}{N}\mathcal N_{pr}\alpha_{rq} + \frac1N\pts*{\bar\beta_t\beta_t-N}\bar\beta_p\beta_q\,,\label{eq:appB-ONFresidual}
\end{equation}
with $\mathcal N_{pq} = \bar\beta_{p}\beta_r\bar\alpha_{rq} + \bar\beta_{q}\beta_{r}\bar\alpha_{rp}$. As for bosons, the last term vanishes only for $\bar\beta_t\beta_t=0$ and $\bar\beta_t\beta_t=N$, so the constraint itself selects the two sectors. The null operator now satisfies $[\bar A_{pq},\mathcal N_{rs}]=0$,
\begin{equation}
	[B_{pq}, \mathcal N_{rs}] = \mathcal N_{ps}\delta_{qr} + \mathcal N_{pr}\delta_{qs}\,,\qquad
	[\mathcal N_{pq}, \mathcal N_{rs}] = -\bar A_{pr}\mathcal N_{qs} - \bar A_{ps}\mathcal N_{qr} - \bar A_{qr}\mathcal N_{ps} - \bar A_{qs}\mathcal N_{pr}\,,
\end{equation}
and
\begin{equation}
	\begin{split}
		[A_{pq}, \mathcal N_{rs}] &= \frac{1}{N}\sq[\Big]{\delta_{qr}\mathcal N_{st}\alpha_{tp} + \delta_{qs}\mathcal N_{rt}\alpha_{tp} - \delta_{pr}\mathcal N_{st}\alpha_{tq} - \delta_{ps}\mathcal N_{rt}\alpha_{tq}}\\
		&\quad+\pts*{1-\frac{\bar\beta_t\beta_t}{N}}\pts*{\delta_{ps}\bar\beta_r\beta_q - \delta_{qs}\bar\beta_r\beta_p + \delta_{pr}\bar\beta_s\beta_q - \delta_{qr}\bar\beta_s\beta_p}\,,
	\end{split}
\end{equation}
whose second line again vanishes in both sectors. On a state of the odd sector annihilated by every $\alpha_{pq}$,
\begin{equation}
	\mathcal N_{pq}\ketv\psi = \pts*{\bar A_{rq}B_{pr} + \bar A_{pq}}\ketv\psi + (p\leftrightarrow q)\,,
\end{equation}
which is the symmetrized form of \cref{eq:specialONB} with $\xi=-1$.

\paragraph{Inversion and adjoint at order $\bm{1/N}$.} Now $[\alpha_{pr}, B_{rq}]=(K-1)\alpha_{pq}$, and the steps of \cref{appx:collONboson} give
\begin{align}
	\alpha_{pq} &= A_{pq} - \frac{1}{N}\pts*{\bar A_{rs}A_{pr}A_{sq} - B_{rq}A_{pr} + B_{rp}A_{qr}} + \mathcal O\pts*{\frac{1}{N^2}}\,,\\
	\bar\beta_{p}\beta_{q} &= B_{pq} - \bar A_{pr}A_{rq} + \frac1N\pts*{\bar A_{pr}\bar A_{st}A_{rs}A_{tq} - \bar A_{pr}B_{sq}A_{rs} + \bar A_{pr}B_{sr}A_{qs}} + \mathcal O\pts*{\frac{1}{N^2}}\,.
\end{align}
For fermions the microscopic adjoint is $\pts*{A_{pq}}^\dagger=\bar A_{qp}=-\bar A_{pq}$ and $\pts*{B_{pq}}^\dagger=B_{qp}$, so that $\pts*{\bar\alpha_{pq}}^\dagger = A_{qp}$ and
\begin{align}
	\pts*{\alpha_{pq}}^\dagger &= \bar\alpha_{qp} - \frac1N\pts*{\bar\alpha_{rq}\bar\alpha_{pt}\alpha_{tr} - \bar\alpha_{rp}\bar\beta_q\beta_r + \bar\alpha_{rq}\bar\beta_p\beta_r} + \mathcal{O}\pts*{\frac1{N^2}}\,,\\
	\pts*{\bar\beta_p\beta_q}^\dagger &= \bar\beta_q\beta_p + \frac1N\pts*{\bar\alpha_{qr}\alpha_{rs}\bar\beta_s\beta_p - \bar\alpha_{sr}\alpha_{rp}\bar\beta_q\beta_s} + \mathcal{O}\pts*{\frac1{N^2}}\,.
\end{align}
With $\bar{\mathcal N}_{pq}=\alpha_{pr}\bar\beta_r\beta_q+\alpha_{qr}\bar\beta_r\beta_p$, the conjugate of the symmetric null operator, the second relation is the same as
\begin{equation}
	\pts*{\bar\beta_p\beta_q}^\dagger = \bar\beta_q\beta_p + \frac1N\bar\alpha_{qr}\bar{\mathcal N}_{rp} - \frac1N\mathcal N_{qr}\alpha_{rp} + \mathcal O\pts*{\frac1{N^2}}\,,
\end{equation}
and the term $\mathcal N_{qr}\alpha_{rp}$ again drops on the quotient. In the vacuum sector,
\begin{gather}
	A_{pq} = \alpha_{pr}\sq*{\pts*{1+\frac{K-1}{N}}\delta_{rq} - \frac1N B_{rq}}\,,\\
	\pts*{\alpha_{pq}}^\dagger=\pts*{1+\frac{K-1}{N}}^{-1}\sq*{\mathbb I-\frac{1}{N+K-1}B}^{-1}_{qr}\bar\alpha_{rp}\,,
\end{gather}
exactly, wherever the matrix in brackets is invertible.

\subsection{\texorpdfstring{$U(N)$}{U(N)} bosons}\label{appx:collUNBosons}
With $[\alpha_{pq},\bar\alpha_{rs}]=\delta_{ps}\delta_{rq}$, the ansatz of \cref{subsec:collUNboson} reproduces the rescaled algebra,
\begin{equation}
	[A_{pq}, \bar A_{rs}] = \delta_{ps}\delta_{rq} + \frac1N\pts*{C_{ps}\delta_{rq} + B_{rq}\delta_{ps}}\,,\qquad [A_{pq}, A_{rs}] = [B_{pq}, C_{rs}] = 0\,,
\end{equation}
together with the remaining relations of \cref{subsec:bilocalalgebras} with $\xi=1$. At order $\frac1N$, the term $\alpha_{pr}\bar\beta_r\beta_q$ of $A_{pq}$ supplies the $\bar\beta\beta$ part of $B$ in $[A,\bar A]$, and $\alpha_{rq}\bar\gamma_r\gamma_p$ supplies the $\bar\gamma\gamma$ part of $C$. On states obeying the flavor constraints
\begin{equation}
	\sum_{p}\bar\beta^k_p\beta^l_p = N\delta^{kl},\qquad \sum_{p}\bar\gamma^k_p\gamma^l_p = N\delta^{kl}
\end{equation}
substituting the ansatz into the rescaled constraints give
\begin{gather}
	\bar A_{pr}A_{rq} - B_{pq} - \frac{1}{N}\pts*{B_{pr}B_{rq} - (K+n_\beta - n_\gamma)B_{pq}} = -\frac{1}{N}\mathcal N_{pr}\alpha_{rq} - \frac{n_\gamma}{N}\bar\beta_p \beta_q\,,\\
	A_{pr}\bar A_{rq} - (K+n_\beta-n_\gamma)\delta_{pq} - C_{pq} - \frac{1}{N}\pts*{C_{pr}C_{rq} + (K+n_\beta-n_\gamma)C_{pq}} = \frac{1}{N}\mathcal N_{rq}\alpha_{pr} - \frac{n_\beta}{N}\bar\gamma_q \gamma_p\,,
\end{gather}
with 
\begin{equation}
	\mathcal N_{pq} = \bar\beta_p\beta_r\bar\alpha_{rq} - \bar\gamma_q\gamma_r\bar\alpha_{pr} - (n_\beta - n_\gamma)\bar\alpha_{pq}\,.
\end{equation}
The flavor constraint is needed to simplify the terms quartic in the auxiliary oscillators, and the residuals do not hold without it.

For a single kind of auxiliary fermion, $n_\beta n_\gamma=0$, and on states obeying the flavor constraint, the null operator satisfies $[\bar A_{pq},\mathcal N_{rs}]=0$ and
\begin{gather}
	[B_{pq}, \mathcal N_{rs}] = \mathcal N_{ps}\delta_{rq}\,,\qquad [C_{pq}, \mathcal N_{rs}] = \mathcal N_{rq}\delta_{ps}\,,\\
	[A_{pq}, \mathcal N_{rs}] = \frac1N\pts*{\delta_{ps}\mathcal N_{rt}\alpha_{tq} + \delta_{qr}\mathcal N_{ts}\alpha_{pt}}\,,\qquad
	[\mathcal N_{pq}, \mathcal N_{rs}] = \bar A_{rq}\mathcal N_{ps} - \bar A_{ps}\mathcal N_{rq}\,.
\end{gather}
The image of $\mathcal N$ is therefore a submodule, as stated in \cref{subsec:nullquotient}. On a state annihilated by every $\alpha_{pq}$, the definition of $\mathcal N$ gives
\begin{equation}
\begin{aligned}
	\mathcal N_{pq}\ketv\psi &= \pts*{\bar A_{rq}B_{pr} - n\bar A_{pq}}\ketv\psi \quad (n>0)\,,\\
	\mathcal N_{pq}\ketv\psi &= -\pts*{\bar A_{pr}C_{rq} - \absv n\bar A_{pq}}\ketv\psi \quad (n<0)\,.
\end{aligned}
\end{equation}
So the null states at the first level are the differences of the two sides of \cref{eq:specialUNB}.

\paragraph{Inversion and adjoint at order $\bm{1/N}$.} Now $[\alpha_{pr}, B_{rq}]=[\alpha_{rq}, C_{pr}]=K\alpha_{pq}$. With $\bar\beta_r\beta_q=B_{rq}-\bar\alpha_{rs}\alpha_{sq}$ and $\bar\gamma_r\gamma_p=C_{pr}-\bar\alpha_{sr}\alpha_{ps}$, the steps of \cref{appx:collONboson} give
\begin{align}
	\alpha_{pq} &= A_{pq} + \frac{1}{N}\pts*{\bar A_{rs}A_{pr}A_{sq} - B_{rq}A_{pr} - C_{pr}A_{rq}} + \mathcal O\pts*{\frac{1}{N^2}}\,,\\
	\bar\beta_{p}\beta_{q} &= B_{pq} - \bar A_{pr}A_{rq} - \frac1N\bar A_{pr}\pts*{\bar A_{st}A_{rs}A_{tq} - B_{sq}A_{rs} - C_{rs}A_{sq}} + \mathcal O\pts*{\frac{1}{N^2}}\,,\\
	\bar\gamma_{q}\gamma_{p} &= C_{pq} - \bar A_{rq}A_{pr} - \frac1N\bar A_{rq}\pts*{\bar A_{st}A_{ps}A_{tr} - B_{sr}A_{ps} - C_{ps}A_{sr}} + \mathcal O\pts*{\frac{1}{N^2}}\,.
\end{align}
The microscopic adjoint is $\pts*{A_{pq}}^\dagger=\bar A_{qp}$, $\pts*{B_{pq}}^\dagger=B_{qp}$ and $\pts*{C_{pq}}^\dagger=C_{qp}$, so that $\pts*{\bar\alpha_{pq}}^\dagger = A_{qp}$ and
\begin{align}
	\pts*{\alpha_{pq}}^\dagger &= \bar\alpha_{qp} - \frac1N\pts*{\bar\alpha_{qr}\bar\alpha_{tp}\alpha_{rt} + \bar\alpha_{rp}\bar\beta_q\beta_r + \bar\alpha_{qr}\bar\gamma_p\gamma_r} + \mathcal{O}\pts*{\frac1{N^2}}\,,\\
	\pts*{\bar\beta_p\beta_q}^\dagger &= \bar\beta_q\beta_p - \frac1N\pts*{\bar\alpha_{qr}\alpha_{rs}\bar\beta_s\beta_p - \bar\alpha_{sr}\alpha_{rp}\bar\beta_q\beta_s} + \mathcal{O}\pts*{\frac1{N^2}}\,,\\
	\pts*{\bar\gamma_q\gamma_p}^\dagger &= \bar\gamma_p\gamma_q - \frac1N\pts*{\bar\alpha_{rp}\alpha_{sr}\bar\gamma_s\gamma_q - \bar\alpha_{rs}\alpha_{qr}\bar\gamma_p\gamma_s} + \mathcal{O}\pts*{\frac1{N^2}}\,.
\end{align}
In terms of the null operator and of its conjugate
\begin{equation}
	\bar{\mathcal N}_{pq} = \alpha_{pr}\bar\beta_r\beta_q - \alpha_{rq}\bar\gamma_r\gamma_p - (n_\beta - n_\gamma)\alpha_{pq}\,,
\end{equation}
the last two relations are the same as
\begin{align}
	\pts*{\bar\beta_p\beta_q}^\dagger &= \bar\beta_q\beta_p - \frac1N\bar\alpha_{qr}\bar{\mathcal N}_{rp} + \frac1N\mathcal N_{qr}\alpha_{rp} + \mathcal O\pts*{\frac1{N^2}}\,,\\
	\pts*{\bar\gamma_q\gamma_p}^\dagger &= \bar\gamma_p\gamma_q + \frac1N\bar\alpha_{rp}\bar{\mathcal N}_{qr} - \frac1N\mathcal N_{rp}\alpha_{qr} + \mathcal O\pts*{\frac1{N^2}}\,,
\end{align}
and as before the terms $\mathcal N\alpha$ drop on the quotient, while $\bar{\mathcal N}$ annihilates the background. In the vacuum sector,
\begin{gather}
	A_{pq} = \alpha_{pr}\sq*{\pts*{1-\frac{K}{N}}\delta_{rq} + \frac1N B_{rq}}\,,\\
	\pts*{\alpha_{pq}}^\dagger=\pts*{1-\frac{K}{N}}^{-1}\sq*{\mathbb I+\frac{1}{N-K}B}^{-1}_{qr}\bar\alpha_{rp}\,.
\end{gather}

\subsection{\texorpdfstring{$U(N)$}{U(N)} fermions}\label{appx:collUNFermions}
For $U(N)$ fermions the auxiliary oscillators are bosons, and the signs of the $\frac1N$ terms change. The ansatz of \cref{subsec:collUNfermion} reproduces
\begin{equation}
	[A_{pq}, \bar A_{rs}] = \delta_{ps}\delta_{rq} - \frac1N\pts*{C_{ps}\delta_{rq} + B_{rq}\delta_{ps}}\,,
\end{equation}
and, on states obeying 
\begin{equation}
	\sum_{p}\bar\beta^k_p\beta^l_p = N\delta^{kl},\qquad \sum_{p}\bar\gamma^k_p\gamma^l_p = N\delta^{kl}
\end{equation}
we obtain the residuals 
\begin{gather}
	\bar A_{pr}A_{rq} - B_{pq} + \frac{1}{N}\pts*{B_{pr}B_{rq} - (K-n_\beta + n_\gamma)B_{pq}} = \frac{1}{N}\mathcal N_{pr}\alpha_{rq} - \frac{n_\gamma}{N}\bar\beta_p \beta_q\,,\\
	A_{pr}\bar A_{rq} - (K-n_\beta+n_\gamma)\delta_{pq} - C_{pq} + \frac{1}{N}\pts*{C_{pr}C_{rq} + (K-n_\beta+n_\gamma)C_{pq}} = -\frac{1}{N}\mathcal N_{rq}\alpha_{pr} - \frac{n_\beta}{N}\bar\gamma_q \gamma_p\,,
\end{gather}
with 
\begin{equation}
	\mathcal N_{pq} = \bar\beta_p\beta_r\bar\alpha_{rq} - \bar\gamma_q\gamma_r\bar\alpha_{pr} + (n_\beta - n_\gamma)\bar\alpha_{pq}\,.
\end{equation}
The null operator satisfies the same relations as for bosons, except
\begin{equation}
	[A_{pq}, \mathcal N_{rs}] = -\frac1N\pts*{\delta_{ps}\mathcal N_{rt}\alpha_{tq} + \delta_{qr}\mathcal N_{ts}\alpha_{pt}}\,,
\end{equation}
and on a state annihilated by every $\alpha_{pq}$,
\begin{equation}
\begin{aligned}
	\mathcal N_{pq}\ketv\psi &= \pts*{\bar A_{rq}B_{pr} + n\bar A_{pq}}\ketv\psi \quad (n>0)\,,\\
	\mathcal N_{pq}\ketv\psi &= -\pts*{\bar A_{pr}C_{rq} + \absv n\bar A_{pq}}\ketv\psi \quad (n<0)\,,
\end{aligned}
\end{equation}
in agreement with \cref{eq:specialUNB} for $\xi=-1$.

\paragraph{Inversion and adjoint at order $\bm{1/N}$.} The commutators $[\alpha_{pr}, B_{rq}]=[\alpha_{rq}, C_{pr}]=K\alpha_{pq}$ are as for bosons, and every $\frac1N$ term changes sign:
\begin{align}
	\alpha_{pq} &= A_{pq} - \frac{1}{N}\pts*{\bar A_{rs}A_{pr}A_{sq} - B_{rq}A_{pr} - C_{pr}A_{rq}} + \mathcal O\pts*{\frac{1}{N^2}}\,,\\
	\bar\beta_{p}\beta_{q} &= B_{pq} - \bar A_{pr}A_{rq} + \frac1N\bar A_{pr}\pts*{\bar A_{st}A_{rs}A_{tq} - B_{sq}A_{rs} - C_{rs}A_{sq}} + \mathcal O\pts*{\frac{1}{N^2}}\,,\\
	\bar\gamma_{q}\gamma_{p} &= C_{pq} - \bar A_{rq}A_{pr} + \frac1N\bar A_{rq}\pts*{\bar A_{st}A_{ps}A_{tr} - B_{sr}A_{ps} - C_{ps}A_{sr}} + \mathcal O\pts*{\frac{1}{N^2}}\,.
\end{align}
The microscopic adjoint takes the same form as for bosons, and
\begin{align}
	\pts*{\alpha_{pq}}^\dagger &= \bar\alpha_{qp} + \frac1N\pts*{\bar\alpha_{qr}\bar\alpha_{tp}\alpha_{rt} + \bar\alpha_{rp}\bar\beta_q\beta_r + \bar\alpha_{qr}\bar\gamma_p\gamma_r} + \mathcal{O}\pts*{\frac1{N^2}}\,,\\
	\pts*{\bar\beta_p\beta_q}^\dagger &= \bar\beta_q\beta_p + \frac1N\bar\alpha_{qr}\bar{\mathcal N}_{rp} - \frac1N\mathcal N_{qr}\alpha_{rp} + \mathcal O\pts*{\frac1{N^2}}\,,\\
	\pts*{\bar\gamma_q\gamma_p}^\dagger &= \bar\gamma_p\gamma_q - \frac1N\bar\alpha_{rp}\bar{\mathcal N}_{qr} + \frac1N\mathcal N_{rp}\alpha_{qr} + \mathcal O\pts*{\frac1{N^2}}\,,
\end{align}
now with $\bar{\mathcal N}_{pq} = \alpha_{pr}\bar\beta_r\beta_q - \alpha_{rq}\bar\gamma_r\gamma_p + (n_\beta - n_\gamma)\alpha_{pq}$, the conjugate of the fermionic null operator. In the vacuum sector,
\begin{gather}
	A_{pq} = \alpha_{pr}\sq*{\pts*{1+\frac{K}{N}}\delta_{rq} - \frac1N B_{rq}}\,,\\
	\pts*{\alpha_{pq}}^\dagger=\pts*{1+\frac{K}{N}}^{-1}\sq*{\mathbb I-\frac{1}{N+K}B}^{-1}_{qr}\bar\alpha_{rp}\,.
\end{gather}

\bibliographystyle{JHEP}

\bibliography{newbackground}

\end{document}